\documentclass[twocolumn,prd,nofootinbib,showpacs,preprintnumbers]{revtex4}
\usepackage{graphicx}
\usepackage{color}
\usepackage{amsmath,amssymb,amsfonts,dsfont}
\usepackage{multirow}
\usepackage{sistyle}
\usepackage{ulem}

\pdfoutput=1 

\newcommand{\lsim}{\mathrel{\mathop{\kern 0pt \rlap
  {\raise.2ex\hbox{$<$}}}
  \lower.9ex\hbox{\kern-.190em $\sim$}}}
\newcommand{\gsim}{\mathrel{\mathop{\kern 0pt \rlap
  {\raise.2ex\hbox{$>$}}}
  \lower.9ex\hbox{\kern-.190em $\sim$}}}

\def \aap  {A\&A}

\def \apjs  {ApJS}
\def \apjl  {ApJL}

\begin{document}


\title{Multimessenger constraints on the dark matter interpretation of the {\it Fermi}-LAT Galactic center excess}

\author{Mattia Di Mauro}
\affiliation{Istituto Nazionale di Fisica Nucleare, via P. Giuria, 1, 10125 Torino, Italy}\email{dimauro.mattia@gmail.com}
\author{Martin Wolfgang Winkler}
\affiliation{Stockholm University and The Oskar Klein Centre for Cosmoparticle Physics, Alba Nova, 10691 Stockholm, Sweden}\email{martin.winkler@su.se}

\begin{abstract}
An excess of $\gamma$ rays in the data measured by the {\it Fermi} Large Area Telescope in the direction of the Galactic center has been reported in several publications. This excess, labeled as the Galactic center excess (GCE), is detected analyzing the data with different interstellar emission models, point source catalogs and analysis techniques.
The characteristics of the GCE, recently measured with unprecedented precision, are all compatible with dark matter particles (DM) annihilating in the main halo of our Galaxy, even if other interpretations are still not excluded.
We investigate the DM candidates that fit the observed GCE spectrum and spatial morphology. We assume a simple scenario with DM annihilating into a single channel but we inspect also more complicated models with two and three channels.
We perform a search for a $\gamma$-ray flux from a list of 48 Milky Way dwarf spheroidal galaxies (dSphs) using state-of-the-art estimation of the DM density in these objects. Since we do not find any significant signal from the dSphs, we put upper limits on the annihilation cross section that result to be compatible with the DM candidate that fits the GCE.
However, we find that the GCE DM signal is excluded by the AMS-02 $\bar{p}$ flux data for all hadronic and semi-hadronic annihilation channels unless the vertical size of the diffusion halo is smaller than 2 kpc -- which is in tension with radioactive cosmic ray fluxes and radio data. Furthermore, AMS-02 $e^+$ data rule out pure or mixed channels with a component of $e^+ e^-$.
The only DM candidate that fits the GCE spectrum and is compatible with constraints obtained with the combined dSphs analysis and the AMS-02 $\bar{p}$ and $e^+$ data annihilates purely into $\mu^+\mu^-$, has a mass of 60 GeV and roughly a thermal cross section.
\end{abstract}

\maketitle

\section{Introduction}
\label{sec:intro}

Several groups have discovered an excess in the $\gamma$-ray data collected by the {\it Fermi} Large Area Telescope ({\it Fermi}-LAT) in the direction of the Galactic center (see, e.g., \cite{Goodenough:2009gk,Hooper:2010mq,Boyarsky:2010dr,Hooper:2011ti,Abazajian:2012pn,Gordon:2013vta,Abazajian:2014fta,Daylan:2014rsa,Calore:2014nla,Calore:2014xka,TheFermi-LAT:2015kwa,TheFermi-LAT:2017vmf,DiMauro:2019frs,Dimaurodata}).
This signal, called the Galactic center excess (GCE), has been detected using different background models, constituted by the flux of point and extended sources, interstellar emission, {\it Fermi} bubbles and an isotropic component, and by performing the analysis with different data selection and analysis techniques.
The GCE has a spectral energy distribution (SED, measured as $E^2 dN/dE$ in units of GeV/cm$^3$/s/sr) that peaks at a few GeV, a spatial morphology that is roughly spherically symmetric and its centroid is located in the dynamical center of the Milky Way \cite{Daylan:2014rsa,Calore:2014xka,Dimaurodata}.

The origin of the GCE is still a mystery.
Refs.~\cite{Bartels:2015aea,Lee:2015fea}, by applying wavelet analysis and non-Poissonian template fitting techniques to {\it Fermi}-LAT data, derived compelling evidence for the existence of a faint population of sources located in the Galactic center with properties that can explain the GCE.
The presence of these sources could be interpreted as a population of millisecond pulsars located around the Galactic bulge.
These results are supported by Refs.~\cite{Macias:2016nev,Bartels:2017vsx} that modeled the Galactic stellar bulge, a possible tracer of pulsars in the center of the Galaxy, using a nuclear bulge and a boxy bulge template. They demonstrate that fitting the GCE with these two templates they obtain a much better fit than using a DM model. This result implies that the GCE is not spherically symmetric since the model used in Refs.~\cite{Macias:2016nev,Bartels:2017vsx} has a boxy shape. 

Very recently, Refs.~\cite{Leane:2019uhc,Chang:2019ars} have shown that the non-Poissonian template fitting method can misattribute un-modeled point sources or imperfections in the IEM to a signal of a faint population of sources or DM.
These results cast serious doubts on the robustness of the results presented in \cite{Lee:2015fea} and the conclusion that the GCE is due to a population of pulsars. 
In addition, Ref.~\cite{Zhong:2019ycb} has applied wavelet analysis, similarly to what has been done in \cite{Bartels:2015aea}, to about 10 years of {\it Fermi}-LAT data using the latest 4FGL catalog released by the {\it Fermi}-LAT Collaboration \cite{Fermi-LAT:2019yla}. They find that the GCE is still present but they do not find any compelling evidence for the existence of a faint population of un-modeled sources.

Outbursts of cosmic rays (CRs) from the Galactic center have been proposed as possible interpretations for the GCE (see, e.g., \cite{Carlson:2014cwa,Petrovic:2014uda,Gaggero:2015nsa}). In these alternative scenarios the GCE is explained by $\gamma$ rays produced through inverse Compton scattering (ICS) of high-energy electrons and positrons on the interstellar radiation field (ISRF) photons or by CR protons interacting with the interstellar gas and producing $\pi^0$, which subsequently decays into $\gamma$ rays.
These mechanisms, however, provide $\gamma$-ray signals not fully compatible with the GCE properties.
For example, the hadronic scenario (i.e., CR protons) predicts a $\gamma$-ray signal that is distributed along the Galactic plane, since the $\pi^0$ decay process is correlated with the distribution of gas present in the Milky Way disk \cite{Petrovic:2014uda}.  
Instead, a leptonic outburst would lead to a signal that is approximatively spherically symmetric but it requires a complicated scenario with at least two outbursts to explain the morphology and the intensity of the excess.

Very recently, Ref.~\cite{Dimaurodata} has provided the most precise results for the GCE properties yet.
They confirm that the GCE SED is peaked at a few GeV and has a high energy tail significantly detected up to about 50 GeV.
The SED changes in normalization by roughly 60\% when using different interstellar emission models (IEMs), data selections and analysis techniques.
The spatial distribution of the GCE is compatible with a dark matter (DM) template modeled with a generalized Navarro-Frenk-White (NFW) density profile with slope $\gamma = 1.2-1.3$. 
The energy evolution of the GCE spatial morphology has been studied with unprecedented precision between $0.6-30$ GeV finding that no change larger than $10\%$ from the $\gamma$ average value, which is 1.25, is detected.
The GCE centroid is compatible with the dynamical center of the Milky Way and its morphology is compatible with a spherical symmetric NFW profile.
In particular, by fitting the DM spatial profile with an ellipsoid they find a major-to-minor axis ratio (aligned along the Galactic plane) between 0.8-1.2 when running the analysis with different IEMs.

The characteristics of the GCE published in Ref.~\cite{Dimaurodata} make $\gamma$ rays from DM particle interactions a viable interpretation.
In fact DM is predicted to be distributed in the Milky Way as a spherically symmetric halo with its centroid located in the dynamical center of the Galaxy. 
Moreover, the signal morphology is expected to be energy independent, i.e.~the value of the NFW slope ($\gamma$) found to fit the GCE morphology data should not vary with energy. 
The GCE SED can be well modeled as $\gamma$ rays produced by DM particles annihilating into $b\bar{b}$ with a thermal annihilation cross section \cite{Daylan:2014rsa,Calore:2014nla}, which is the proper cross section to explain the observed density of DM in the Universe \cite{Aghanim:2018eyx}. 
All these characteristics make the GCE very appealing for the DM interpretation.

If DM is the origin of the GCE, $\gamma$ rays should be emitted from these elusive particles also in Milky Way dwarf spheroidal galaxies (dSphs).
dSphs are among the most promising targets for the indirect search of DM with $\gamma$ rays because gravitational observations indicate that they have a high DM density, i.e.~a large mass-to-luminosity ratio of the order of $100-1000$ (see, e.g., \cite{Abdo_2010}). 
In addition, since they do not contain many stars or gas, they have an environment with predicted low astrophysical backgrounds.
All the analyses performed so far in the direction of known dSphs (see, e.g., \cite{Abdo_2010,Ackermann:2015zua,Fermi-LAT:2016uux,Calore:2018sdx,Hoof:2018hyn,2019MNRAS.482.3480P}) have not provided any detection of $\gamma$ rays and, as a consequence, they could provide tight constraints on the DM interpretation of the GCE.

The indirect search of DM is performed also with CR antiparticles, such as positrons ($e^+$) and antiprotons ($\bar{p}$), which are among the rarest cosmic particles in the Galaxy.
$e^+$ and $\bar{p}$ fluxes have been precisely measured by the AMS-02 experiment on the International Space Station up to almost 1 TeV \cite{PhysRevLett.117.091103,PhysRevLett.122.041102}.
Very recently, the AMS-02 Collaboration has released the data for several CR species with 7 years of data including new measurements for cosmic $e^+$ and $\bar{p}$ fluxes \cite{AGUILAR2020}.

$e^+$ mainly originate from secondary production, due to the spallation reactions of CRs with interstellar gas atoms, and from PWNe (see, e.g., \cite{DiMauro:2014iia,DiMauro:2019yvh,Manconi:2020ipm}). 
No clear signal of DM can be claimed with CR $e^+$ because of the large uncertainty mainly due to the possible PWN contribution \cite{Manconi:2020ipm}. 
In fact, recent observations of ICS halos detected in $\gamma$ rays around close pulsars (see, e.g., \cite{Abeysekara:2017science,DiMauro:2019yvh}) have provided clear evidences that PWNe inject $e^{\pm}$ in the interstellar space.
However, it is still not clear which fraction of pulsar spin-down energy is converted into $e^{\pm}$ and what is exactly the acceleration process that takes place in these sources (see, e.g., \cite{DiMauro:2019hwn,Manconi:2020ipm}).
The flux data of these particles can provide very tight constraints for leptonic DM, i.e.~annihilating or decaying into the $e^+ e^-$, $\mu^+\mu^-$ and $\tau^+\tau^-$ channels, (see, e.g., \cite{Bergstrom:2013jra,DiMauro:2015jxa}).

Different groups have found an excess of $\bar{p}$, with respect to the secondary production, in the data collected after 4 years of mission by AMS-02 between 5-20 GeV \cite{PhysRevLett.117.091103}. 
Its significance was determined as $3-5\sigma$ depending on the analysis technique used and the $\bar{p}$ production cross sections employed in the analysis.
The excess was interpreted in terms of DM particles with a mass of $60-80$ GeV annihilating into $b\bar{b}$~\cite{Cuoco:2017rxb,Cui:2016ppb,Cuoco:2019kuu,Cholis:2019ejx}. Possible links to the GCE were considered. Very recently, Ref.~\cite{Heisig:2020nse} investigated the presence of the $\bar{p}$ excess by fully including in the analysis the uncertainties on the $\bar{p}$ cross sections~\cite{Reinert:2017aga}, CR propagation and correlations in the AMS-02 systematic errors.
They find that when including all these sources of uncertainties the global significance is reduced to below $1\sigma$. As a result the constraints for the DM interpretation of the GCE for the hadronic channels, i.e.~DM annihilating into quarks, might be strong also using $\bar{p}$ CR data.

In this paper we investigate the DM interpretation of the GCE with a combined analysis of the targets that are the most promising for the search in $\gamma$ rays, i.e.~the Galactic center and dSphs, and using the flux data of AMS-02 for positrons and antiprotons which are among the rarest CRs. It is the first time ever that such an analysis for DM is performed at the same time in different astrophysical targets and cosmic particles and with a consistent model for the DM density distribution and coupling parameters.
We first determine the DM density in the Galaxy using at the same time the GCE surface brightness data reported in \cite{Dimaurodata} and the results recently obtained in Ref.~\cite{2019JCAP...10..037D} from observations of the rotation curve of the Milky Way.
We fit the GCE spectrum and find the relevant DM parameters, mass, annihilation cross section and branching ratio, in case of annihilation into single, double and triple channels.
Then, we search for a DM signal in a combined analysis of {\it Fermi}-LAT data in the direction of 48 dSphs reported in \cite{2019MNRAS.482.3480P}. We include in the analysis the uncertainty for the DM density of these objects. Since we do not find any significant flux, we put upper limits for the annihilation cross section for the cases that best fit the GCE spectrum.
We also search for a DM signal in the latest AMS-02 measurements of $e^+$ and $\bar{p}$ data \cite{AGUILAR2020}. For $\bar{p}$ we use an analysis, as in Ref.~\cite{Heisig:2020nse}, that accounts for the uncertainties on the $\bar{p}$ cross sections, CR propagation and correlations between AMS-02 data.
Instead, for CR $e^+$, given the current uncertainty in the possible flux of these particles from PWNe, we derive constraints on a DM contribution with a conservative and an optimistic approach. In the former we require the sum of secondary background and DM signal not to overshoot the AMS-02 data, while in the latter we include a possible pulsar contribution through an analytic function similar to the approach in Ref.~\cite{Bergstrom:2013jra}.
Finally, we compare the DM candidates that fit well the GCE data with the constraints found from dSphs, $\bar{p}$ and $e^+$ and provide the channels and coupling parameters that satisfy all the above observations.

The paper is organized as follow: in Sec.~\ref{sec:model} we explain the model we use for the calculation of the $\gamma$-ray, $\bar{p}$ and $e^+$ flux from DM and for the secondary production. 
In Sec.~\ref{sec:gammarayGCEres} we estimate the DM density in the Galaxy using the GCE data and latest rotation curve data of the Milky Way. We fit the GCE with $\gamma$ rays from DM using different annihilation models reporting the best-fit values of the relevant DM coupling parameters.
In Sec.~\ref{sec:dwarfres} we perform a DM search from dSphs in the {\it Fermi}-LAT data and produce limits for the annihilation cross section that we compare with the DM candidates compatible with the GCE SED. In Sec.~\ref{sec:antip} and \ref{sec:pos} we will investigate the compatibility of the DM interpretation of the GCE with $\bar{p}$ and $e^+$ flux data from AMS-02.
Finally, in Sec.~\ref{sec:conclusions} we draw our conclusions.


\section{Model for cosmic particle production from dark matter}
\label{sec:model}

\subsection{$\gamma$-ray flux from dark matter}
\label{sec:DMgammaray}

\subsubsection{Prompt emission}
\label{sec:prompt}

The $\gamma$-ray emission from DM particle interactions is usually calculated including two components.
The first one is the so-called prompt emission that is due to the direct production of $\gamma$ rays through an intermediate annihilation channel. 
The prompt emission is calculated as follows:
\begin{equation}
\frac{dN}{dE} = \frac{1}{2} \frac{r_{\odot}}{4\pi}  \left( \frac{\rho_{\odot}}{M_{\rm{DM}}} \right)^2 \bar{\mathcal{J}} \times  \langle \sigma v \rangle \sum_f Br_f   \left( \frac{dN_\gamma}{dE} \right)_f,
\label{eq:fluxprompt}
\end{equation}
where $M_{\rm{DM}}$ is the DM mass, $\rho_{\odot}$ is the local DM density, $r_{\odot}$ is the distance of the Earth from the center of the Galaxy, $\langle \sigma v \rangle$ defines the annihilation cross section times the relative velocity, averaged over the Galactic velocity distribution function.
Moreover, $\bar{\mathcal{J}}$ is the geometrical factor averaged over the viewing solid angle $\Delta \Omega$ of our region of interest (ROI) that is $40^{\circ}\times40^{\circ}$ centered in the Galactic center as in Ref.~\cite{Dimaurodata}.

$(dN_\gamma/d E)_f$ is the $\gamma$-ray spectrum from DM annihilation for a specific annihilation channel labeled as $f$ and $Br_f$ is its branching ratio. 
We take $(dN_\gamma/d E)_f$ from Ref.~\cite{Cirelli:2010xx} where this quantity has been calculated using the Pythia Monte Carlo code (version 8.162).
In particular we consider the tables reported at this webpage \url{http://www.marcocirelli.net/PPPC4DMID.html} for the case where electroweak corrections are also included. 

$\bar{\mathcal{J}}$ is calculated as the integral performed along the line of sight (l.o.s., $s$) of the squared DM density distribution $\rho$ divided for $\Delta \Omega$:
\begin{equation}
\bar{\mathcal{J}} = \frac{1}{\Delta \Omega}  \int_{\Delta \Omega} d\Omega \int_{l.o.s.}  \frac{ds}{r_{\odot}} \left( \frac{\rho(r(s,\Omega))}{\rho_{\odot}} \right)^2 .
\label{eq:geom}
\end{equation}
We parametrize $\rho$ with a generalized NFW (gNFW) DM density function \cite{1997ApJ...490..493N}:
\begin{equation}
\rho_{\rm{gNFW}} = \frac{\rho_s}{\left( \frac{r}{r_s} \right)^{\gamma} \left( 1 + \frac{r}{r_s} \right)^{3-\gamma}} ,
\label{eq:NFW}
\end{equation}
or with an Einasto profile \cite{1965TrAlm...5...87E}:
\begin{equation}
\rho_{\rm{Einasto}} = \rho_s \exp{ \left( -\frac{2}{\alpha} \left[ \left(\frac{r}{r_s}\right)^{\alpha} -1 \right]  \right) },
\label{eq:Einasto}
\end{equation}
or using a cored Burkert profile \cite{1995ApJ...447L..25B}:
\begin{equation}
\rho_{\rm{Burkert}} = \frac{\rho_s}{\left( 1+  \frac{r}{r_s} \right) \left[ 1 + \left(\frac{r}{r_s}\right)^{2} \right]}.
\label{eq:Burkert}
\end{equation}
The parameters $\rho_s$ and $r_s$ are the normalization and scale radius of the DM density profile, which has to be found calibrating $\rho$ on the observed distribution of DM in the Galaxy.
The results will be given with the gNFW and Einasto profiles since, as we will demonstrate in Sec.~\ref{sec:DMdensity} the Burkert profile is not adequate to fit the rotation curve and the GCE surface brightness data.

\subsubsection{Inverse Compton scattering emission}
\label{sec:ICS}

In case DM particles annihilate into leptonic channels, i.e. $e^+ e^-$, $\mu^+ \mu^-$ and $\tau^+ \tau^-$, there is a secondary production of $\gamma$ rays that becomes relevant.
This involves $e^{\pm}$ produced from the prompt emission and that subsequently generate $\gamma$ rays through ICS on the ISRF photons.
This component is particularly relevant for the Galactic center where the density of the starlight and dust components of the ISRF are roughly a factor of 10 higher than the local one (see, e.g., \cite{Porter_2008}).

The flux of $\gamma$ rays for ICS at energy $E$ is calculated as \cite{Cirelli:2010xx,Blanchet:2012vq,DiMauro:2015tfa}:
\begin{eqnarray}
&&\frac{dN}{dE} (E) =  \frac{r_{\odot}}{4\pi}  \left( \frac{\rho_{\odot}}{M_{\rm{DM}}} \right)^2 \int_{\Delta \Omega} d\Omega \int_{l.o.s.}  \frac{ds}{r_{\odot}} \times \nonumber  \\
&\times&  \int_{E}^{M_{\rm{DM}}} dE_e \, \mathcal{N}_e(E_e,{\bf r}(s,\Omega)) \mathcal{P} (E,E_e,{\bf r}(s,\Omega)),
\label{eq:fluxIC}
\end{eqnarray}
where $\mathcal{N}_e(E_e,{\bf r}(s,\Omega))$ is the density of $e^{\pm}$ produced with energy $E_e$ from DM at a position ${\bf r}$ and $\mathcal{P} (E,E_e,{\bf r})$ is the power of $\gamma$ rays produced for ICS on the ISRF.
$\mathcal{P}$ is defined as:
\begin{eqnarray}
\mathcal{P} (E,E_e,{\bf r}) &=& \frac{3\sigma_T c \, m_e^2 c^4}{4 E_e^2} \int_{1/(4\gamma^2)}^1 dq \, \left( 1 - \frac{m_e c^2}{4q E_e(1-\epsilon)} \right)\nonumber \\
&\times& n(\epsilon(q,{\bf r})) \frac{\mathcal{G}(q)}{q},
\label{eq:PIC}
\end{eqnarray}
where $\sigma_T$ is the Thomson cross section, $m_e$ is the electron rest mass,  $n(\epsilon)$ is the ISRF spectrum with photon energy $\epsilon$, $\Gamma=4\epsilon\gamma/(m_e c^2)$ and $q=\epsilon / (\Gamma (\gamma m_e c^2 - \epsilon))$.
$\mathcal{G}(q)$ is calculated from the Klein-Nishina cross section is defined as:
\begin{equation}
\mathcal{G}(q) = 2 q \log{q} + (1+2q)(1-q) + \frac{\psi^2 (1-q)}{2(1-\psi)},
\end{equation}
where $\psi = E/E_e$.

In order to find $\mathcal{N}_e$ we solve the equation for the propagation of $e^{\pm}$ in the Galactic diffusive halo. The propagation for $e^{\pm}$, that is dominated by energy losses for ICS and synchrotron radiation and the diffusion
on the irregularities of the Galactic magnetic field, is modeled as:
\begin{equation}
\label{eq:transport}
 \partial_t \mathcal{N}_e  - \mathbf{\nabla} \cdot \left\lbrace K(E)  \mathbf{\nabla} \mathcal{N}_e \right\rbrace + \partial_E \left\lbrace b(E) \mathcal{N}_e \right\rbrace = \mathcal{Q}(E, \mathbf{r}),
\end{equation}
where $b(E)$ represents the energy losses, $K(E)$ the diffusion,
and $Q(E, \mathbf{r})$ the source term for the production of $e^{\pm}$ from DM.
Other processes usually taken into account for CR nuclei are negligible for the propagation of $e^{-}$ (see, e.g., \cite{Evoli:2016xgn}).
Assuming homogeneous energy losses and diffusion in the Galaxy, the solution of the propagation equation is found as:
\begin{equation}
\mathcal{N}_e(E_e,{\bf r}) = \int_{E_e}^{M_{\rm{DM}}} dE_s \int_{dV} d^3 {\bf r} \, \mathcal{G}(E_e,{\bf r} \leftarrow E_s,{\bf r}_s) \mathcal{Q}(E_s,{\bf r}_s),
\label{eq:solNe}
\end{equation}
where $\mathcal{G}(E_e,{\bf r} \leftarrow E_s,{\bf r}_s)$ is the Green function which accounts for the probability that $e^{\pm}$ emitted at an initial Galactic position ${\bf r}_s$ and with an energy $E_s$ is detected at a final position ${\bf r}$ and energy $E_e$. Since the boundaries of the propagation zone do virtually not affect the solution for $e^\pm$ in the Galactic center region, we can employ the free Green function:
\begin{equation}
\mathcal{G}(E_e,{\bf r} \leftarrow E_s,{\bf r}_s) = \frac{1}{b(E_e) (\pi \lambda^2)^{3/2}} \exp{\left( -\frac{({\bf r} - {\bf r_s})^2}{\lambda^2} \right)}.
\label{eq:G}
\end{equation}
$\lambda^2$ is the propagation length for $e^{\pm}$ affected by energy losses and diffusion:
 \begin{equation}
\lambda^2 (E,E_s) = 4 \int_{E_s}^{E_e} dE' \frac{K(E')}{b(E')}.
\label{eq:lambda}
\end{equation}
The source term $\mathcal{Q}(E_s,{\bf r}_s)$ for DM is calculated as:
 \begin{equation}
\mathcal{Q}(E_s,{\bf r}_s) = \left( \frac{\rho({\bf r_s})}{\rho_{\odot}} \right)^2 \sum_f Br_f   \left( \frac{dN_e}{dE_s} \right)_f,
\label{eq:Q}
\end{equation}
where $(dN_e/dE_s)_f$ is the spectrum of $e^{\pm}$ produced from DM particle interactions and it depends on the specific annihilation channel assumed and labeled in the equation with $f$.

In our ROI centered in the Galactic center, we can neglect diffusion because the propagation of $e^{\pm}$ is dominated by energy losses since the starlight and infrared components of the ISRF are a factor of about 30 and 8 larger than in the local Galaxy respectively \cite{Porter_2008}.
Therefore, for the calculation of the ICS $\gamma$-ray flux produced in our ROI we can neglect diffusion.
Assuming that $K(E)$ is parametrized as $K(E)=K_0 E^{\delta}$, the typical timescale of diffusion is calculated as $\tau \sim L^2/K(E) \sim 500\cdot E^{-\delta}$ Myr with $\delta\approx 0.40$ and $K_0 = 3\cdot 10^{28}$ cm$^3$/s \cite{Heisig:2020nse}. Instead, the energy losses in the Galactic center region have a characteristic time scale $\tau \sim E/(b(E)) \sim 10\cdot E^{-0.7}$ Myr. At 10 GeV the energy loss $\tau$ is thus a factor of about 100 smaller than the one for diffusion confirming thus that we can neglect diffusion.
In this scenario, Eq.~\ref{eq:fluxIC} for the ICS flux simplifies to the following expression \cite{Cirelli:2009vg}:
\begin{eqnarray}
\frac{dN}{dE} &=& \frac{r_{\odot}}{4\pi}  \left( \frac{\rho_{\odot}}{M_{\rm{DM}}} \right)^2 \bar{\mathcal{J}} \times  \langle \sigma v \rangle \sum_f Br_f \times \nonumber  \\
&\times& \int_{E_e}^{M_{\rm{DM}}} dE_e \frac{ \mathcal{P} (E,E_e) \mathcal{Y}_f(E_e) }{ b(E_e) },
\label{eq:fluxICsimplified}
\end{eqnarray}
where $\mathcal{Y}_f(E_e)$ is defined as $\mathcal{Y}(E_e) = \int^{M_{\rm{DM}}}_{E_e} (dN_e/dE_e)_f$.

\begin{figure}
\includegraphics[width=0.49\textwidth]{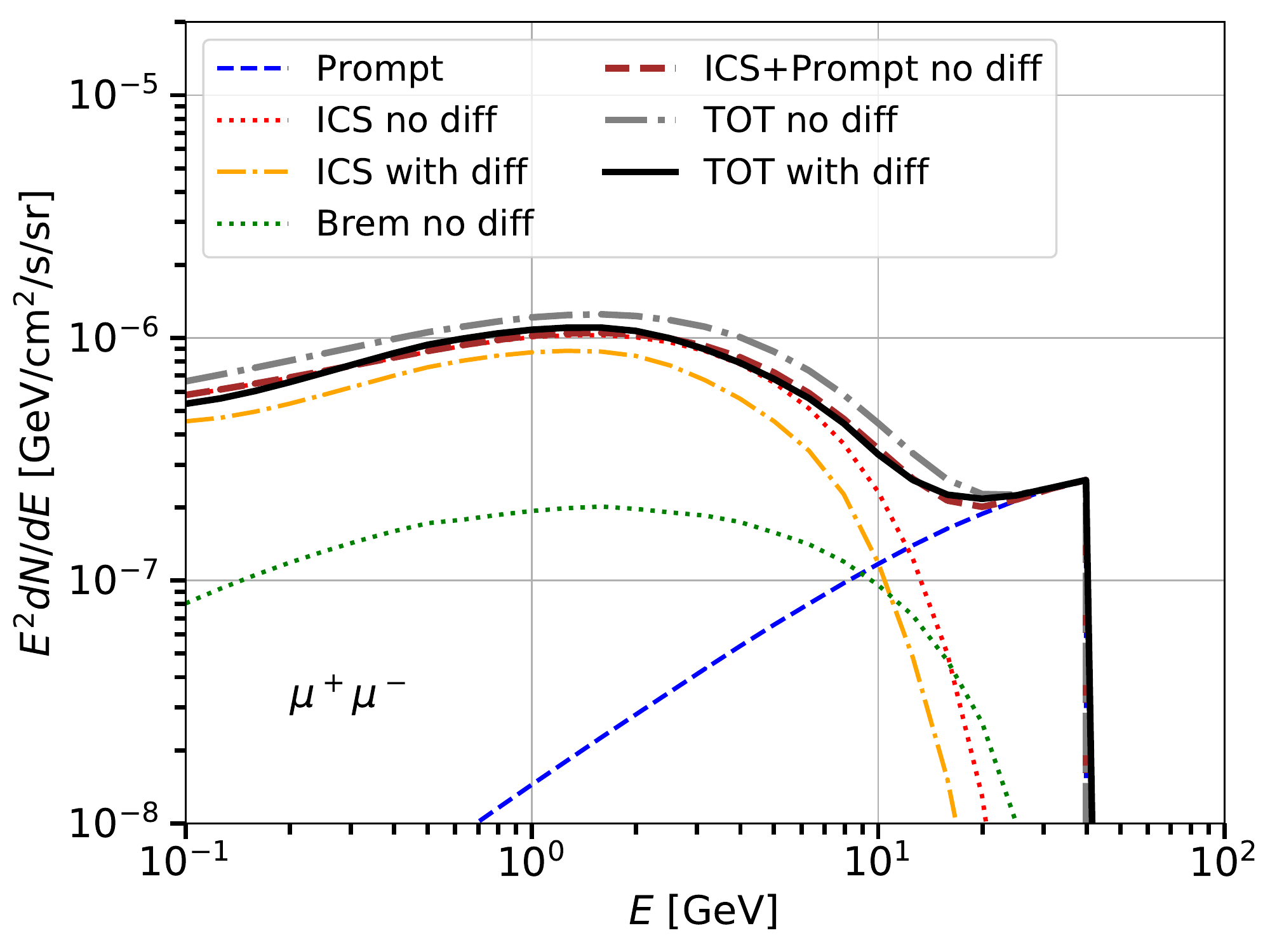}
\includegraphics[width=0.49\textwidth]{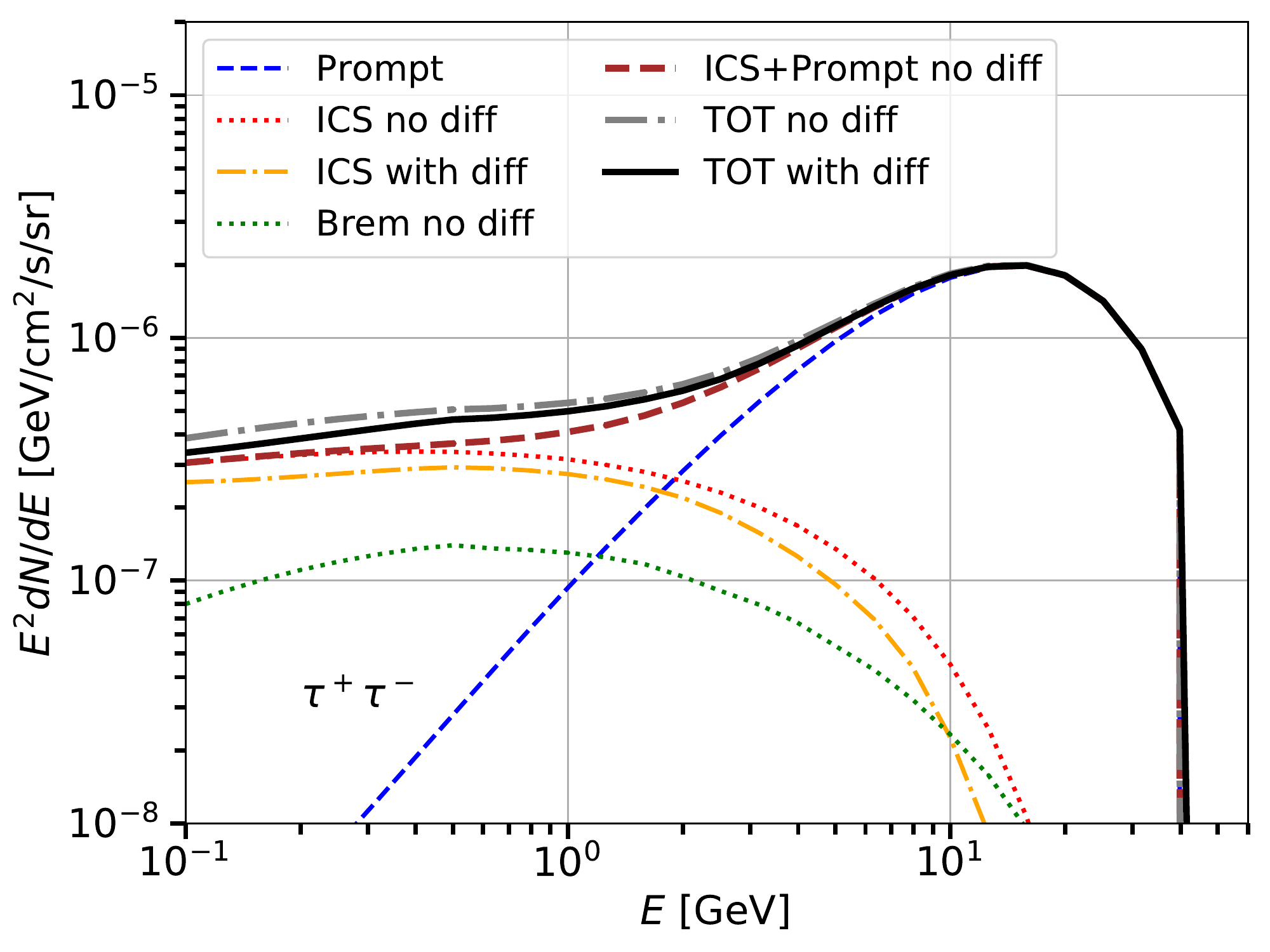}
\includegraphics[width=0.49\textwidth]{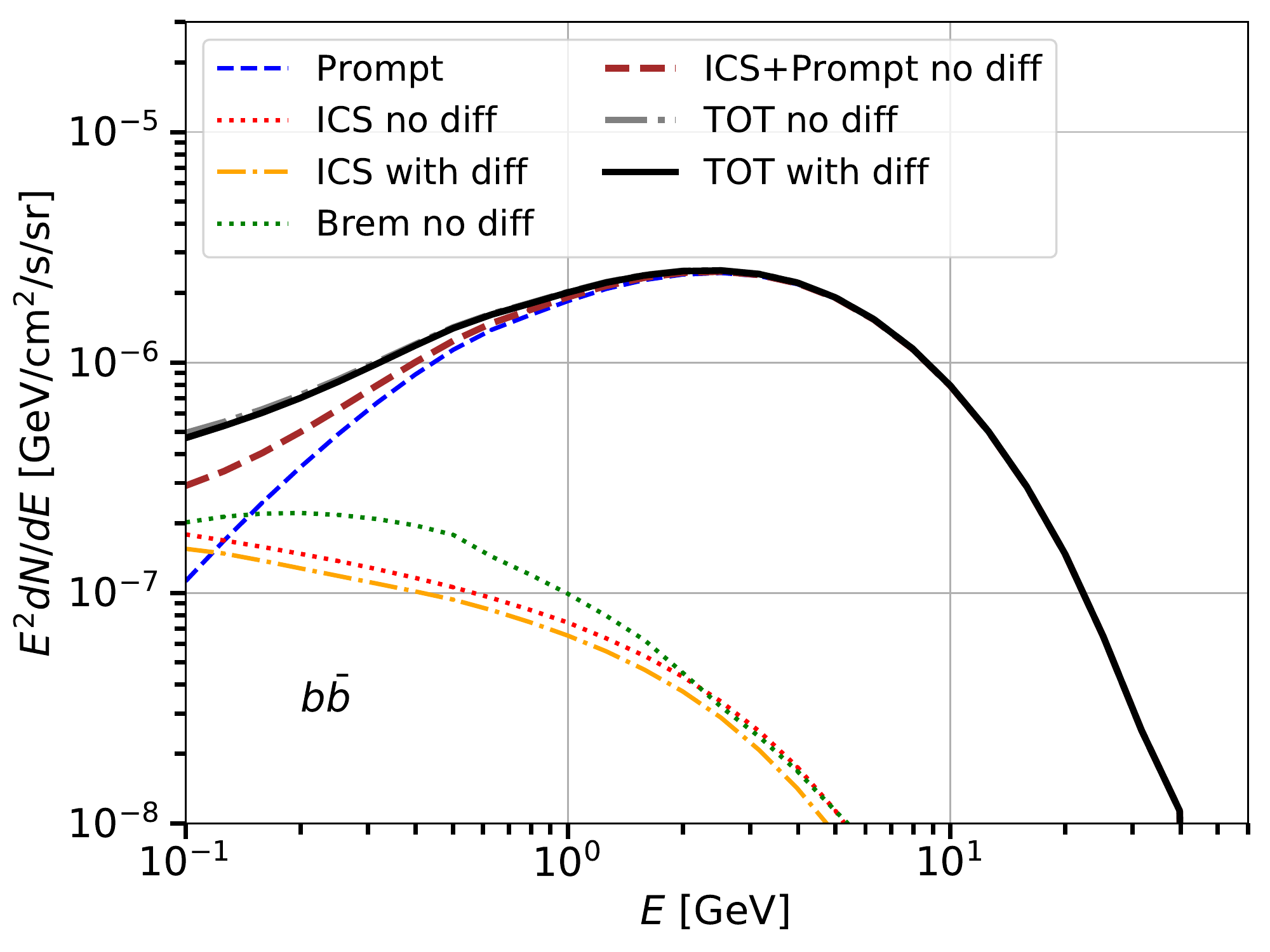}
\caption{Flux of $\gamma$ rays produced for prompt (blue dashed line), Bremsstrahlung (dotted green line) and ICS emission from DM particles annihilating into $\mu^+ \mu^-$, $\tau^+ \tau^-$ and $b\bar{b}$, from top to bottom panels, for $M_{\rm{DM}}= 50$ GeV and $\langle \sigma v \rangle = 3\times 10^{-26}$ cm$^3$/s. We present two cases for the ICS emission with (orange dot-dashed line) and without (red dotted line) accounting for diffusion. We also display the total emission (prompt plus ICS and Bremsstrahlung), with (solid black line) and without (grey dot-dashed line) diffusion and the case with ICS, calculated without diffusion, plus prompt emission (dashed brown line). The case with prompt emission and ICS, calculated without diffusion, is the model we use in the analysis.}
\label{fig:promptICS}
\end{figure}

In order to demonstrate further that diffusion can be neglected, we calculate the $\gamma$-ray emission for ICS including and neglecting diffusion.
We assume the ISRF model
for the Galactic center as in Ref.~\cite{Porter_2008} and the parametrization of diffusion as in Ref.~\cite{Heisig:2020nse}.
We perform the calculation for two leptonic channels $\mu^+ \mu^-$ and $\tau^+ \tau^-$ and the hadronic channel $b\bar{b}$ and for a DM mass and cross section of 50 GeV and $3\times 10^{-26}$ cm$^3$/s. These are roughly the DM parameters that best fit the GCE spectrum (see Sec.~\ref{sec:onechannels}). 
We show in Fig.~\ref{fig:promptICS} the result of this calculation.  
The case with the $\mu^+ \mu^-$ channel is the one, among the three shown, for which the difference between the case with and without diffusion is more evident in the total flux because the ICS component gives the largest contribution with respect to the prompt emission.
Instead, for the $b\bar{b}$ channel since the ICS component is negligible with respect to the prompt one, the effect of diffusion has a minimal effect in the total flux.
The inclusion of diffusion has the effect of reducing the ICS flux by a renormalization factor that changes at most of about $20-25\%$ the flux for ICS.
Since the numerical calculation of Eq.~\ref{eq:fluxIC}, that includes diffusion, is very time consuming and its addition does not change significantly the predictions for the $\gamma$-ray flux, we decide to neglect this process in the calculation. 
Therefore, we will use Eq.~\ref{eq:fluxICsimplified} in our analysis.
We will discuss in Sec.~\ref{sec:fitGCE} how the inclusion of the diffusion can affect our results.

\subsubsection{$\gamma$ rays from bremsstrahlung}

There is an additional secondary production of $\gamma$ rays from DM that is associated with the Bremsstrahlung process. This involves $e^{\pm}$, produced from DM particles annihilation, interacting with interstellar gas, in the neutral, ionized and molecular forms, and generating photons typically at X-ray and $\gamma$-ray energies.
The calculation of this contribution follows the one in Eq.~\ref{eq:fluxICsimplified}, if the diffusion is not taken into account, where the $\gamma$-ray power for ICS is substituted with the one for Bremsstrahlung.
For this latter quantity we consider the approximated form in Ref.~\cite{1999ApJ...513..311B}.

We show in Fig.~\ref{fig:promptICS} the contribution of Bremsstrahlung $\gamma$ rays to the total DM contribution for the $\mu^+ \mu^-$ and $\tau^+ \tau^-$ and $b\bar{b}$ channels, assuming for the interstellar gas an average density of $1$ cm$^{-3}$. This value is justified by the density of interstellar gas in the inner few kpc from the Galactic center. In particular the distribution of gas on the Galactic plane in the inner 3-5 kpc is between 1-3 cm$^{-3}$. However, the gas density decreases as an exponential function with scale radius of about 0.1-0.2 kpc \cite{Evoli:2016xgn}. For the DM mass and cross section we assume 50 GeV and $3\times 10^{-26}$ cm$^3$/s. Bremsstrahlung contributes mostly at energies below 1 GeV to the total flux. The addition of this mechanisms does not have a significant effect since for $\mu^+ \mu^-$ ($\tau^+ \tau^-$) it is a factor of about 5 (3) smaller than the ICS one and for $b\bar{b}$ it is much smaller than the prompt emission for most of the energies considered.
In addition, the effect that Bremsstrahlung brings to the total flux is opposite with respect to the addition of diffusion. Therefore, the combination of adding Bremsstrahlung emission and the diffusion process in the calculation has the net effect of producing a difference in the total flux that is minimal with respect to the case where we include prompt and ICS emission only without accounting for diffusion.
There is an additional reason to assume that Bremsstrahlung is not contributing significantly to the GCE. The $\gamma$-ray flux for Bremsstrahlung would be associated with the distribution of the interstellar gas distribution that is elongated on the Galactic plane. However, there is no evidence that there is an asymmetry on the Galactic plane of the GCE.

To summarize, we decide to perform the calculation for the $\gamma$-ray flux from DM by including prompt and ICS emissions and without accounting for the diffusion mechanism. The ISRF is modeled as in the model of Ref.~\cite{Porter_2008} for the Galactic center region.

\subsection{Antiprotons and positrons flux from dark matter}

\subsubsection{Dark matter and astrophysical source terms}

DM annihilation can induce a primary flux of $\bar{p}$ and $e^+$. 
The source term which denotes the differential production rate of $i=\bar{p},e^+$ per volume, time and energy reads exactly as in Eq.~\ref{eq:Q} where
$(dN_i/dE )_f$ is the spectrum of antiprotons/ positrons for the annihilation channel. We again take $(dN_i/dE )_f$ from Ref.~\cite{Cirelli:2010xx}.
In addition, there is an astrophysical antimatter background which originates from the scattering of CR protons and nuclei on the interstellar matter. The source term for this so-called secondary production reads:
\begin{equation}
  \mathcal{Q}^{\rm{sec}}_i = \sum\limits_{j,k} 4\pi \int dE^\prime \left(\frac{d\sigma_{jk\rightarrow i}}{d E}\right) \,n_k \;\Phi_j(E^\prime),
\end{equation}
where $\Phi_j$ denotes the flux of the progenitor species $j$, while $n_k$ stands for the number density of the target nucleus $k$ in the Galactic disc. We can set $j,k=p,\rm{He}$ since contributions from heavier nuclei are strongly suppressed. Furthermore, we will approximate the proton and helium fluxes as spatially constant in the Galactic disc. This simplification leaves local secondary fluxes virtually unaffected since the radial dependence of progenitor fluxes is effectively absorbed into the propagation parameters.

The differential $\bar{p}$ production cross sections $d \sigma_{ij\rightarrow \bar{p}}/d E$ are taken from Ref.~\cite{Kappl:2014hha,Winkler:2017xor} and include the full modeling of prompt $\bar{p}$ emission as well as displaced $\bar{p}$ production via hyperon and $\bar{p}$ decays (see Refs.~\cite{diMauro:2014zea,Korsmeier:2018gcy} for other recent cross section parametrizations). Since production cross sections are only known to a few percent precision, they comprise an important source of systematic error in the modeling of $\bar{p}$ fluxes which needs to be incorporated in DM searches. These uncertainties and their full correlations have been parameterized in Ref.~\cite{Winkler:2017xor} and will be included in our analysis as well.

In the case of cosmic $e^+$, secondary production contributes strongly to the astrophysical background contribution below 10 GeV while at higher energies the cumulative flux from Galactic pulsar wind nebulae (PWNe) liekely dominates. 
Since the contribution of PWNe to the $e^+$ data is still not well constrained, we are interested in providing conservative constraints to the DM contribution. Therefore, we use the $e^+$ production cross section parameterization of Ref.~\cite{Kamae:2006bf} which yields the lowest secondary flux among the parametrizations in the literature (see Ref.~\cite{Delahaye:2008ua}).

\subsubsection{Antimatter propagation}

The propagation of antimatter follows a transport equation analogous to Eq.~\eqref{eq:transport}. Besides diffusion and energy losses, we, however, also include reacceleration by magnetic shock waves as well as annihilation processes in the Galactic disc. Convective winds will be neglected since they are not preferred by recent CR analyses (see e.g.~\cite{Weinrich:2020cmw,Heisig:2020nse}). 

We solve the transport equation within the two-zone diffusion model~\cite{Maurin:2001sj,Donato:2001ms,Maurin:2002ua} which assumes that diffusion occurs homogeneously and isotropically in a cylinder of radius $R$ and half-height $L$ around the Galactic disc. The disc itself is taken to exhibit a thickness $2h=0.2\:\text{kpc}$ and to contain a constant number density of hydrogen and helium, $n_{\text{H}}=0.9\:\text{cm}^{-3}$ and $n_{\text{He}}=0.1\:\text{cm}^{-3}$. With these assumptions, the transport equation becomes:
\begin{align}\label{eq:diffeq}
&  -K \Delta \mathcal{N}_i 
+ 2 h \delta(z) \big[\partial_E (b_{\text{disc}}  \mathcal{N}_i -K_{EE} \:\partial_E \mathcal{N}_{i})+ \Gamma_\text{ann}\,\mathcal{N}_{i}\big] \nonumber\\
&+ \partial_E (b_{\text{halo}} \mathcal{N}_i)   =2h\delta(z)\mathcal{Q}_i^{\text{sec}} + \mathcal{Q}_i^{\text{prim}}\,.
\end{align}
The extension of the disc in vertical direction (z-direction) has been neglected. Processes confined to the disc were multiplied by $2 h \delta(z)$ for proper normalization.

The diffusion coefficient $K$ is modeled as a broken power law in the rigidity $\mathcal{R}$~\cite{Genolini:2017dfb}
\begin{equation}\label{eq:diffusion_coefficient}
K = K_0\, \beta^\eta \left(\frac{\mathcal{R}}{\rm{GV}}\right)^\delta \left(1+\left(\frac{\mathcal{R}}{\mathcal{R}_b}\right)^{\Delta\delta/s}\right)^{-s},
\end{equation}
with power law index $\delta$ below the break position $\mathcal{R}_b$ and $\delta+\Delta\delta$ above. The parameter $s$ describes the smoothness of the break. The diffusion break is required to account for observed spectral breaks in the proton and nuclear cosmic ray spectra~\cite{Aguilar:2015ooa,Aguilar:2015ctt} but plays a subleading role in the energy range accessible to antimatter searches.\footnote{A break in the diffusion term can be linked to the transition from diffusion on CR self-generated turbulence at low rigidity to diffusion on external turbulence at high rigidity~\cite{Blasi:2012yr}.} We also allow for a free scaling of $K$ with the velocity $\beta$. While $\eta=1$ in the original two-zone diffusion model, recent studies discovered a significant improvement in the fit to secondary nuclear cosmic rays if $\eta$ is taken as a free parameter~\cite{DiBernardo:2009ku,Maurin:2010zp,Genolini:2019ewc,Weinrich:2020cmw}. Physically, an increase of the diffusion coefficient (negative $\eta$) towards low rigidity is motivated by wave damping on cosmic rays~\cite{Ptuskin:2005ax}.

Reacceleration by Alfv\'en waves is modeled as diffusion in momentum space via the term~\cite{Maurin:2002ua}
\begin{align}
K_{EE}=\frac{4}{3}\frac{V_a^2}{K}\frac{p^2}{\delta(4-\delta)(4-\delta^2)},
\end{align}
where $V_a$ stands for the Alfv\'en velocity. Energy losses in the Galactic disc arise from Coulomb interactions, ionization, bremsstrahlung and reacceleration, such that $
 b_{\text{disc}}= b_{\text{coul}} + b_{\text{ion}} + b_{\text{brems}}  + b_{\text{reac}}
 $. We extract $b_{\text{coul}}$, $b_{\text{ion}}$, $b_{\text{brems}}$ from~\cite{Strong:1998pw} and $b_{\text{reac}}$ from~\cite{Maurin:2002ua}. 

For $e^+$, we also need to include the energy loss term $b_{\text{halo}}= b_{\text{ic}} + b_{\text{synch}}$ which accounts for inverse Compton scattering and synchrotron emission in the Galactic halo as described in Sec.~\ref{sec:ICS}. We use for the ICS calculation the full Klein Nishina formalism and the ISRF model as in Ref.~\cite{Porter_2008}.

Annihilation in the Galactic disc is only relevant for antiprotons. The annihilation rate $\Gamma_{\rm{ann}}$ is taken from Ref.~\cite{Protheroe:1981gj,Tan:1983de}.\footnote{The antiproton annihilation cross section was interpolated between the two parameterizations as in~\cite{Kappl:2011jw}.} Furthermore, we consider inelastic (non-annihilating) scattering of antiprotons with the interstellar matter through inclusion of a tertiary source term as in~\cite{Donato:2001ms}.

The spatial part of the antiproton transport equation can be solved analytically. For secondary antiprotons, whose source term is located in the Galactic disc, one obtains\footnote{We neglected the radial boundary $R$ which is justified since $R\ll L$ for the propagation configuration we consider in this work.}\cite{Donato:2001ms,Maurin:2001sj}
\begin{align}\label{eq:energyeq}
&\left(2 h \Gamma_\text{ann} +  \frac{2K}{L}\right)\mathcal{N}_{\bar{p}}
 +  2 h \partial_E \left(b_{\text{disc}} \,\mathcal{N}_{\bar{p}} -K_{EE} \:\partial_E \mathcal{N}_{\bar{p}} \right) \nonumber \\
&=  2 h (\mathcal{Q}^{\text{sec}}_{\bar{p}}+\mathcal{Q}^{\text{ter}}_{\bar{p}}),
\end{align}
with the tertiary source term as defined in~\cite{Donato:2001ms}. This equation needs to be solved numerically. Since the primary antiproton source term contains an additional spatial dependence on the dark matter profile, the solution for antiprotons from dark matter requires a Bessel expansion in the radial coordinate. The procedure has been described in full detail in~\cite{Barrau:2001ev,Donato:2003xg}.

An approximate solution of the transport equation for positrons was already given in~Eq.~\ref{eq:solNe}. The latter considers only diffusion and halo energy losses, while neglecting reacceleration as well as positron interactions with matter in the Galactic disc. When determining the local cosmic ray positron flux we include the vertical boundary of the diffusion zone (while the radial boundary can still be neglected). The free Green function given in~Eq.~\ref{eq:G} gets modified and one obtains~\cite{Baltz:1998xv}
\begin{align}
&\mathcal{G}(E_e,{\bf r} \leftarrow E_s,{\bf r}_s) = \frac{1}{b(E_e) (\pi \lambda^2)^{3/2}} \sum\limits_{n=-\infty}^{\infty}(-1)^n\nonumber\\
&\times \exp{\left( -\frac{(x -  x_s)^2+(y -  y_s)^2+(z -  z_{sn})^2}{\lambda^2} \right)},
\end{align}
with ${\bf r}=\{x,y,z\}$ and
\begin{equation}
  z_{sn}= 2 n L + (-1)^n z_s.
\end{equation}
The solution in Eq.~\ref{eq:solNe} with the Green function as defined above holds for both, primary and secondary positrons. In a next step, one can include reacceleration and disc energy losses for positrons through the pinching method described in Ref.~\cite{Boudaud:2016jvj}. We note that reacceleration and disc losses only affect the low-energy range ($E\lesssim 3\:\text{GeV}$). Both effects are mostly relevant for the spectrum of secondary positrons which is more strongly peaked towards low energy compared to primary positrons from DM. We have therefore implemented the pinching method for the secondary positron background but employ the high-energy approximation as written above for primary positrons.

Finally, on their passage through the heliosphere, CR are affected by the magnetic field of the sun. Diffusion, drifts, convection, and adiabatic energy losses are the dominant effects. In the force field approximation~\cite{Gleeson:1968zza} solar modulation is described by a single time-dependent parameter, the Fisk potential $\phi$, which is universal among CR species. In this work we will employ an improved force field approximation which additionally allows us to include charge breaking effects due to CR drifts. During a positive solar polarity phase, positively charged particles access the heliosphere on direct trajectories along the poles. Negatively charged particles enter by inward drift along the current sheet which gives rise to additional energy losses~\cite{Kota:1979,Jokipii:1981}. In order to incorporate these effects, the Fisk potential is written as a rigidity-dependendent function of the form~\cite{Cholis:2015gna,Cholis:2020tpi}
\begin{equation}\label{eq:forcefield2}
 \phi= \phi_0 + \phi_1\,\frac{\text{GV}}{\mathcal{R}}\,.
\end{equation}
The second term on the right-hand-side models the increased energy loss along the current sheet faced by particles whose charge sign is opposite to the solar polarity. It is taken to vanish for particles with charge sign equal to the polarity (in which case the standard force field approximation is recovered). For the AMS-02 data taking period, which (mostly) refers to a positive polarity phase, we will take $\phi_1=0$ for positrons, but non-zero for antiprotons.

\section{Dark matter interpretation of the Galactic center excess}
\label{sec:gammarayGCEres}

\subsection{Dark matter density}
\label{sec:DMdensity}
One of the main ingredients to calculate $\gamma$-ray fluxes from DM is its density distribution in the Galaxy that enters through the geometrical factor $\bar{\mathcal{J}}$ (see Eq.~\ref{eq:fluxICsimplified}).
We use the surface brightness data of the GCE reported in Ref.~\cite{Dimaurodata} and the recent results from the rotation curve of the Milky Way from Ref.~\cite{2019JCAP...10..037D} to estimate the values of the DM density parameters. We employ the results obtained in this section for the estimation of $\bar{\mathcal{J}}$ for $\gamma$ rays but also for the calculation of the $\bar{p}$ and $e^+$ production from DM.

We derive the predicted surface brightness from DM calculating the $\gamma$-ray flux for different angular distance from the center of the Galaxy, i.e.~in Eq.~\ref{eq:fluxICsimplified} the geometrical factor becomes a function of the angular distance from the Galactic center.
We test the three DM density profiles reported in Sec.~\ref{sec:prompt}: gNFW, Einasto and Burkert.
For the gNFW and Einasto we fit the values of $\gamma$ and $\alpha$, respectively. For both profiles we fix $r_s=20$ kpc since the surface brightness data are at small angular distances from the Galactic center and $r_s$ is thus unconstrained. In fact, we check that by using different values for $r_s$ the results do not change.
Finally, for the Burkert profile we leave free to vary $r_s$ since the slope is fixed.
The values of normalization of the DM density profile cannot be derived with this method since, in the flux calculation, $\rho_s$ is completely degenerate with the annihilation cross section.

\begin{figure}
\includegraphics[width=0.49\textwidth]{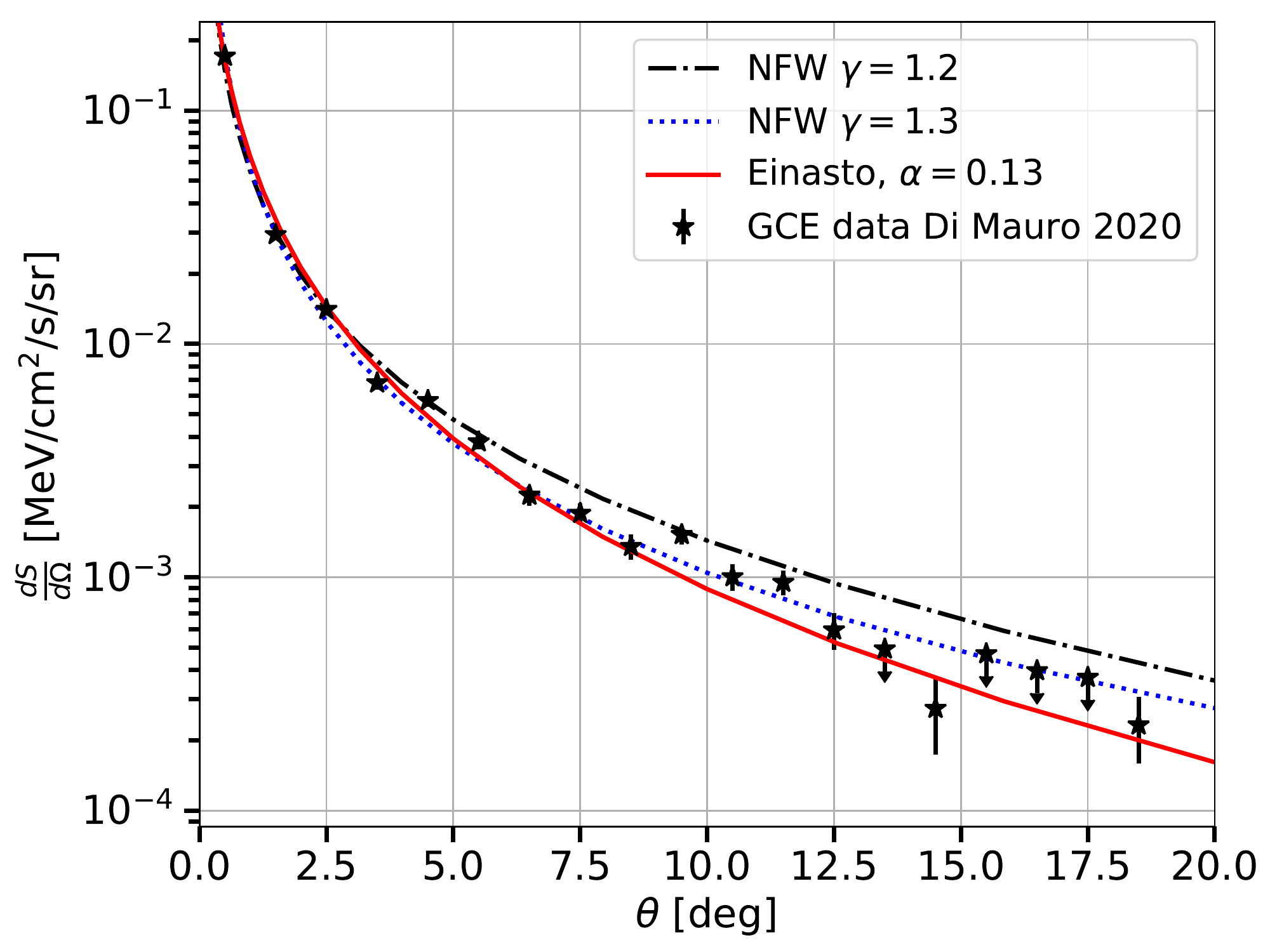}
\caption{Result of the fit to the GCE surface brightness data (black data points)  \cite{Dimaurodata} with a DM signal calculated for a gNFW profile with $\gamma=1.2$ (dot-dashed black line) and $\gamma=1.3$ (dotted blue line) and an Einasto profile with $\alpha=0.13$ (red solid line).}
\label{fig:SBfit}
\end{figure}

We find the best-fit values of $\gamma$ for the gNFW, $\alpha$ for the Einasto and $r_s$ for the Burkert profile by fitting the predicted DM flux to the GCE data for the surface brightness recently measured in Ref.~\cite{Dimaurodata} between $0^{\circ}$ to $20^{\circ}$ from the Galactic center.
The result of the fit is that the $\gamma$ parameter for the gNFW profile must be between $1.2-1.3$ consistently to what found in Ref.~\cite{Dimaurodata,Calore:2014nla,Daylan:2014rsa}.
The goodness of fit with gNFW is in terms of the reduced $\chi^2$ ($\tilde{\chi}^2$) $3.9$ and $2.0$ for $\gamma=1.2$ and 1.3, respectively.
The Einasto profile provides a good fit to the GCE data with $\alpha=0.13$ and $\tilde{\chi}^2=1.9$. Finally, the Burkert profile gives a very poor fit with $\tilde{\chi}^2=12.8$. The Burkert profile gives a flat surface brightness in the inner few degrees from the Galactic center where instead the data are very peaked. In addition, the best-fit for $r_s$ is about 0.26 kpc that is a too small value if compared to the observed DM density in the outer part of the Galaxy (see, e.g., \cite{Cirelli:2010xx}).
Therefore, we decide to consider the following three cases to bracket the possible uncertainty of the DM density profile: gNFW with $\gamma=1.2$ and $1.3$ and Einasto with $\alpha=0.13$.
We show the best-fit we obtain with these three models in Fig.~\ref{fig:SBfit} compared to the GCE data for the surface brightness obtained with the {\tt Baseline} IEM.
In particular we can observe that all the three cases provide a good fit to the GCE data. DM is able to fit properly the peaked data in the inner few degrees from the Galactic center but also the extended tail beyond $5^{\circ}$.
We also test the same analysis using the surface brightness data obtained in Ref.~\cite{Dimaurodata} with other IEMs and we find very similar results for $\gamma$ and $\alpha$.

\begin{table}
\begin{center}
\begin{tabular}{|c|c|c|c|c|c|}
\hline
\hline
DM density  & slope ($\gamma$/$\alpha$) & $\rho_s$ [GeV/cm$^3$] & $r_s$ [kpc] & $\bar{\mathcal{J}}$ & label \\
\hline
\multicolumn{5}{|c|}{$\rho_{\odot}=0.300$ GeV/cm$^3$ $M_{200}=5.50\cdot 10^{11}$ $M_{\odot}$ } & \\
\hline
gNFW  &  1.20  &  0.416  &  12.87  &  111.5 & {\tt MIN} \\
gNFW  &  1.30  &  0.314  &  14.18  &  155.3 &  \\
Einasto  &  0.13  &  0.376  &  7.25  &  288.9 &  \\
\hline
\multicolumn{5}{|c|}{$\rho_{\odot}=0.345$ GeV/cm$^3$ $M_{200}=5.90\cdot 10^{11}$ $M_{\odot}$ } & \\
\hline
gNFW  &  1.20  &  0.587  &  11.57  &  166.1 &  \\
gNFW  &  1.30  &  0.449  &  12.67  &  231.0 & {\tt MED}  \\
Einasto  &  0.13  &  0.569  &  6.35  &  449.3 &  \\
\hline
\multicolumn{5}{|c|}{$\rho_{\odot}=0.390$ GeV/cm$^3$ $M_{200}=6.30\cdot 10^{11}$ $M_{\odot}$ } & \\
\hline
gNFW  &  1.20  &  0.851  &  10.20  &  246.8 &  \\
gNFW  &  1.30  &  0.649  &  11.20  &  339.1 &  \\
Einasto  &  0.13  &  0.864  &  5.51  &  686.7 & {\tt MAX} \\
\hline
\hline
\end{tabular}
\caption{This table summarizes the best-fit for the DM density parameters for each case considered in the paper.
We list nine cases that result from choosing three different DM density profiles and three measurements for the local DM density and $M_{200}$ from Ref.~\cite{2019JCAP...10..037D}. We report the value of the slope ($\gamma$ for the gNFW and $\alpha$ for Einasto), $\rho_s$, $r_s$ and the value of the geometrical factor $\bar{\mathcal{J}}$ calculated for an ROI $40^{\circ}\times40^{\circ}$ centered in the Galactic center.}
\label{tab:models}
\end{center}
\end{table}

We derive the normalization $\rho_s$ and the scale radius $r_s$ of the DM density profile using the results for the local DM density and total DM mass published in Ref.~\cite{2019JCAP...10..037D}.
The authors analyze precise circular velocity curve measurements of the Milky Way for distances between $5-25$ kpc from the Galactic centre obtained by Gaia DR2 \cite{2019ApJ...871..120E}.
They explore several Galactic mass models that differ in the distribution of baryons and DM in order to use the rotation curve data to constrain the local DM density.
Using this technique they find that the local DM density varies between $\rho_{\odot} = [0.30,0.39]$ GeV/cm$^3$. Instead, the DM mass is provided through the quantity $M_{200}$, defined as the mass contained within the radius $r_{200}$ such that the energy density is 200 times larger than the critical energy density of the Universe. $M_{200}$ is found to vary between $[5.5,6.3]\times 10^{11} M_{\odot}$ for $r_{200} = [175,180]$ kpc.

We use the following cases reported in Ref.~\cite{2019JCAP...10..037D}: gNFW with $\gamma=1.2$, $\rho_{\odot} = 0.30$ GeV/cm$^3$ and $M_{200} = 5.5\times 10^{11} M_{\odot}$ (see Tab.~2 of \cite{2019JCAP...10..037D}) and $\gamma=1.3$, $\rho_{\odot} = 0.39$ GeV/cm$^3$ and $M_{200} = 6.3\times 10^{11} M_{\odot}$ (see Tab.~3 of \cite{2019JCAP...10..037D}).
We introduce a scenario that is an average of the previous two and defined as $\rho_{\odot}=0.345$ GeV/cm$^3$ and $M_{200} = 5.9\times 10^{11} M_{\odot}$.
We also use the Einasto profile with $\alpha=0.13$, that provides a good fit to the GCE data, with the above cited three set of values for $\rho_{\odot}$ and $M_{200}$.
We employ the parameters reported for the previous three set of $\rho_{\odot}$ and $M_{200}$ to estimate the value of $r_s$ and $\rho_s$. In particular these two parameters are found by fixing the local DM density to the values $\rho_{\odot} = [0.300,0.345,0.390]$ GeV/cm$^3$ and by integrating the DM density as $\int_0^{r_{200}} d^3r \rho(\rho_s,r_s)$ such that $M_{200}$ is equal to $M_{200} = [5.5,5.9,6.3]\times 10^{11} M_{\odot}$, respectively for each $\rho_{\odot}$ value.

We report in Tab.~\ref{tab:models} the best-fit values for $\rho_s$ and $r_s$ that we find applying this technique to the three DM density models used in the paper.
Since we assume three DM density profiles and we consider three possible choices of the quantities $\rho_{\odot}$ and $M_{200}$, we end up with nine possible scenarios for the parametrization of $\rho$.
We calculate for each of the nine cases the value of $\bar{\mathcal{J}}$ using Eq.~\ref{eq:geom}.
As expected with a larger value of the local DM density also the value for the geometrical factor is larger.
In particular by looking to the gNFW with $\gamma=1.3$ case, $\bar{\mathcal{J}}$ changes from 155 to 339 by varying $\rho_{\odot}$ from 0.30 to 0.39 GeV/cm$^3$. This increase is proportional to the variation of $\rho^2_{\odot}$.
Moreover, by changing the DM density profile from gNFW with $\gamma=1.2$ to $\gamma=1.3$ and Einasto, the geometrical factor increases by a factor of 1.4 and 2.7, respectively.
We can thus choose three of the nine cases as representative of the variation of $\bar{\mathcal{J}}$ due to the modeling of the DM density, and in particular in its local density and functional form.
These are the cases gNFW with $\gamma=1.2$ and $\rho_{\odot}=0.300$ GeV/cm$^3$, labeled as {\tt MIN}, $\gamma=1.3$ and $\rho_{\odot}=0.345$ GeV/cm$^3$, named as {\tt MED}, and 
Einasto with $\rho_{\odot}=0.390$ GeV/cm$^3$ with {\tt MAX}.
The value of the geometrical factor varies by a factor of 6.2 between the {\tt MIN} and the {\tt MAX} models. 

The variation we consider in this paper for $\rho$ encompasses the systematic on the choice of the DM density profile and the local DM density.
However, some of the literature papers find an even larger variation because of estimate of the local DM density that is beyond our range of $0.30-0.39$ GeV/cm$^3$. For example, Refs.~\cite{2019JCAP...09..046K,Benito:2020lgu} report values larger than 
$0.40$ GeV/cm$^3$. 
Using the results of these latter papers would have the consequence of providing smaller values of annihilation cross section with respect to the ones reported in this paper.
Our results on the systematic of the DM density distribution are similar to the ones obtained recently in Ref.~\cite{Benito:2019ngh} with the Milky Way rotation curve data. In particular, their variation of the DM density parameters produce a systematics on the value of the geometrical factor similar to what we estimate using the models {\tt MIN}, {\tt MED} and {\tt MAX}.

We only assume annihilating DM because fitting the GCE surface brightness data with this model provides DM density profiles compatible with expectations from N body simulations. Instead, the calculation of the geometrical factor for decaying DM would be proportional to $\rho$. Therefore, in order to fit well the GCE data, values around $\gamma \sim 2.4$ are required. These are much larger than the N-body simulation predictions that give $\gamma \sim 1$.

\subsection{Fitting the Galactic center excess SED}
\label{sec:fitGCE}

In this section we fit the GCE SED measured in Ref.~\cite{Dimaurodata} in order to find the best-fit DM mass and annihilation cross section.
We use Eq.~\ref{eq:fluxICsimplified} to calculate the $\gamma$-ray flux for the prompt and ICS emission.

\subsubsection{Single channel case}
\label{sec:onechannels}

\begin{figure}
\includegraphics[width=0.49\textwidth]{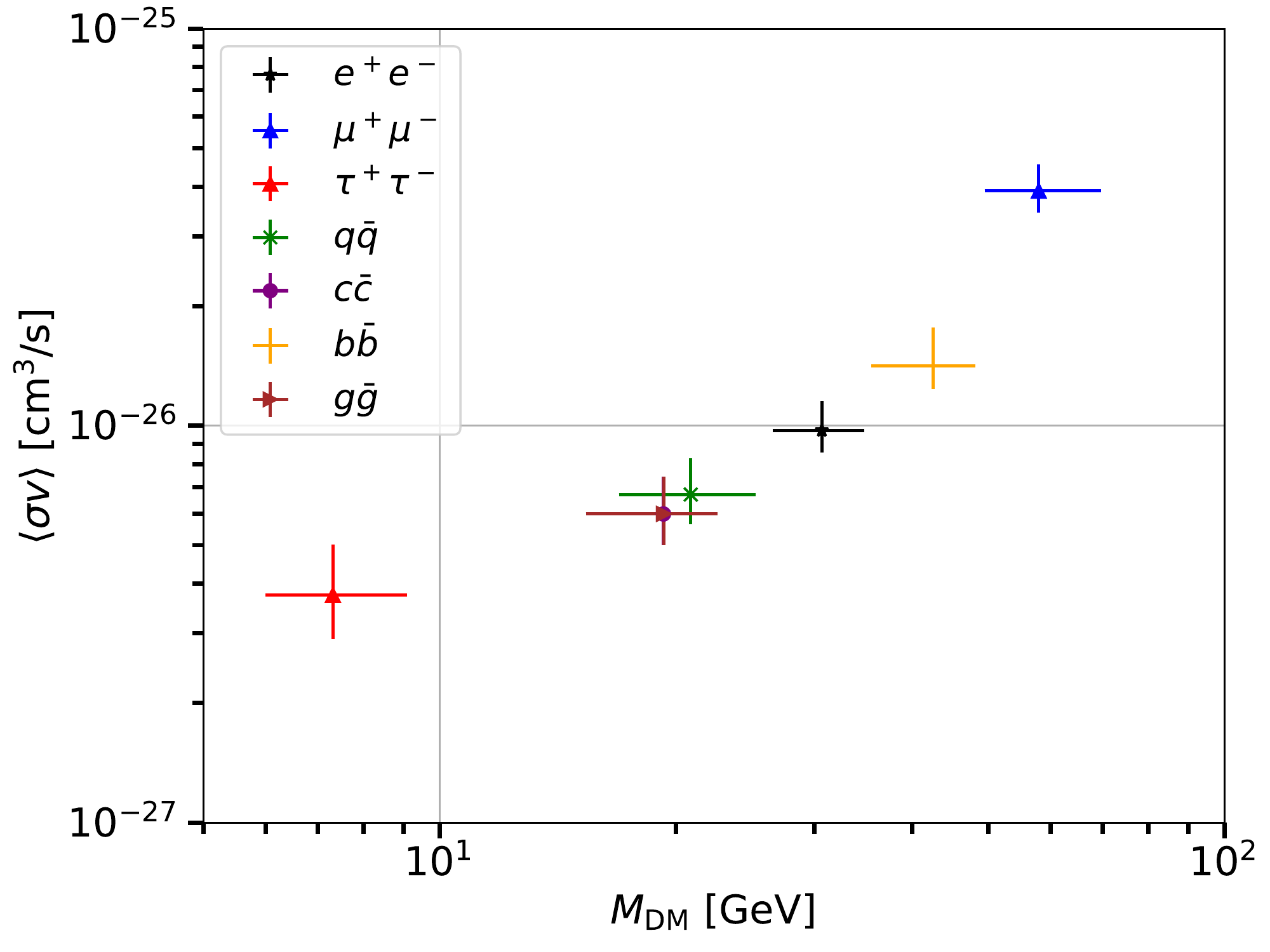}
\caption{Best-fit for the DM parameters $M_{\rm{DM}}$ and $\langle \sigma v \rangle$ obtained by fitting the GCE data in Ref.~\cite{Dimaurodata}. The values of these data points are reported in Tab.~\ref{tab:singlefit}. The green data point labeled with $q\bar{q}$ denotes a DM annihilation channel into the light quarks $u,d,s$.}
\label{fig:Msigmasingle}
\end{figure}

\begin{table}
\begin{center}
\begin{tabular}{|c|c|c|c|}
\hline
\hline
Channel  & $M_{\rm{DM}}$ [GeV] & $\langle \sigma v \rangle$ [$\times 10^{-26}$ cm$^3$/s]  &  $\chi^2(\tilde{\chi}^2)$\\
\hline
$e^+e^-$  & $30^{+4}_{-4}$ & $1.13^{+0.21}_{-0.12}$ & $161.61 \, (5.39)$ \\
$\mu^+\mu^-$  & $ 58^{+11}_{-9}$ & $3.9^{+0.5}_{-0.6}$ & $164.12 \, (5.47)$ \\
$\tau^+\tau^-$  & $ 7.2^{+1.9}_{-1.2}$ & $0.43^{+0.15}_{-0.10}$ & $1178.40 \, (39.3)$ \\
\hline
$q\bar{q}$  & $ 21^{+4}_{-4}$ & $0.77^{+0.19}_{-0.12}$ & $208.89 \, (6.96)$ \\
$c\bar{c}$  & $ 20^{+3}_{-5}$ & $0.70^{+0.16}_{-0.11}$ & $214.11 \, (7.14)$ \\
$b\bar{b}$  & $ 42^{+6}_{-7}$ & $1.41^{+0.35}_{-0.18}$ & $176.47 \, (5.88)$ \\
$gg$  & $19^{+3}_{-4}$ & $0.70^{+0.16}_{-0.11}$ & $214.14 \, (7.14)$ \\
\hline
\hline
\end{tabular}
\caption{This table reports the best-fit for the DM parameters $M_{\rm{DM}}$ and $\langle \sigma v \rangle$ derived by fitting the GCE data obtained in Ref.~\cite{Dimaurodata} with different IEMs. The errors on $M_{\rm{DM}}$ and $\langle \sigma v \rangle$ represent the variation of the best-fit values due to the systematic on the IEMs. We also display the value of the $\chi^2$ ($\tilde{\chi}^2$).}
\label{tab:singlefit}
\end{center}
\end{table}

\begin{figure}
\includegraphics[width=0.45\textwidth]{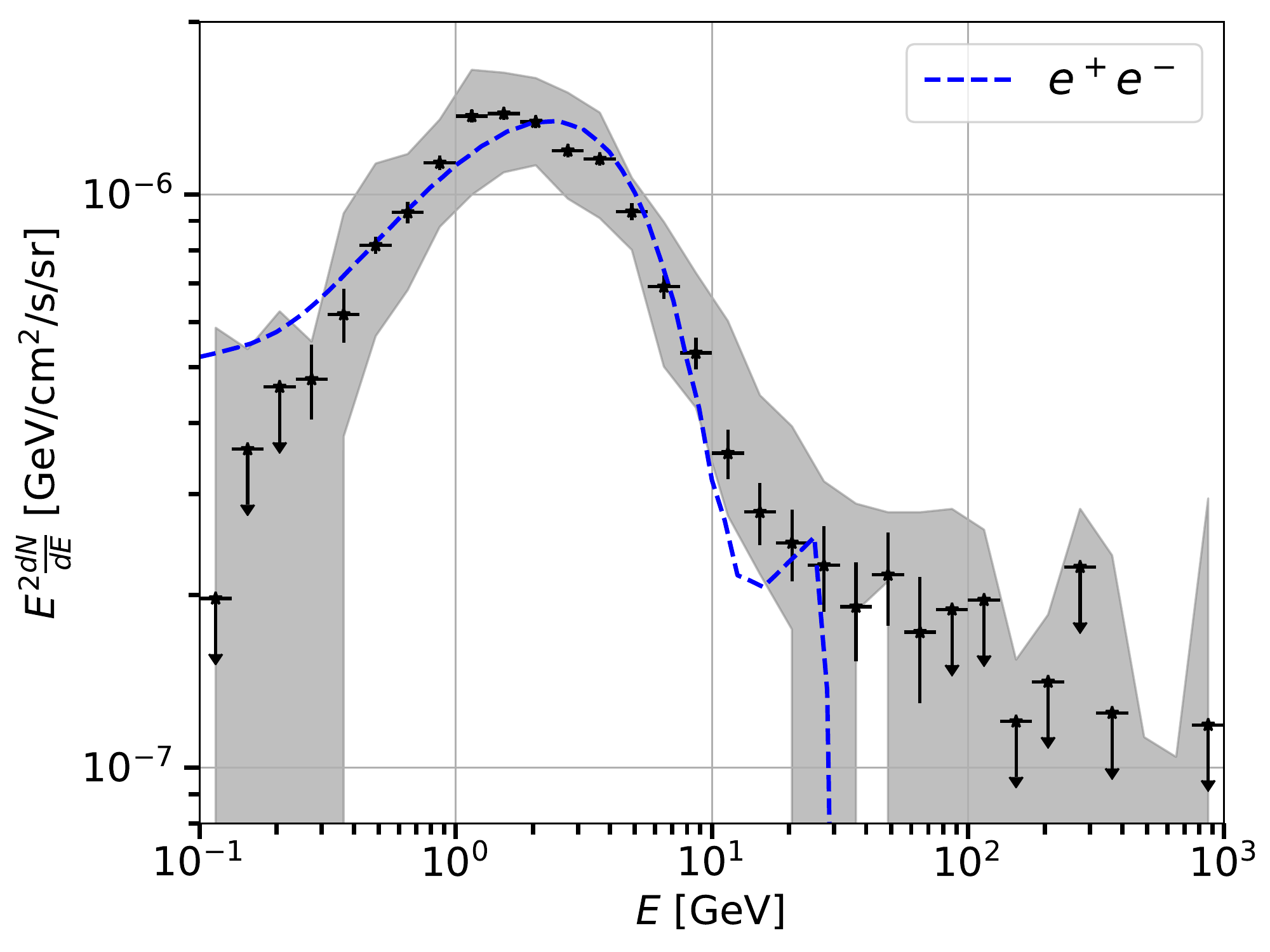}
\includegraphics[width=0.45\textwidth]{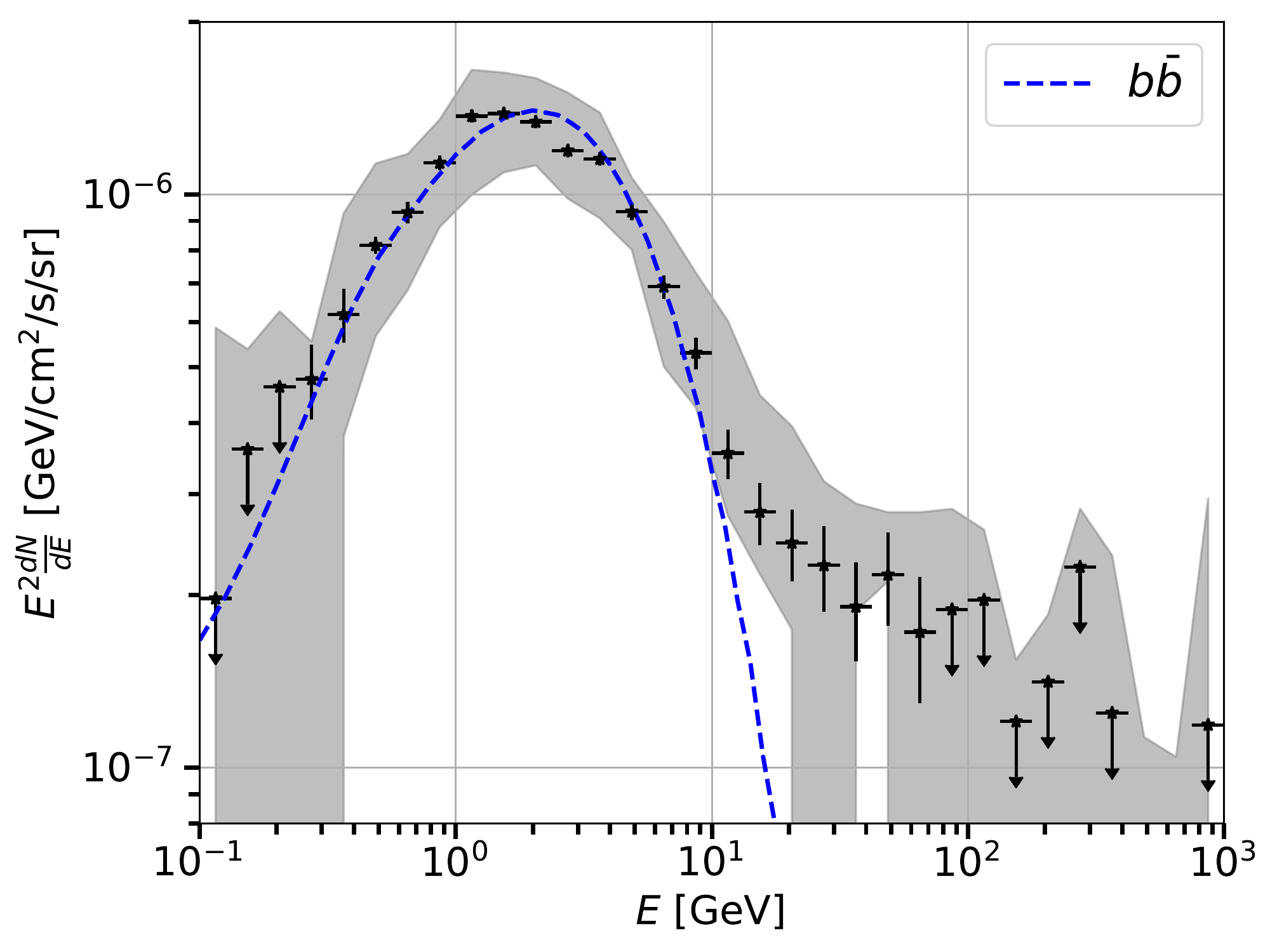}
\caption{Best-fit $\gamma$-ray flux obtained with $e^+e^-$ (top panel) and $b\bar{b}$ (bottom panel) annihilation channels (blue dashed line) compared to the GCE data (black data points) reported in Ref.~\cite{Dimaurodata}. The grey band takes into account the variation in the GCE data found by performing the analysis with different analysis techniques and IEMs.}
\label{fig:singlechannel}
\end{figure}

First we assume the simplest scenario with DM particles annihilating into a single channel (i.e.~$Br=1$).
We consider the following channels: leptonic ($e^{+}e^{-}$, $\mu^{+}\mu^{-}$, $\tau^{+}\tau^{-}$), quarks $q\bar{q}$ ($q = u,d,s$ denotes a light quark), $c\bar{c}$, $b\bar{b}$ and gluon Gauge bosons $gg$.
All the plots and $\chi^2$ values are found by fitting the GCE data obtained in Ref.~\cite{Dimaurodata} with the {\tt Baseline} IEM.
The case with the $t\bar{t}$ quark, the Gauge bosons $Z^0Z^0$, $W^+W^-$ and the Higgs bosons $hh$ provide very poor fits to the GCE flux since the masses of these particles are higher than at least 80 GeV and the GCE flux peaks at much smaller energies. Therefore, we decide to avoid reporting the results we obtain with these channels.

In Tab.~\ref{tab:singlefit} and Fig.~\ref{fig:Msigmasingle} we show the results for the best fit of $M_{\rm{DM}}$ and $\langle \sigma v \rangle$.
The errors represent the variation of the DM parameters derived by fitting the GCE SED data obtained with the different IEMs in Ref.~\cite{Dimaurodata}.
The annihilation channels that provide the best match with the data, with increasing values of the chi-square ($\chi^2$), are: $e^{+}e^{-}$, $\mu^{+}\mu^{-}$, $b\bar{b}$, $q\bar{q}$, $c\bar{c}$, $g\bar{g}$ and $\tau^{+}\tau^{-}$.
The reduced chi-square $\tilde{\chi}^2=\chi^2/d.o.f.$ obtained for the quarks channels $b\bar{b}$, $c\bar{c}$, $q\bar{q}$ is between 6 and 7 while for the $e^{+}e^{-}$ and $\mu^{+}\mu^{-}$ ones is about 5.4. The channel $\tau^+\tau^-$, instead, provides a much poorer fit with $\tilde{\chi^2}=39.3$. Therefore, this latter channel alone is not able to explain sufficiently well the GCE SED. 
The cases $c\bar{c}$, $q\bar{q}$ and $g\bar{g}$ provide very similar results for the DM parameters and goodness of fit. In fact, the intrinsic $\gamma$-ray spectrum $dN_{\gamma}/dE$ is very similar for these channels (see Fig.~3 of \cite{Cirelli:2010xx}).
The results obtained for the single channel are similar to the ones published, for example, in Refs.~\cite{Calore:2014nla,Daylan:2014rsa}.

As shown in Sec.~\ref{sec:DMgammaray} the inclusion of diffusion process and Bremsstrahlung emission in the calculation can slightly affect the results. 
In particular, considering the DM masses we find from the GCE SED, these two ingredients would change in opposite directions the predictions, i.e.~the inclusion of diffusion (Bremsstrahlung) decreases (increases) the predictions for the $\gamma$-ray flux.
The variations for the best-fit values of $M_{\rm{DM}}$ and $\langle \sigma v \rangle$ depends on the specific value of the gas density considered in the analysis. Assuming a value of about 1 cm$^{-3}$ the changes in the best-fit values for the DM coupling parameters are minimal.

We show the results obtained with $e^{+}e^{-}$ and $b\bar{b}$ annihilation channels in Fig.~\ref{fig:singlechannel}.
In particular the flux for the $e^{+}e^{-}$ channel is dominated by the ICS contribution that has a peak at about a few GeV. Instead, for the $b\bar{b}$ channel the SED is  mainly due to the prompt emission.
As expected, the peak of the prompt emission for the $b\bar{b}$ channel is at about a factor of 10 smaller energy than the DM mass.
The values of $\tilde{\chi^2}$ are larger than 1 for all channels meaning that the fit is not sufficiently good. For the hadronic channels the reason is that the $\gamma$-ray flux has a strong softening above roughly 1/10 of the DM mass. Therefore, the $\gamma$-ray flux above 10 GeV is much smaller than the GCE data (see bottom panel of Fig.~\ref{fig:singlechannel} for the $b\bar{b}$ channel).
Instead, the leptonic channels $e^{+}e^{-}$ and $\mu^{+}\mu^{-}$ gives a larger contribution above 10 GeV thanks to interplay between the ICS and prompt emission. However, the SED from DM is systematically above the data between $0.1-0.4$ GeV. 
Therefore, even if the DM contribution in a single channel scenario represents well the GCE SED at the peak where the excess is more significantly detected, all the channels are not able to reproduce well enough the low or high-energy tails at the same time.

We also test a possible variation of the ICS contribution that is particularly relevant for the leptonic channels. 
In order to do so, we add an additional free parameter that renormalizes the ISRF density.
The fit for the $e^+e^-$ channel improves significantly with a $\Delta \chi^2=25$ and a renormalization of the ISRF of 0.70.
The best-fit values for $M_{\rm{DM}}$ and $\langle \sigma v \rangle$ become $28.5$ GeV and  $1.3 \times 10^{-26}$ cm$^3$/s.
The improvement with $\mu^+\mu^-$ gives $\Delta \chi^2=10$, a renormalization of the ICS flux of 1.4, $M_{\rm{DM}}=56$ GeV and $\langle \sigma v \rangle = 2.8 \cdot 10^{-26}$ cm$^3$/s.
Instead, the $\tau^+\tau^-$ channel continues to provide a poor fit to the GCE SED.
The $b\bar{b}$ channel fit improves by $\Delta \chi^2=20$ with a renormalization of the ICS flux of 4.5 but the best-fit values for $M_{\rm{DM}}$ and $\langle \sigma v \rangle$ remain unchanged with respect to Tab.~\ref{tab:singlefit}.
Variations of the ISRF density of the order of $30\%$ from the one in Ref.~\cite{Porter_2008} are possible considering the current uncertainties in modeling the ISRF in the center of the Milky Way (see, e.g., the differences between the mode in Ref.~\cite{Porter_2008} and \cite{Vernetto:2016alq}).

\subsubsection{Two and three channels cases}
\label{sec:twochannels}

In this section we investigate a more complicated scenario where DM particles annihilate into two or three annihilation channels.
In order to account for these cases we use a branching ratio $Br$ that multiplies the annihilation cross section of the first channel, as in Eq.~\ref{eq:fluxICsimplified}, while the second channel is multiplied by $1-Br$.
For example, a case with $Br = 0.7$ for the $\mu^+\mu^- - b\bar{b}$ case implies that $\langle \sigma v \rangle$ is multiplied for 0.7 for the former and 0.3 for the latter channel as follow:
\begin{equation}
\frac{dN_{\gamma}}{dE} = Br \frac{dN_{\mu^+\mu^-}}{dE} + (1-Br) \frac{dN_{b\bar{b}}}{dE}
\label{eq:eqBr}
\end{equation}
The procedure we use to find the DM coupling parameters is the same applied for the single channel in the previous section.
In Tab.~\ref{tab:twochannels} we show the best-fit values for the DM parameters $M_{\rm{DM}}$, $\langle \sigma v \rangle$ and $Br$ found by fitting the GCE flux data obtained with the {\tt Baseline} IEM. Instead in Tab.~\ref{tab:twochannelsIEM} we show the uncertainties for the same parameters derived when we fit the DM flux to the GCE data obtained with different IEMs as in Ref.~\cite{Dimaurodata}.
We do not consider here DM annihilating into $q\bar{q}$ and $g\bar{g}$ since it gives very similar results to $c\bar{c}$ (see Tab.~\ref{tab:singlefit}).

\begin{figure}
\includegraphics[width=0.49\textwidth]{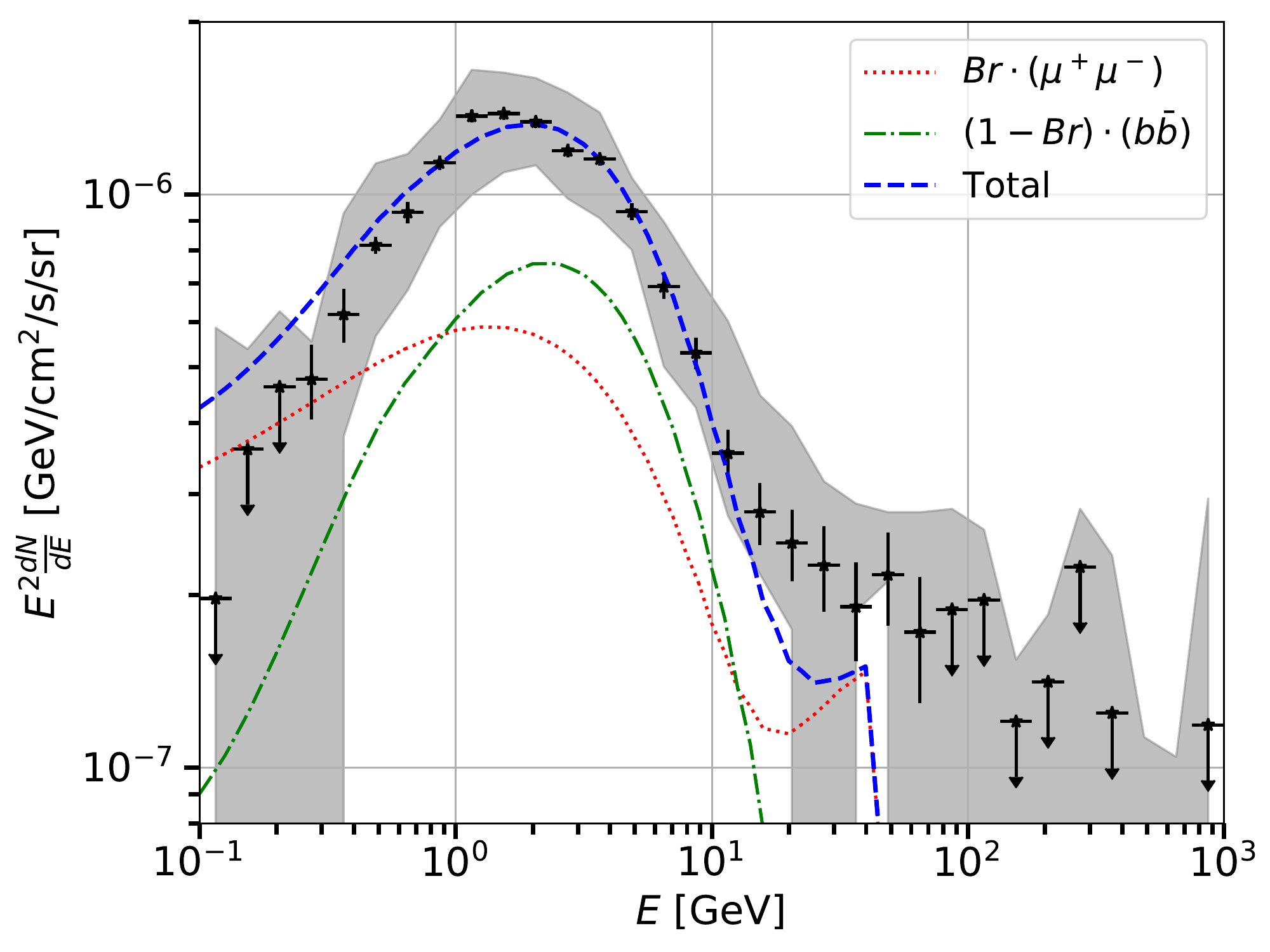}
\includegraphics[width=0.49\textwidth]{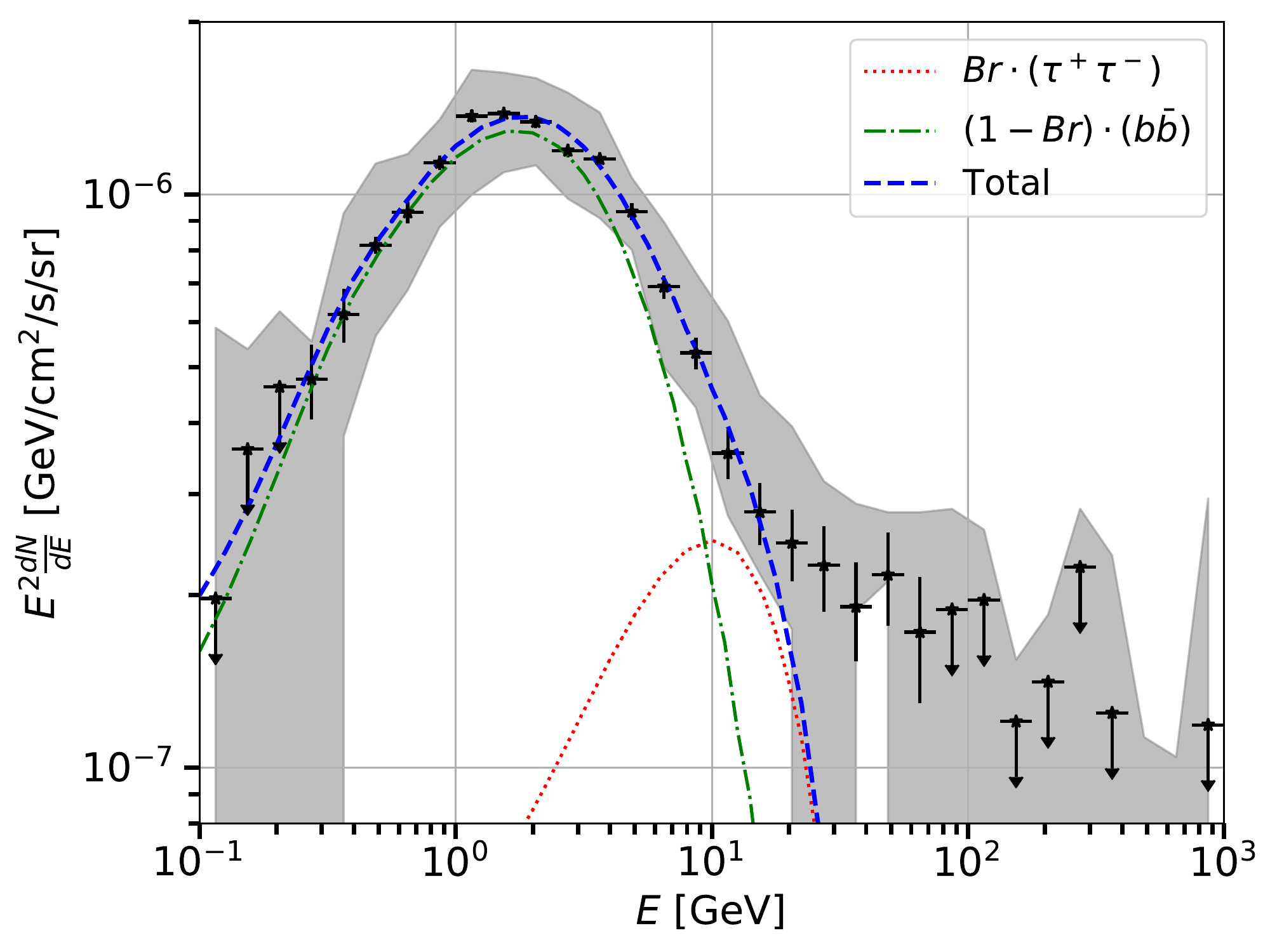}
\includegraphics[width=0.49\textwidth]{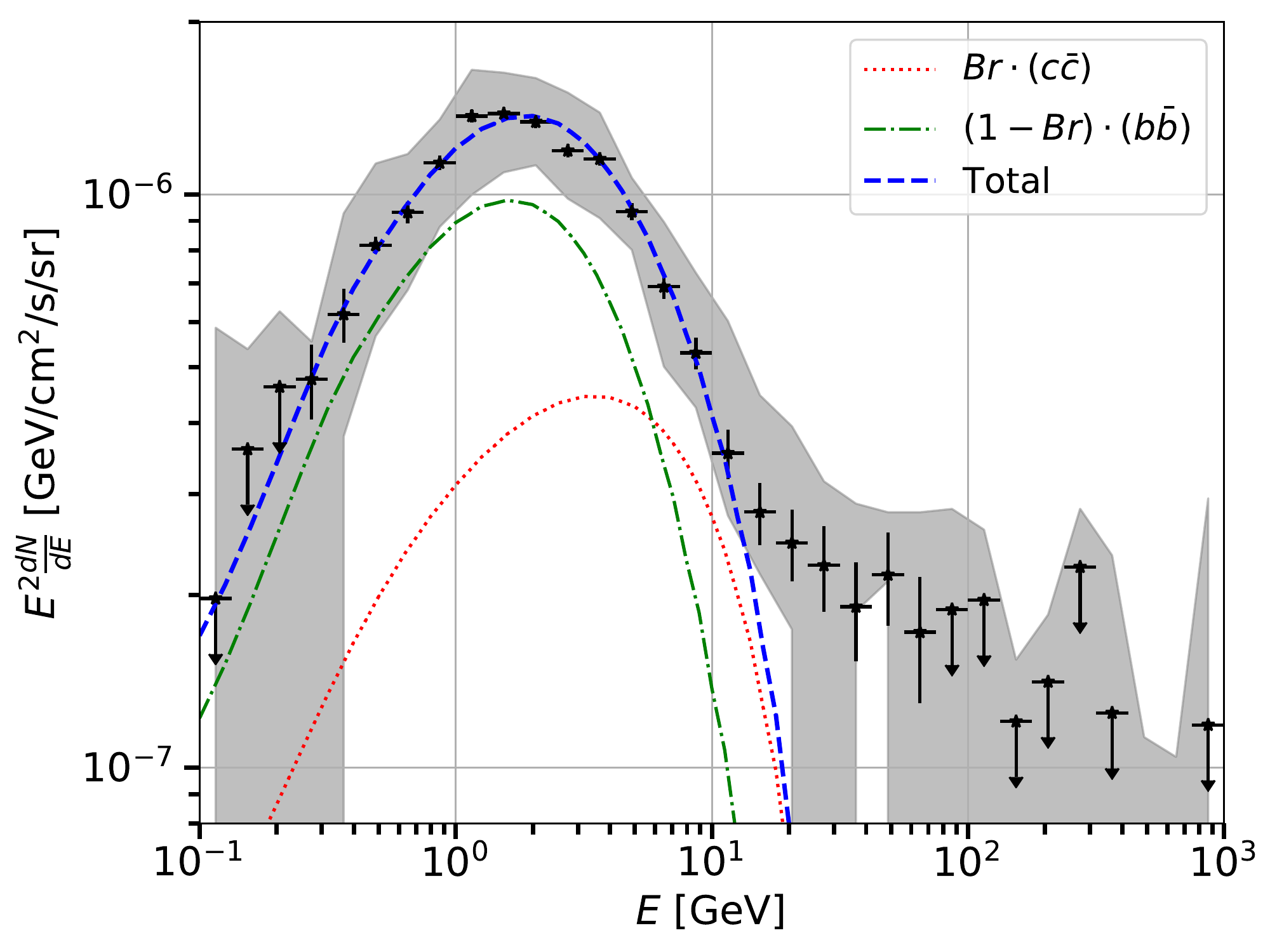}
\caption{Flux of $\gamma$ rays from DM particle annihilating into two channels. We show the contribution of both channels and the total flux compared to the GCE flux data.}  
\label{fig:twochannels}
\end{figure}

The DM candidates that provide the largest improvement in the goodness of fit with respect to Sec.~\ref{sec:onechannels} are $\mu^+\mu^- - b\bar{b}$ and $\tau^+\tau^- - b\bar{b}$ with $\Delta \chi^2$ of 74 and 82, respectively.
These values of $\Delta \chi^2$ are associated with the additional parameter $Br$ and they imply $8.4$ and $9.0\sigma$ significance for the two channels with respect to the single one.
The DM parameters required to fit the GCE flux data are $M_{\rm{DM}} \sim 50$ (35) GeV, $\langle \sigma v \rangle \sim 3 \times 10^{-26}$ ($1.4 \times 10^{-26}$) cm$^3$/s and $Br \sim 0.7$ (0.2) for the $\mu^+\mu^- - b\bar{b}$ ($\tau^+\tau^- - b\bar{b}$) DM candidate.
Other cases provide a significant improvements such as $c\bar{c} - b\bar{b}$, $e^+e^- - b\bar{b}$ and $e^+e^- - c\bar{c}$ at the $7.7$, $5.5 \sigma$ level.
In Fig.~\ref{fig:twochannels} we show the best fit we obtain for $\mu^+\mu^- - b\bar{b}$, $\tau^+\tau^- - b\bar{b}$ and $c\bar{c} - b\bar{b}$.
In particular we see that the two channels provide a better fit to the GCE flux data because the total contribution of $\gamma$-ray from DM cover also the energies between 10-30 GeV where the single channel was not able to contribute significantly.
Instead, the channels $\mu^+\mu^- - \tau^+\tau^-$, $\mu^+\mu^- - c\bar{c}$ and $\tau^+\tau^- - c\bar{c}$, do not provide any improvement in the fit since the branching ratio value is 0 or 1, i.e.~they provide a fit with the same $\chi^2$ of the single channel with $\mu^+\mu^-$ or $c\bar{c}$.

\begin{figure}
\includegraphics[width=0.49\textwidth]{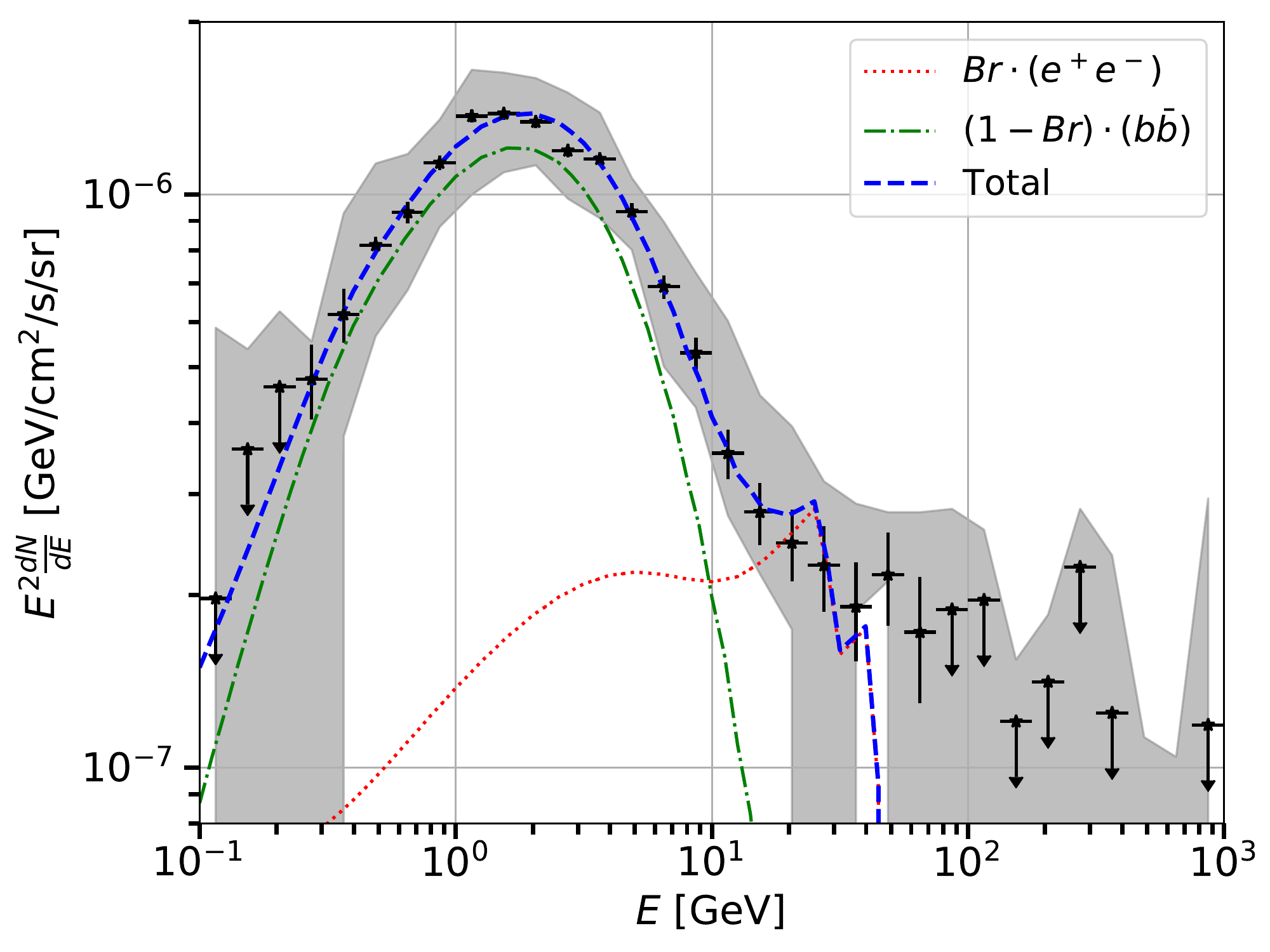}
\caption{Same as Fig.~\ref{fig:twochannels} leaving free to vary also a renormalization of the ICS contribution. For the case reported in this figure a renormalization of the ICS emission of 0.1 is found from the fit.}  
\label{fig:twochannelsICS}
\end{figure}

We also test a possible variation of the ISRF density that could change the ICS contribution.
We perform a fit to the GCE flux by adding a free parameter for the ICS component.
We find that the goodness of fit improves significantly for the  $e^+e^- - c\bar{c}$, $e^+e^- - b\bar{b}$ and $\mu^+\mu^- - b\bar{b}$ with a ISRF renormalization with respect to the model in Ref.\cite{Porter_2008} of 0.33, 0.10 and 0.10, respectively.
The best-fit values found for the ICS renormalization are equivalent of reducing the starlight density in the inner part of the Milky Way. Values of $0.1-0.3$ makes the ISRF density we use for the Galactic center similar to the local one \cite{Porter_2008,Vernetto:2016alq}.
The $\tilde{\chi}^2$ we find for these three cases are 3.1, 1.8 and 1.8 so the fit improves significantly ($\Delta\chi^2=24,60,40$). The best-fit values for the DM parameters we obtain in this case are reported in the bottom block of Tab.~\ref{tab:twochannels}.
We show the DM candidate $e^+e^- - b\bar{b}$ that best fits the GCE SED in Fig.~\ref{fig:twochannelsICS}. We can see that the fit improves significantly, with respect to the case with renormalization equal to 1, because with a fainter ICS flux, the low energy flux is more compatible with the GCE data and the prompt emission for the $e^+e^-$ channel reproduced very well the flux above 10 GeV that is difficult to fit in the models tested before.

\begin{table*}
\begin{center}
\begin{tabular}{|c|c|c|c|c|c|c|}
\hline
\hline
Channel 1  & Channel 2  & $M_{\rm{DM}}$ & $\langle \sigma v \rangle$ & $Br$  &$\chi^2 (\tilde{\chi}^2)$ & $\Delta \chi^2 ($sign.$)$ \\
\hline
  &  & [GeV] & [$10^{-26}$ cm$^3$/s] &  &  & \\
\hline
$e^+e^-$ & $\mu^+\mu^-$ &  $32.66 \pm 0.66$ & $1.32 \pm 0.07$& $0.64 \pm 0.05$ & $126.6(4.37)$  &  $18(4.1\sigma)$ \\
$e^+e^-$ & $\tau^+\tau^-$ &  $27.07 \pm 0.58$ & $0.95 \pm 0.01$& $0.84 \pm 0.03$ & $113.7(3.92)$  &  $31(5.4\sigma)$  \\
$e^+e^-$ & $c\bar{c}$ &  $24.30 \pm 0.57$ & $0.79 \pm 0.02$& $0.50 \pm 0.05$ & $112.3(3.87)$  &  $32(5.5\sigma)$  \\
$e^+e^-$ & $b\bar{b}$ &  $34.73 \pm 0.89$ & $1.10 \pm 0.03$& $0.50 \pm 0.07$ & $112.9(3.89)$   &  $32(5.5\sigma)$  \\
\hline
$\mu^+\mu^-$ & $\tau^+\tau^-$ &  $55.23 \pm 0.72$ & $3.77 \pm 0.05$& $1.00 \pm 0.00$ & $164.1(5.66)$   &  $0(0\sigma)$  \\
$\mu^+\mu^-$ & $c\bar{c}$ &  $55.22 \pm 0.72$ & $3.77 \pm 0.05$& $1.00 \pm 0.01$ & $164.1(5.66)$   &  $0(0\sigma)$  \\
$\mu^+\mu^-$ & $b\bar{b}$ &  $47.82 \pm 0.92$ & $2.42 \pm 0.14$& $0.65 \pm 0.05$ & $90.5(3.12)$   &  $74(8.4\sigma)$  \\
\hline
$\tau^+\tau^-$ & $c\bar{c}$ &  $18.57 \pm 0.27$ & $0.56 \pm 0.01$& $0.00 \pm 0.04$ & $214.1(7.38)$   &  $0(0\sigma)$  \\
$\tau^+\tau^-$ & $b\bar{b}$ &  $35.93 \pm 0.98$ & $1.32 \pm 0.03$& $0.20 \pm 0.02$ & $82.0(2.83)$   &  $82(9.0\sigma)$  \\
\hline
$c\bar{c}$ & $b\bar{b}$ &  $33.79 \pm 1.48$ & $1.11 \pm 0.05$& $0.32 \pm 0.04$ & $115.1(3.97)$   &  $61(7.7\sigma)$  \\
\hline
\hline
\hline
$e^+e^-$ & $c\bar{c}$ &  $20.00 \pm 0.55$ & $1.00 \pm 0.18$ & $0.56 \pm 0.03$ &  $88.9(3.07)$   &  $56(7.1\sigma)$ \\
$e^+e^-$ & $b\bar{b}$ &  $35.96 \pm 0.81$ & $2.30 \pm 0.17$ & $0.56 \pm 0.03$ & $51.7(1.78)$   &  $79(8.4\sigma)$ \\
$\mu^+\mu^-$ & $b\bar{b}$ &  $38.01 \pm 0.95$ & $3.64 \pm 0.21$ & $0.70 \pm 0.02$ & $50.7(1.75)$   &  $58(7.2\sigma)$ \\
\hline
\hline
\end{tabular}
\caption{This table reports the best-fit for the DM parameters $M_{\rm{DM}}$, $\langle \sigma v \rangle$ and $Br$ derived by fitting the GCE data in Ref.~\cite{Dimaurodata} obtained with the {\tt Baseline} IEM. The annihilation cross section multiples $Br$ for channel 1 and $(1- Br)$ for channel 2 as reported in Eq.~\ref{eq:eqBr}. We also display the value of the $\chi^2$ ($\tilde{\chi}^2$) and in last column the difference of $\chi^2$ (significance) between the case of the two channel and the single channel reported in Tab.~\ref{tab:singlefit}. The last three rows represent the results we find if we leave free to vary the ISRF density with a renormalization factor with respect to the model in Ref.\cite{Porter_2008}. The best-fit values for this renormalization factor is for the three DM candidates, from top to bottom: 0.33, 0.10 and 0.10.}
\label{tab:twochannels}
\end{center}
\end{table*}

\begin{table}
\begin{center}
\begin{tabular}{|c|c|c|c|c|}
\hline
\hline
Channel 1  & Channel 2  & $M_{\rm{DM}}$ & $\langle \sigma v \rangle$ & $Br$ \\
\hline
  &  & [GeV] & [$10^{-26}$ cm$^3$/s] &  \\
\hline
$e^+e^-$  &  $\mu^+\mu^-$ &  $43.3_{-15.9}^{+16.2}$  &  $2.35_{-1.44}^{+1.70}$  &  $0.42_{-0.42}^{+0.58}$ \\
$e^+e^-$  &  $\tau^+\tau^-$ &  $27.4_{-4.1}^{+4.8}$  &  $0.97_{-0.10}^{+0.18}$  &  $0.82_{-0.20}^{+0.16}$ \\
$e^+e^-$  &  $c\bar{c}$ &  $27.8_{-7.8}^{+6.9}$  &  $0.89_{-0.19}^{+0.26}$  &  $0.73_{-0.31}^{+0.27}$ \\
$e^+e^-$  &  $b\bar{b}$ &  $36.7_{-3.8}^{+6.5}$  &  $1.19_{-0.20}^{+0.34}$  &  $0.41_{-0.23}^{+0.30}$ \\
\hline
$\mu^+\mu^-$  &  $\tau^+\tau^-$ &  $57.4_{-7.9}^{+12.3}$  &  $3.88_{-0.56}^{+0.67}$  &  $0.99_{-0.05}^{+0.01}$ \\
$\mu^+\mu^-$  &  $c\bar{c}$ &  $48.5_{-8.5}^{+11.5}$  &  $3.02_{-0.52}^{+0.75}$  &  $0.87_{-0.13}^{+0.13}$ \\
$\mu^+\mu^-$  &  $b\bar{b}$ &  $53.0_{-7.1}^{+9.2}$  &  $2.83_{-0.54}^{+1.00}$  &  $0.71_{-0.15}^{+0.21}$ \\ 
\hline
$\tau^+\tau^-$  &  $c\bar{c}$ &  $19.4_{-4.8}^{+3.2}$  &  $0.59_{-0.09}^{+0.15}$  &  $0.03_{-0.03}^{+0.10}$ \\
$\tau^+\tau^-$  &  $b\bar{b}$ &  $34.9_{-4.1}^{+5.5}$  &  $1.35_{-0.14}^{+0.36}$  &  $0.23_{-0.11}^{+0.10}$ \\
\hline
$c\bar{c}$  &  $b\bar{b}$ &  $34.1_{-3.8}^{+4.3}$  &  $1.15_{-0.14}^{+0.31}$  &  $0.32_{-0.17}^{+0.16}$ \\
\hline
\hline
\hline
\end{tabular}
\caption{Same as Tab.~\ref{tab:twochannels} but for the fit performed on the GCE data obtained in Ref.~\cite{Dimaurodata} with different IEMs. Therefore, the errors on the DM parameters are due to the variation in the results obtained by fitting the GCE SED data obtained in Ref.~\cite{Dimaurodata} with a variation of the choice for the interstellar emission.}
\label{tab:twochannelsIEM}
\end{center}
\end{table}

We finally test whether three annihilation channels improve further the fit.\footnote{For DM particles annihilating in three channels, $Br_1$ multiplies the annihilation cross section for channel 1, $Br_2$ multiplies the annihilation cross section for channel 2 and $1-Br_1-Br_2$ multiplies the annihilation cross section for channel 3.}
We consider all the possible combinations of the single channels reported before.
We do not find any significant improvement with respect to the two channel cases. In particular, the DM candidate with the largest improvements are $\mu^+\mu^- - \tau^+\tau^- - b\bar{b}$ with best-fit parameters $M_{\rm{DM}}=40$ GeV, $\langle \sigma v \rangle = 1.76 \times 10^{-26}$ cm$^3$/s, $Br_{1}=0.3$, $Br_{2}=0.1$ and $\mu^+\mu^- - \tau^+\tau^- - b\bar{b}$ for $M_{\rm{DM}}=40$ GeV, $\langle \sigma v \rangle = 1.9\times 10^{-26}$ cm$^3$/s, $Br_1=0.50$, $Br_2=0.15$.
These DM candidates improve the fit by $3.1\sigma$ and $2.8\sigma$ significance with respect to the two channel case.

\section{Dwarf Spheroidal galaxies constraints on the Galactic center excess}
\label{sec:dwarfres}

In this section we investigate whether the DM candidates that explain GCE would generate a detectable signal in the analysis of data from dSphs.
We consider for this scope the list of 48 dSphs published in \cite{2019MNRAS.482.3480P} and the best-fit values and errors for the geometrical factors reported in Tab.~1 and A2. We exclude from the list the satellites of the Andromeda galaxy.
We also test the sample of 41 dSphs used in Ref.~\cite{Fermi-LAT:2016uux}. We select all the objects listed in Tab.~1 of Ref.~\cite{Fermi-LAT:2016uux} except for the ones labeled as ``Ambiguous Systems''. We take the best-fit and errors for the geometrical factors as in Tab.~1 for sources with a measured $\mathcal{J}$ factor while for the others we use the value predicted by the $\mathcal{J}$ factor-distance relation in Eq.~2 of their paper and assuming an error on $\log_{10}{(\mathcal{J})}$ of 0.6.
The differences between the sample of dSphs in the two above cited references are in the list of objects and the estimated geometrical factors. 
Bootes III is not considered in Ref.~\cite{2019MNRAS.482.3480P} while for Tucana III only upper limits for the geometrical factor are reported. Thus this latter object is not included in the analysis for the sample of Ref.~\cite{2019MNRAS.482.3480P}. The objects Aquarius II, Carina II, Cetus, Leo T are not listed in Ref.~\cite{Fermi-LAT:2016uux} while Segue 2 has the chemical signatures of a dSph, but exhibits a low velocity dispersion and so has not been considered.
There are also differences in the best-fit values of the geometrical factor that, however, is for most of the objects well within the $1\sigma$ errors.
We find similar results using the two samples at the $15-20\%$ level in the relevant mass range for the DM interpretation of the GCE (see Sec.~\ref{sec:resdsphs}), i.e.~for $M_{\rm{DM}}\in[10,100]$ GeV.

\subsection{Data selection and analysis technique}
\label{sed:datanalysis}

We select the same exposure time of the GCE analysis \cite{Dimaurodata}, i.e.~eleven years\footnote{Mission Elapsed Time (MET): 239557417 s $-$ 586490000 s} of Pass 8 data (data processing P8R3). 
We select SOURCEVETO class events\footnote{SOURCEVETO is an event class recently created by the {\it Fermi}-LAT team to maximize the acceptance while minimizing the irreducible cosmic-ray background contamination. In fact, SOURCEVETO class has the same contamination level of P8R2\_ULTRACLEANVETO\_V6 class while maintaining the acceptance of P8R2\_CLEAN\_V6 class.}, passing the basic quality filter cuts\footnote{DATA\_QUAL$>$0 \&\& LAT\_CONFIG==1}, and their corresponding P8R3\_SOURCEVETO\_V2 response functions, as in Ref.~\cite{Dimaurodata}.
We choose energies between 0.3 to 1000 GeV and apply a cut to zenith angles $<100^\circ$ between 0.3 to 1 GeV and $<105^\circ$ above 1 GeV in order to exclude the Earth Limb's contamination.
We model the background with sources reported in the 10-year Source Catalog (4FGL)\footnote{\url{https://arxiv.org/pdf/2005.11208.pdf}} which is an extension of the 8-year Source Catalog (4FGL-DR2) \cite{Abdollahi_2020}\footnote{\url{https://fermi.gsfc.nasa.gov/ssc/data/access/lat/10yr_catalog/}}.
We also include the latest released IEM, namely {\tt gll\_iem\_v07.fits}\footnote{A complete discussion about this new IEM can be found at \url{https://fermi.gsfc.nasa.gov/ssc/data/analysis/software/aux/4fgl/Galactic_Diffuse_Emission_Model_for_the_4FGL_Catalog_Analysis.pdf}}, and its corresponding isotropic template {\tt iso\_P8R3\_SOURCEVETO\_V3\_v1.txt}.
We analyze the $12\times12$ deg$^2$ regions of interest (ROI) centered in the dSphs position and choose pixel size of $0.08$ deg. We include in the background model sources located in a region $16\times16$ deg$^2$ in order to include also sources at most $2^{\circ}$ outside our ROI.
We will run the analysis with different choices of some of the assumptions done above to see how the results change. In particular, we change the lower bound of the energy range to $0.5$ GeV, we select ULTRACLEANVETO data, and select a larger ROI of $15\times15$ deg$^2$.

The analysis of the DM search in our sample of dSphs follows the one performed in the past by the {\it Fermi}-LAT Collaboration on these sources (see, e.g., \cite{Ackermann:2015zua}) or more recently in the direction of Andromeda and Triangulum galaxies \cite{DiMauro:2019frs}.
We provide a general overview and we refer to Refs.~\cite{Ackermann:2015zua,DiMauro:2019frs} for a complete description.
We use the public \textit{Fermipy} package (version 0.19.0) to perform a binned analysis with eight bins per energy decade. 
\textit{Fermipy} is a python wrapper of the official {\tt Fermitools}, for which we use version 1.3.8.

\begin{itemize}
\item {\it ROI optimization}. A baseline fit is performed on each ROI including sources in the 4FGL-DR2 catalog, IEM and isotropic template.
A refinement of the model is run by relocalizing all the sources in the model. We check that the new positions are compatible with the ones reported in the 4FGL catalog. Then, we search for new sources with a Test Statistic\footnote{The Test Statistic ($TS$) is defined as twice the difference in maximum log-likelihood between the null hypothesis (i.e., no source present) and the test hypothesis: $TS = 2 ( \log\mathcal{L}_{\rm test} -
  \log\mathcal{L}_{\rm null} )$~\cite{1996ApJ...461..396M}.} ($TS$) $TS>25$ and distant at least $1^{\circ}$ from the center of the ROI. A final fit is then performed, where all the SED parameters of the sources, normalization and spectral index of the IEM and normalization of the isotropic component are free to vary. With this first step we thus have a background model that represents properly the $\gamma$-ray emission in the ROI. In fact, in all the ROIs considered the residuals found by performing a $TS$ map are at most at the level of $\sqrt{TS}\sim 2-3$.
\item {\it DM SED}. The DM source associated with each dSph is added in the center of the ROI as a point source, since their predicted angular extension is for most of them smaller than the {\it Fermi}-LAT PSF (see, e.g.~\cite{2019MNRAS.482.3480P}). A fit is then performed. The SED for the dSphs is calculated by performing a fit energy bin by energy bin. Specifically, the SED run gives for each energy bin the value of the likelihood as a function of the DM energy flux. With the SED information we can thus test every possible spectrum for the source of interest.
\item {\it Conversion from energy flux to DM space}. Specific DM candidates are tested. We use the SED information obtained in step two to calculate, for every annihilation channel, the likelihood as a function of annihilation cross section and DM mass values. For a given DM annihilation channel and mass the theoretical DM SED shape is fixed and for different values of $\langle \sigma v \rangle$ we extract the correspondent likelihoods from the SED data.
\item {\it Extracting the $TS$ for the detection of DM or upper limits for $\langle \sigma v \rangle$}. The DM detection $TS$ is found by finding the minimum of the likelihood in $\langle \sigma v \rangle$ and $M_{\rm{DM}}$ space and comparing it with the likelihood of the null hypothesis, i.e. the one of the optimized ROI fit without the DM emission. The upper limits of $\langle \sigma v \rangle$ are instead calculated in the following way. For a fixed DM mass, we take the likelihood profile as a function of $\langle \sigma v \rangle$ ($\mathcal{L}(\langle \sigma v \rangle)$). We then can calculate the upper limits for $\langle \sigma v \rangle$ by finding the minimum of $\mathcal{L}(\langle \sigma v \rangle)$ and calculating the $\langle \sigma v \rangle$ that worsens the best-fit likelihood value by $\Delta \mathcal{L} = 2.71/2$, which is associated with the one-sided $95\%$ CL upper limits. This is the same procedure used in several other papers where the frequentist approach is employed (see, e.g.~\cite{Fermi-LAT:2016uux}). In finding the $TS$ or the upper limits for $\langle \sigma v \rangle$ we add to the Poissonian term of the likelihood a factor that takes into account the uncertainty on the $\mathcal{J}$ factor (see Eq.~3 in \cite{Ackermann:2015zua}) taken from \cite{2019MNRAS.482.3480P}.
\end{itemize}

\subsection{Results for the detection and upper limits for $\langle \sigma v \rangle$}
\label{sec:resdsphs}
\begin{figure}
\includegraphics[width=0.49\textwidth]{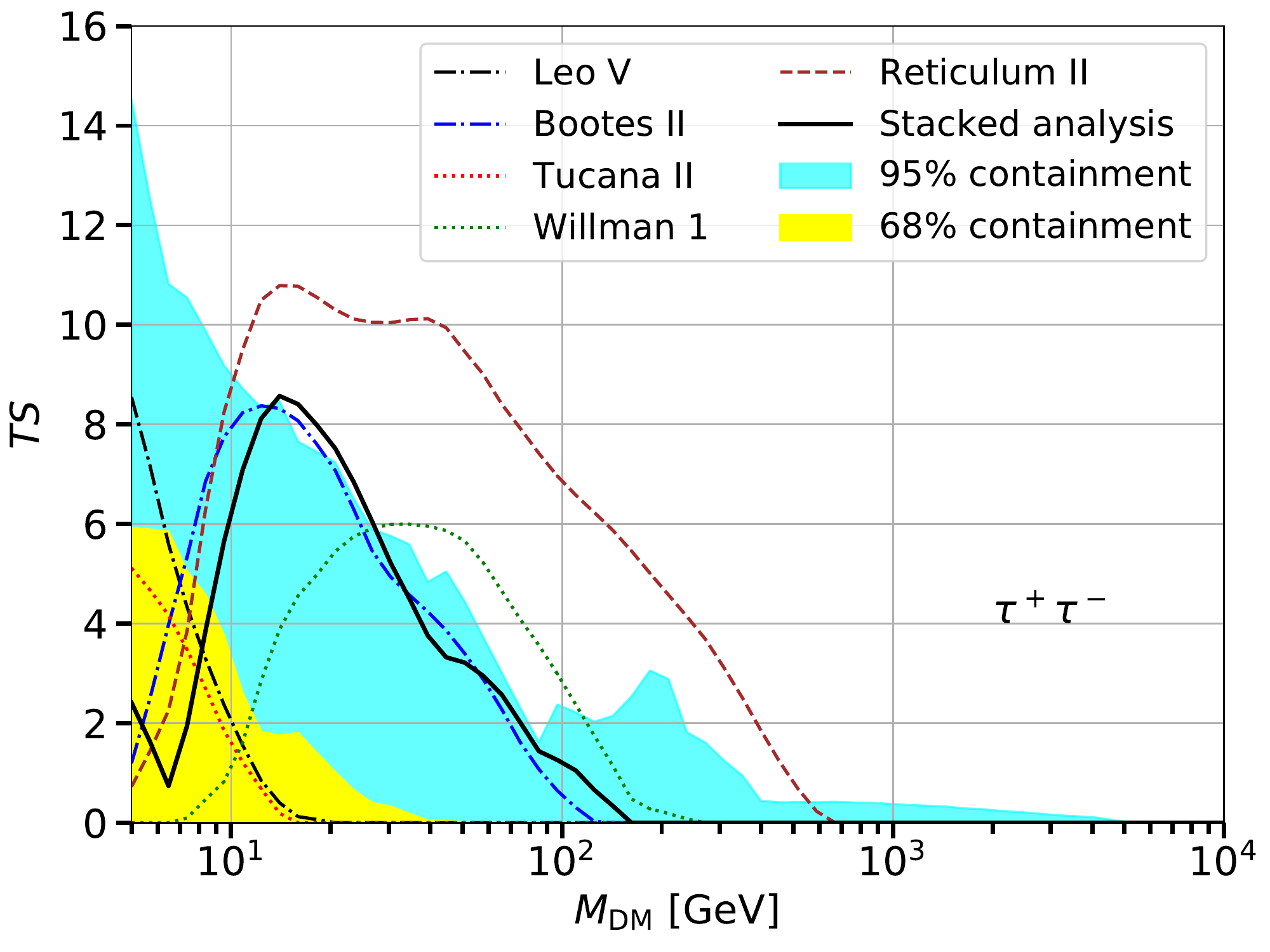}
\includegraphics[width=0.49\textwidth]{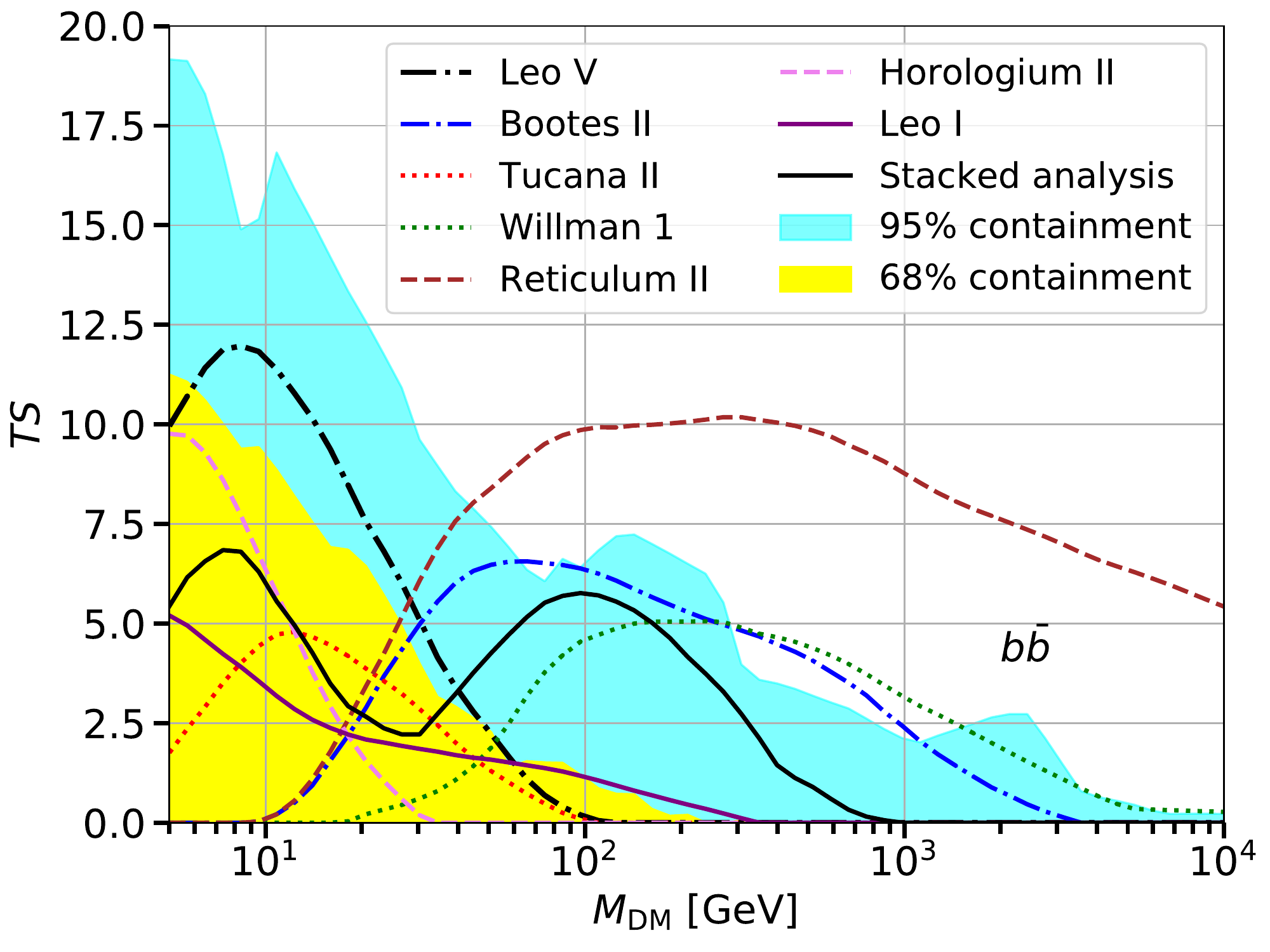}
\caption{$TS$ as a function of mass for the dSphs detected with the highest significance. We also show the $TS$ for the joint likelihood analysis on the dSphs sample and the 68$\%$ and 95$\%$ containment bands for the random direction runs. We show the results for the $\tau^+\tau^-$ (top panel) and $b\bar{b}$ annihilation channels (bottom panel).}
\label{fig:TSdwarfs}
\end{figure}

In this section we report the results for the search of DM in the directions of the dSphs in our sample.
First, we calculate the $TS$ of each individual source. We show in Fig.~\ref{fig:TSdwarfs} the objects for which we find the highest detection significance: Leo V, Tucana II, Willman 1, Reticulum II, Horologium II and Bootes I.
Among the dSphs selected the one detected with the highest $TS$ is Reticulum II with a mass of 300 (40) GeV, $\langle \sigma v \rangle = 1.5\times 10^{-26}$ ($9\times 10^{-27}$) cm$^3$/s for the $b\bar{b}$ ($\tau^+\tau^-$) annihilation channel and detected with a $TS\sim10$, which corresponds to a p-value of $2.2\times 10^{-3}$ ($4.4\times 10^{-3}$) local, i.e.~pre-trials, significance of $\sim 2.8\sigma$ ($2.6\sigma$)\footnote{In order to convert the $TS$ into the p-value and the detection significance, we have considered the analysis in 4800 random directions and derived the $TS$ distribution of the detection of the dSphs.}.
These $TS$ are below the reference value of 25 that is usually used by the {\it Fermi}-LAT Collaboration to include a source in the catalogs. 
In order to verify more precisely if our findings are significant or not, we run the same analysis in 100 random directions in each ROI.
The analysis pipeline is run exactly as before but the dSphs emission is searched in other directions where we do not expect to detect any signal from DM. These simulations provide thus the expected signal in case of the null hypothesis, i.e.~that there is no emission from DM in dSphs.
In Fig.~\ref{fig:TSdwarfs} we show the 68$\%$ and 95$\%$ containment bands for the $TS$ for the runs in the 100 random directions.
The $TS$ profiles found for most of the dSphs are compatible with the results of the random directions except for Reticulum II, Bootes II and Willman 1. 
Once we have the likelihood profile for each dSph as a function of DM mass and annihilation cross section, we can sum all of them together and get the joint combined likelihood profile for the entire sample of dSphs.
The result for the $TS$ as a function of mass for the joint likelihood analysis is not completely contained inside the $95\%$ containment band of the random direction runs.

\begin{figure}
\includegraphics[width=0.49\textwidth]{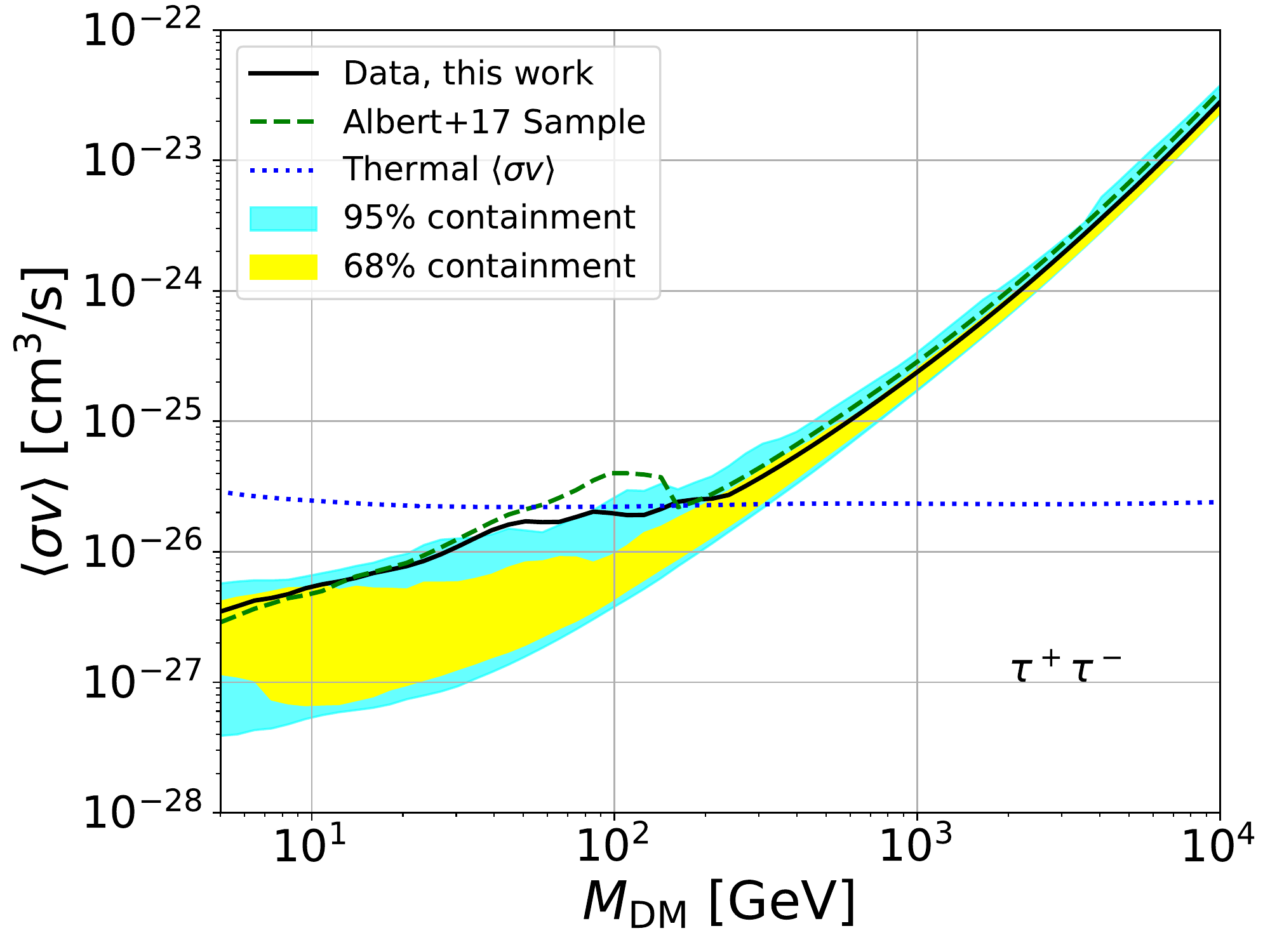}
\includegraphics[width=0.49\textwidth]{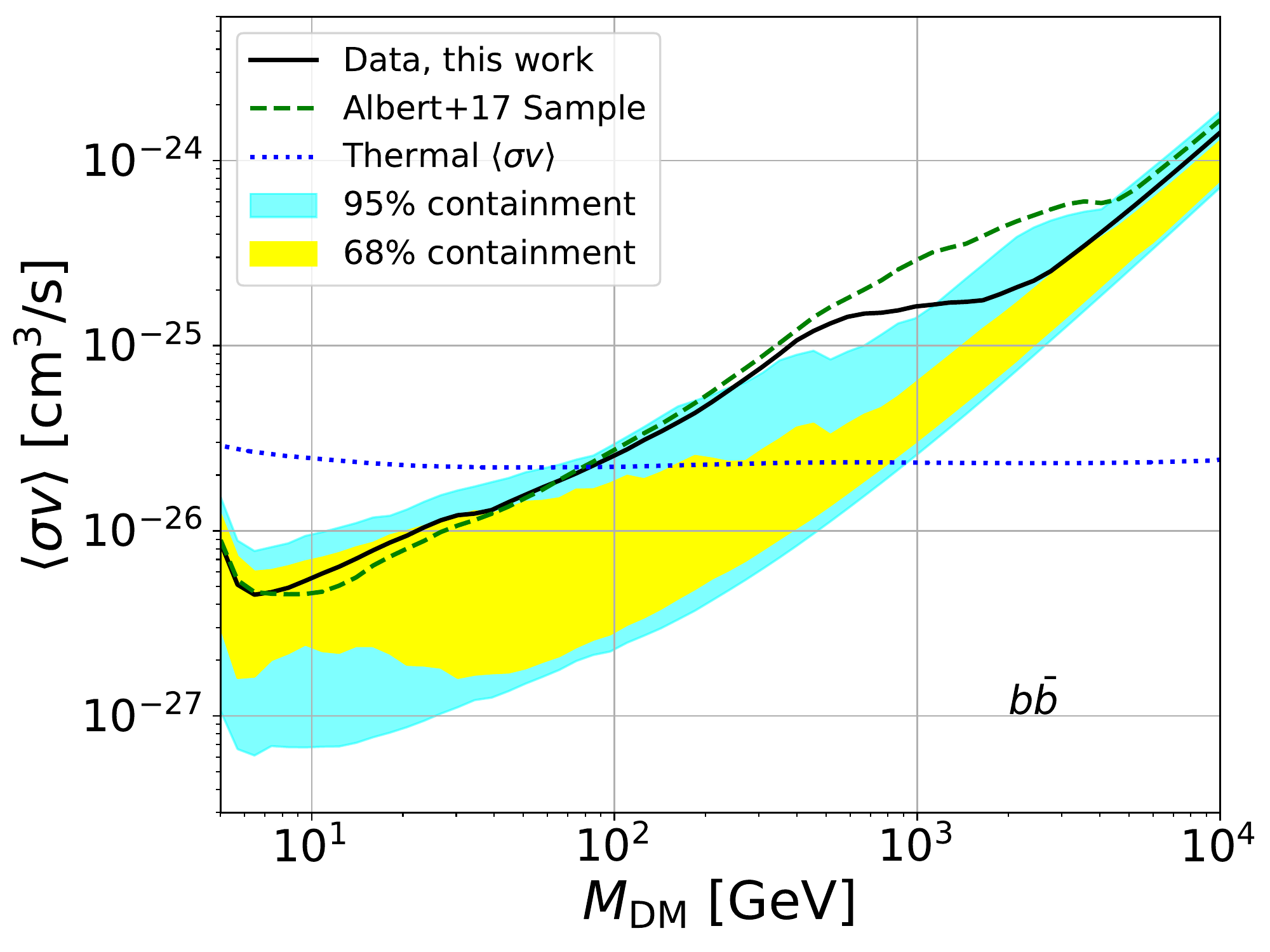}
\caption{$95\%$ CL upper limits for $\langle \sigma v \rangle$ for the $\tau^+\tau^-$ (top panel) and $b\bar{b}$ (bottom pannel) annihilation channels found with the dSphs sample in Ref.~\cite{2019MNRAS.482.3480P} (black solid line) and Ref.~\cite{Fermi-LAT:2016uux}  (Albert+17, green dashed line). We also show the $68\%$ (yellow band) and $95\%$ (cyan band) containment band for the limits obtained in random directions (read the main text for further details). We report the thermal cross section taken from Ref.~\cite{Steigman:2012nb}.}
\label{fig:sigmadwarfs}
\end{figure}

\begin{figure}
\includegraphics[width=0.49\textwidth]{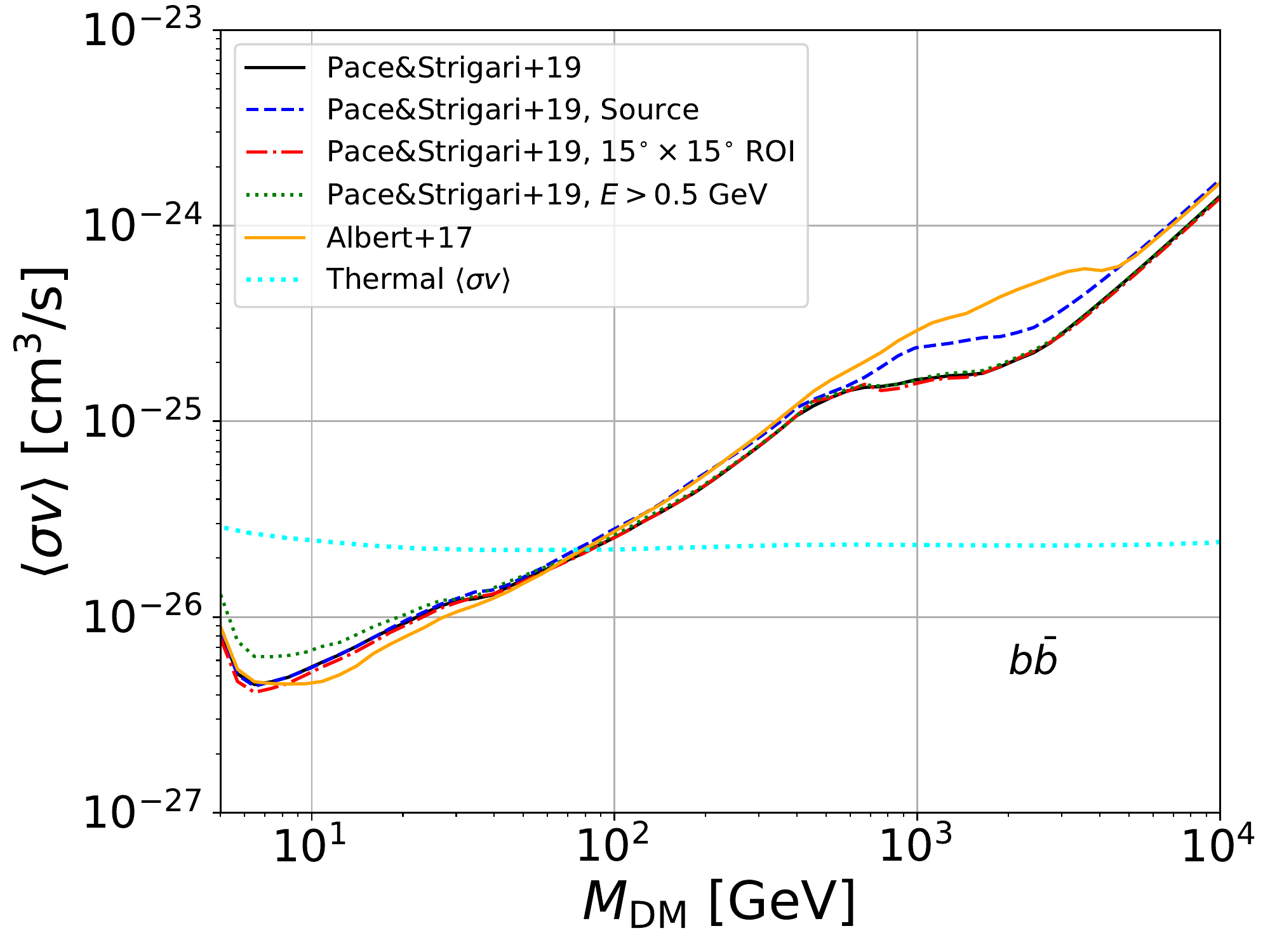}
\caption{$95\%$ CL upper limits for $\langle \sigma v \rangle$ for the $b\bar{b}$ annihilation channel for our baseline analysis (Pace\&Strigari+19 \cite{2019MNRAS.482.3480P}, black solid). We also show the limits obtained with the SOURCE IRFs (dashed blue), with a wider ROI of $15^{\circ}\times15^{\circ}$ (red dot-dashed), selecting data above 0.5 GeV (green dotted), and using the dSphs sample from Ref.~\cite{Fermi-LAT:2016uux} (Albert+17, orange solid).}
\label{fig:sigmadwarfssys}
\end{figure}

Since the signal detected from each individual dSph and for the stacked sample does not seem to be significant, we calculate upper limits for the 
annihilation cross section.
We display them in Fig.~\ref{fig:sigmadwarfs} for the $b\bar{b}$ and $\tau^+\tau^-$ annihilation channels.
The $95\%$ CL upper limits are below the thermal cross section up to roughly 100 GeV for both channels.
We also display the upper limits obtained with the list of dSphs in Ref.~\cite{Fermi-LAT:2016uux} and using the geometrical factors reported in that publication.
The results obtained with dSphs in Ref.~\cite{Fermi-LAT:2016uux} are similar to the one found with our reference sample.
We also show the $68\%$ and $95\%$ containment bands for the limits obtained in 100 random directions.
These expected limits in case of no detection are wider at low mass where the LAT is more sensitive and could pick up residuals due to faint sources or mismodeling of the IEM. Moreover, the $68\%$ containment band is much narrower than the $95\%$ one, as expected.
The limits found for the dSphs are compatible with the $95\%$ containment band for both the $b\bar{b}$ and $\tau^+\tau^-$ annihilation channels.
Instead, the observed limits are significantly higher than the $68\%$ containment band between about $50-2000$ GeV for $b\bar{b}$ and $10-200$ GeV for $\tau^+\tau^-$ because at these DM masses there is a small signal in the joint likelihood analysis as shown in Fig.~\ref{fig:TSdwarfs}.

In Fig.~\ref{fig:sigmadwarfssys} we show the ULs obtained for different assumptions of our analysis. In particular we perform the analysis with the SOURCE IRFs, with a wider ROI of $15^{\circ}\times15^{\circ}$, selecting data above 0.5 GeV, and using the dSphs sample from Ref.~\cite{Fermi-LAT:2016uux} (Albert2017). The results are similar for all the cases reported and in the DM mass range 1-100 GeV that is the relevant one for the DM interpretation of the GCE. This implies that our results do not change significantly making different choices of the data analysis or using a different dSphs sample.

Our results for the upper limits with dSphs are similar at the $20-30\%$ level with recently published in Refs~\cite{Calore:2018sdx,Hoof:2018hyn} where different list of sources and analysis techniques have been applied.

\subsection{Combining the Galactic center excess with dSphs limits}

If DM is responsible for the GCE, an interesting question arises about its compatibility with the non detection of a signal from dSphs.
In order to answer this question, we compare the coupling parameters of the DM candidates that explain the GCE with the limits found from dSphs.
We test the one/two/three channels cases that provide the best fits to the GCE SED: $b\bar{b}$ and $\mu^+\mu^-$, $\tau^+\tau^- - b\bar{b}$ and $\mu^+\mu^- - \tau^+\tau^- - b\bar{b}$. 
We take the values of the masses, annihilation cross sections and branching ratio from Tabs.~\ref{tab:singlefit} and \ref{tab:twochannelsIEM} that contain the systematic due to the choice of the IEM.
For the first time in literature the limits for $\langle \sigma v \rangle$ for dSphs are calculated assuming specific models with two and three annihilation channels.
This is done with the same procedure explained in Sec.~\ref{sec:fitGCE} but assuming for the intrinsic $\gamma$-ray spectrum from DM $dN_{\gamma}/dE$ the specific DM two or three channel branching ratios.

\begin{figure*}
\includegraphics[width=0.49\textwidth]{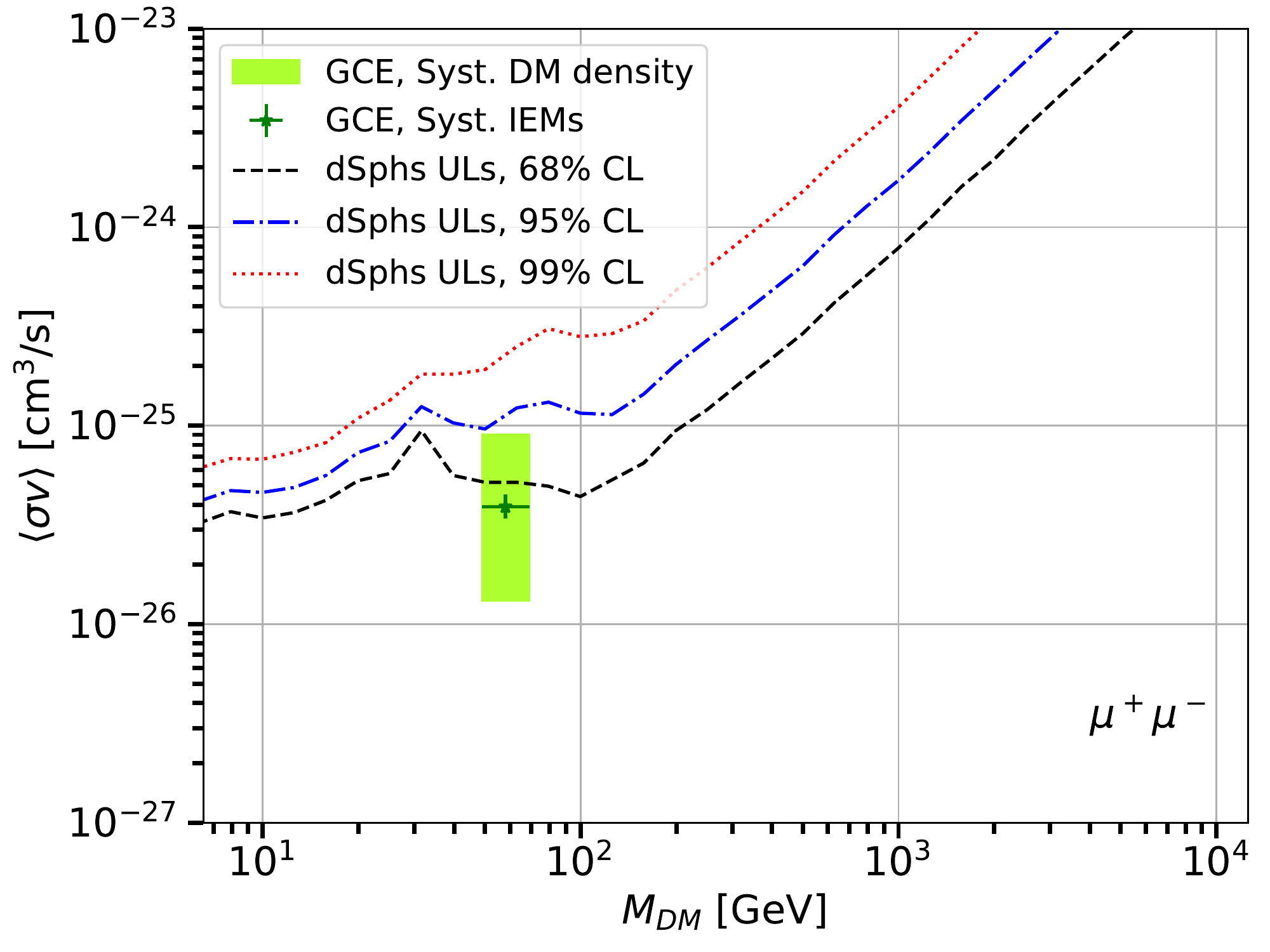}
\includegraphics[width=0.49\textwidth]{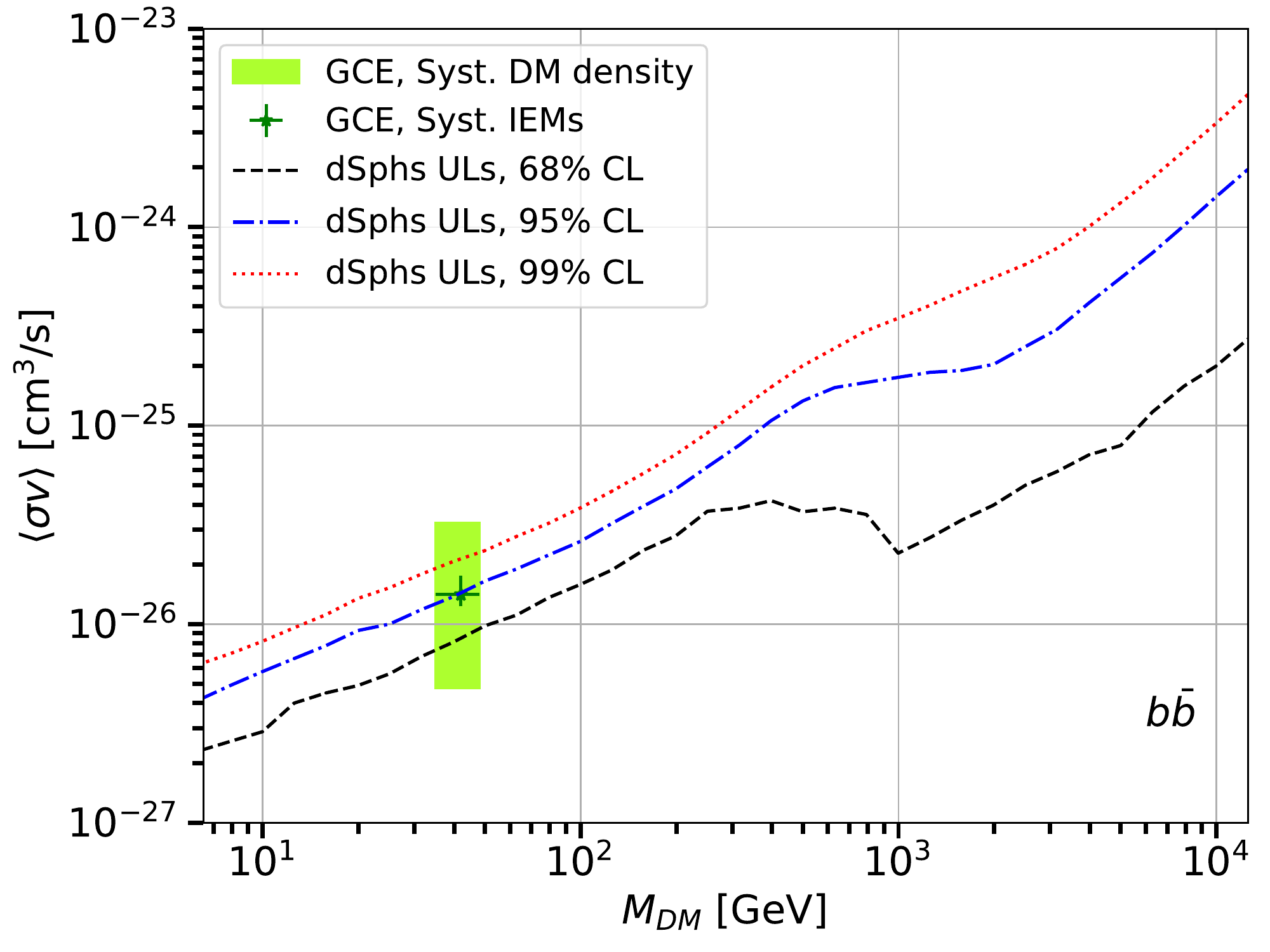}
\includegraphics[width=0.49\textwidth]{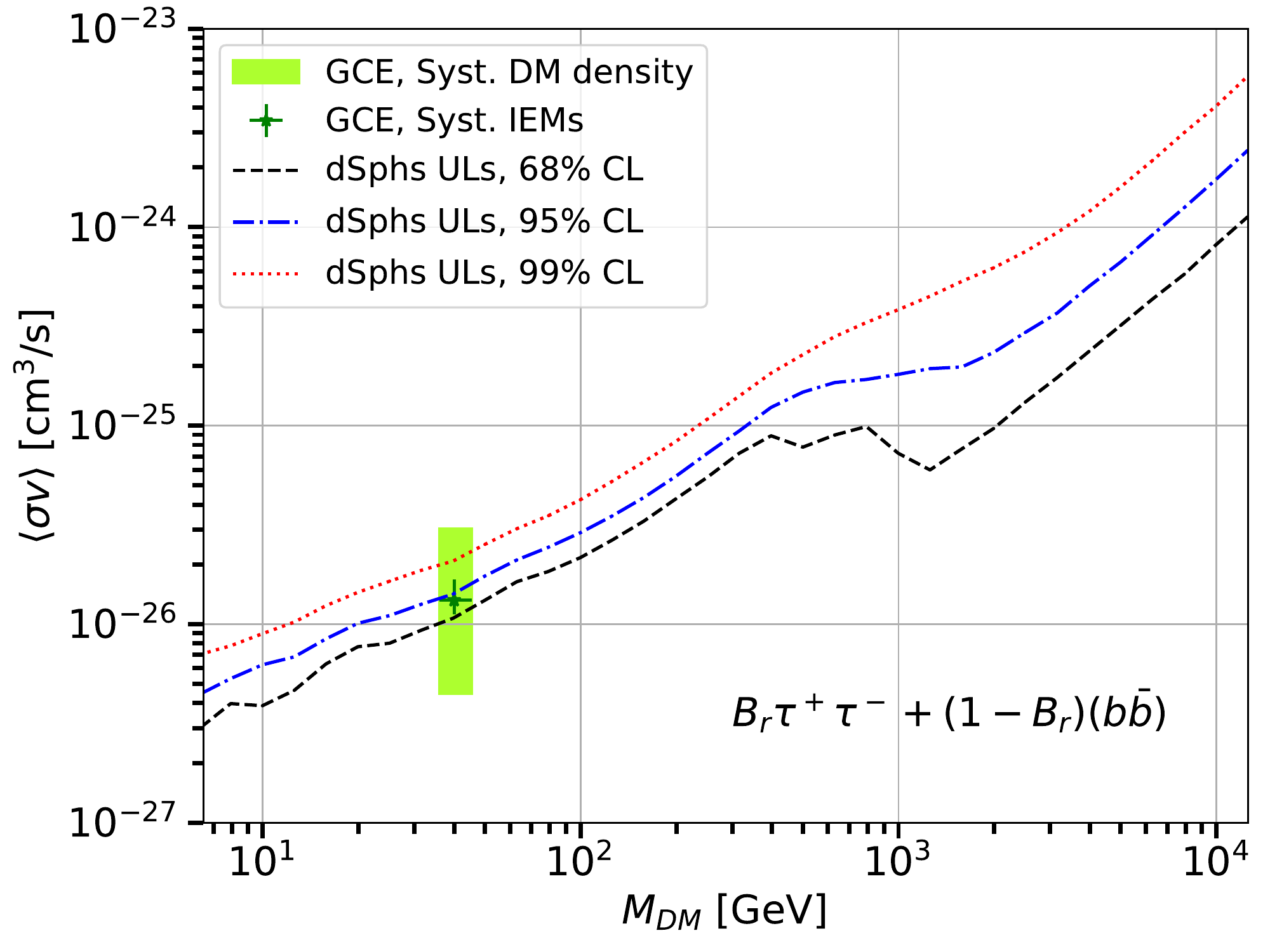}
\includegraphics[width=0.49\textwidth]{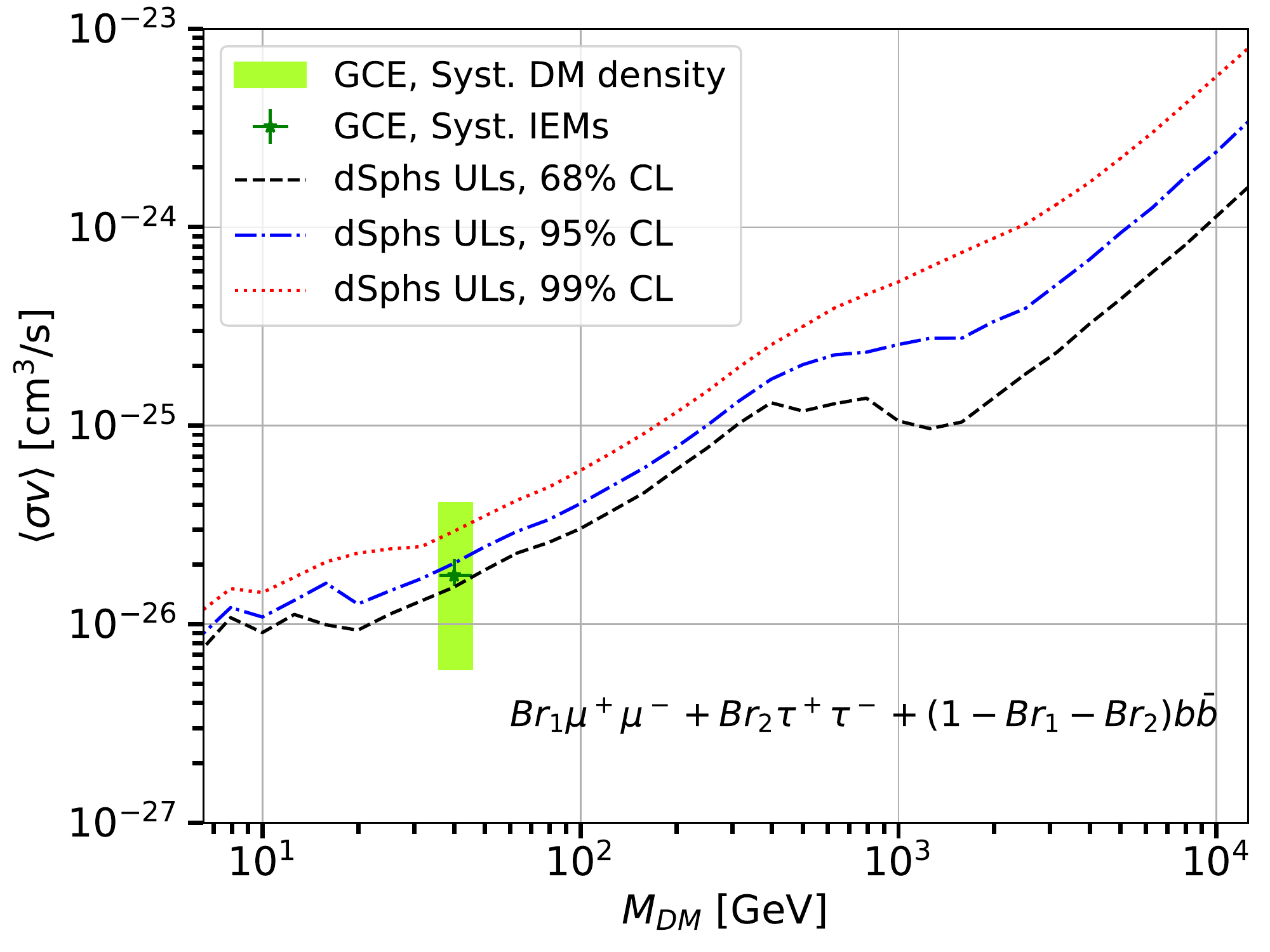}
\caption{Comparison between the $95\%$ (red dotted), $90\%$ (blue dot-dashed) and $68\%$ (black dashed) CL upper limits for $\langle \sigma v \rangle$ obtained from the analysis of the dSphs in Ref.~\cite{2019MNRAS.482.3480P} and the DM candidate that fit the GCE flux data obtained in with different IEMs (green data point). We also display with a green band the variation in $\langle \sigma v \rangle$ due to the modeling of the DM density in the inner part of the Galaxy (see Tab.~\ref{tab:models}). We display DM annihilating into $b\bar{b}$ and $\mu^+\mu^-$, $\tau^+\tau^- - b\bar{b}$ and $\mu^+\mu^- - \tau^+\tau^- - b\bar{b}$ channels.}
\label{fig:GCEdwarfs}
\end{figure*}

We show the result of this analysis in Fig.~\ref{fig:GCEdwarfs}.
The GCE DM candidate obtained with the $\mu^+\mu^-$ is below the limits, even in the $68\%$ CL level case, which is the strongest.
However, we have to stress that in the calculation of the $\gamma$ rays from the GCE we have included both the ICS and prompt emission while for the flux from dSphs we have accounted only for the prompt emission. 
For DM with a mass of 60 GeV the peak of the emission is at about a few GeV and it is mainly due to ICS on starlight (see top panel of Fig.~\ref{fig:twochannels}). Since the stellar light in dSphs is orders of magnitude smaller than in the Milky Way, the ICS contribution is negligible with respect to the prompt emission.
Instead, the annihilation channels $b\bar{b}$, $\tau^+\tau^- - b\bar{b}$ and $\mu^+\mu^- - \tau^+\tau^- - b\bar{b}$ are dominated by the prompt $\gamma$-ray emission from the $b\bar{b}$ annihilation channel. Thus the effect of the diffusion in the ICS calculation for dSphs, that we do not take into account in our calculation, is negligible.
For all these channels the properties of the DM candidate that explains the GCE in the MED DM model is roughly at the $95\%$ CL upper limits of the dSphs limits. This implies a tension at about $2\sigma$ significance.
However, considering the variation in $\langle \sigma v \rangle$ obtained by considering the MIN and MAX models, the GCE interpretation of DM is compatible with the $68\%$ CL upper limits of the dSphs, that implies there is no tension. 

\section{Constraints on dark matter using AMS-02 $\bar{p}$ data}
\label{sec:antip}
Messengers that have provided tight constraints on DM in the past are $\bar{p}$ CRs. It is thus very interesting to investigate the compatibility of the DM interpretation of the GCE with the newest $\bar{p}$ flux data collected in 7 years of mission by AMS-02~\cite{AGUILAR2020}. This is particularly true since a tentative DM signal has previously been found in the AMS-02 $\bar{p}$ data~\cite{Cuoco:2016eej,Cui:2016ppb} which was argued to be compatible with the GCE~\cite{Cuoco:2017rxb,Cholis:2019ejx}. On the other hand, it was noted that the significance of the $\bar{p}$ excess is drastically reduced, once uncertainties in the production of secondary antiprotons~\cite{Reinert:2017aga,Cuoco:2019kuu} and the correlations in the AMS-02 systematic errors~\cite{Boudaud:2019efq,Heisig:2020nse} are properly included.

We will perform our $\bar{p}$ analysis mostly following the approach described in~\cite{Reinert:2017aga,Heisig:2020nse}. The main aspects shall briefly be described below. In a first step, the high energy break in the diffusion coefficient (Eq.~\eqref{eq:diffusion_coefficient}) is fixed by a fit to the primary proton, helium, carbon, nitrogen and oxygen fluxes fluxes measured by AMS-02~\cite{Reinert:2017aga}. The break parameters take the values $\mathcal{R}_b=275\:\text{GV}$, $\Delta\delta=0.157$ and $s=0.074$. Since the high energy break is practically irrelevant for the $\bar{p}$ spectrum in the energy range covered by data, uncertainties in the break parameters can be neglected for our purposes. 

The Fisk potential parameter $\phi_0$ for the AMS-02 data taking period from a combined fit to the AMS-02 \cite{PhysRevLett.114.171103} and Voyager \cite{Stone150} proton data falls in the range $\phi_0=0.60 - 0.72\:\text{GV}$. The uncertainty encompasses different parameterizations of the interstellar proton flux, while statistical errors are negligible~\cite{Reinert:2017aga}. For the sake of a conservative approach we adopt the upper value $\phi_0=0.72\:\text{GV}$ in the following. The diffusion coefficient parameters $K_0$, $\delta$, $\eta$ and the Alfv\'en velocity $V_a$ are determined within a joined fit to the AMS-02 $\bar{p}$ and B/C data~\cite{AGUILAR2020}.

In addition we allow the solar modulation parameter $\phi_1$ which accounts for charge-breaking effects to float (Eq.~\eqref{eq:forcefield2}). In order to constrain it we also include the $\bar{p}$ flux ratio between AMS-02 and PAMELA~\cite{Adriani:2012paa} in our fit since PAMELA was run in a phase of opposite solar polarity compared to AMS-02 (see~\cite{Reinert:2017aga} for details). The last remaining propagation parameter, the vertical half-height of the diffusive zone $L$, cannot be determined within our fits due to a well-known degeneracy with the diffusion coefficient (which applies to stable secondary CRs). Based on an analysis of radioactive CRs it has recently been determined as $L=4.1^{+1.3}_{-0.8}\:\text{kpc}$ for the propagation configuration we are employing (QUAINT model in~\cite{Weinrich:2020ftb}). We will, therefore, mostly focus on $L=2-6\:\text{kpc}$ roughly corresponding to the $2\sigma$ range in the following. However, we will also test values of $L$ down to $1.5\:\text{kpc}$ constituting a $3\sigma$ deviation from the preferred value. We note that the value of $L$ significantly affects the DM-induced flux of $\bar{p}$ because for larger $L$ more DM annihilations occur within the diffusion zone and more space is available for propagation to the earth. As a result in order to have the same $\bar{p}$ flux from DM for a larger $L$, a smaller annihilation cross section is required.

The secondary $\bar{p}$ production is modeled through the cross section parameterization derived in Ref.~\cite{Winkler:2017xor}. The latter was obtained by a comprehensive analysis of $\bar{p}$ production in fixed target and collider physics experiments. A potential asymmetry between $\bar{n}$ and $\bar{p}$ production as well as exotic channels including hyperons have been taken into account. The secondary production is subject to uncertainties a the level of $5-10\%$ which have also been derived in~\cite{Winkler:2017xor}. These consist of a (fully correlated) normalization uncertainty $\mathcal{N}_{\bar{p}}=1\pm 0.06$ as well as uncertainties with a finite correlation length due to smooth variations in the cross section. We will include $\mathcal{N}_{\bar{p}}$ as a fit parameter and map the remaining uncertainties into the $\bar{p}$ flux through a covariance matrix (following the procedure described in~\cite{Winkler:2017xor,Cuoco:2019kuu,Boudaud:2019efq}). Similarly, we will include uncertainties in the boron production through the covariance matrix derived in~\cite{Winkler:2017xor}.

The AMS-02 $\bar{p}$ and $B/C$ data exhibit few-percent-level precision over wide rigidity ranges. Unless for low and high rigidities systematic errors dominate over statistical errors. In this light it is unfortunate that correlations in the systematic errors have so far not been provided by the AMS collaboration. We will estimate the correlations in the AMS-02 systematic errors in the $\bar{p}$ and B/C data following the approach of Ref.~\cite{Heisig:2020nse}. The dominant systematics come from uncertainties in the CR absorption cross sections within the detector material which are modeled within the Glauber-Gribov theory in~\cite{Heisig:2020nse}. We will also investigate the sensitivity of our results with respect to the inclusion of correlations in the AMS data.

\begin{figure*}
\includegraphics[width=0.49\textwidth]{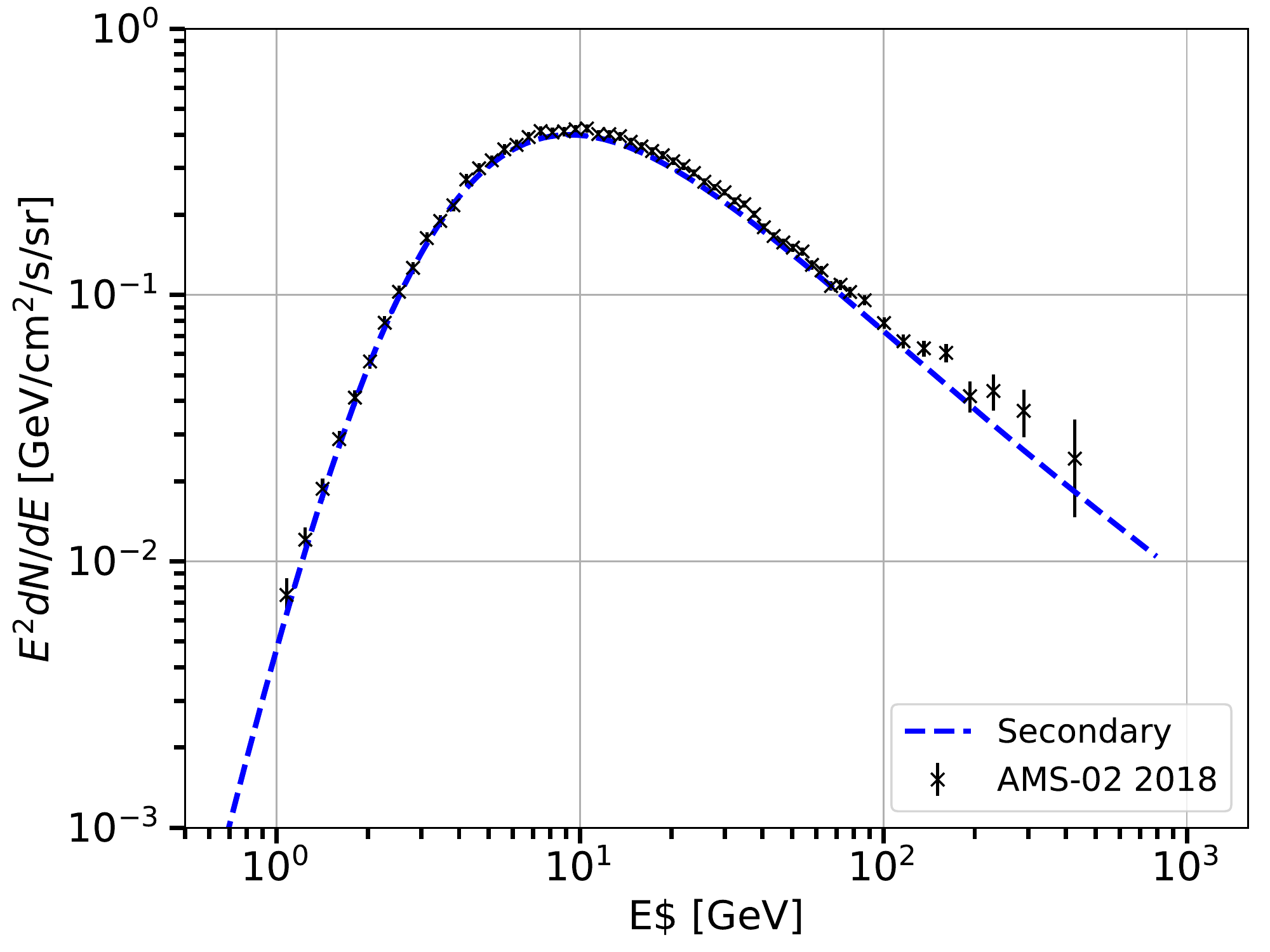}
\includegraphics[width=0.49\textwidth]{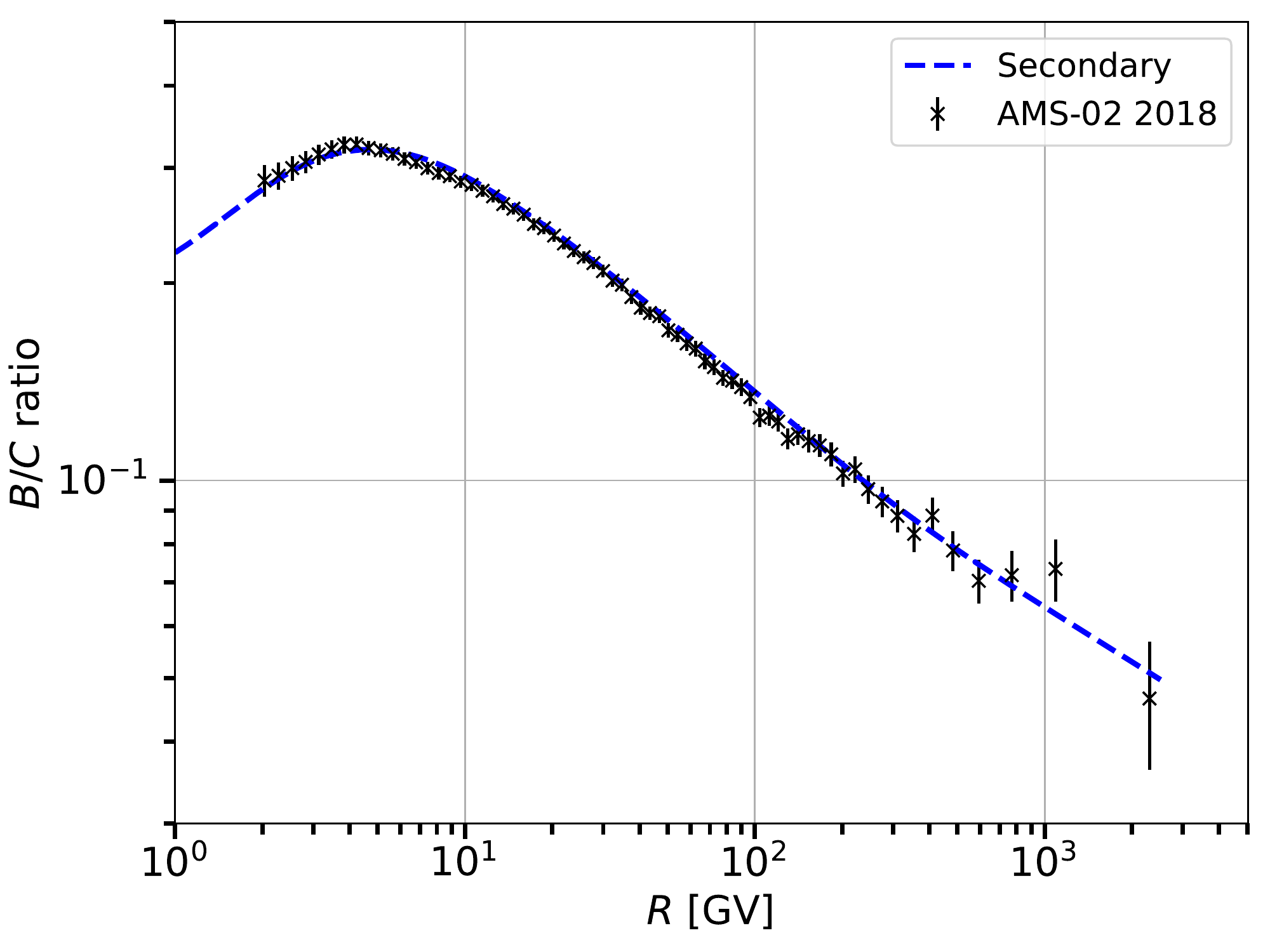}
\includegraphics[width=0.49\textwidth]{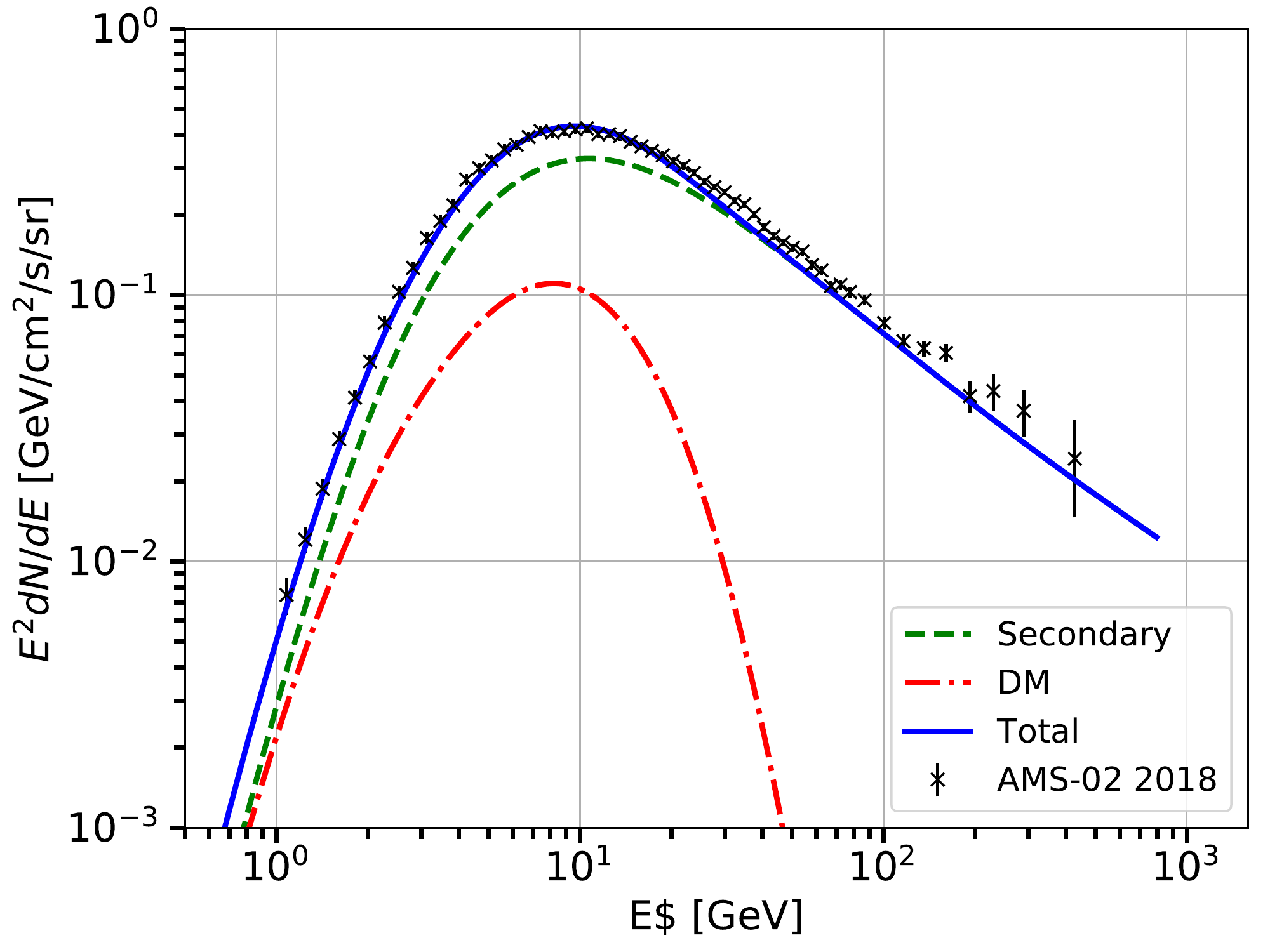}
\includegraphics[width=0.49\textwidth]{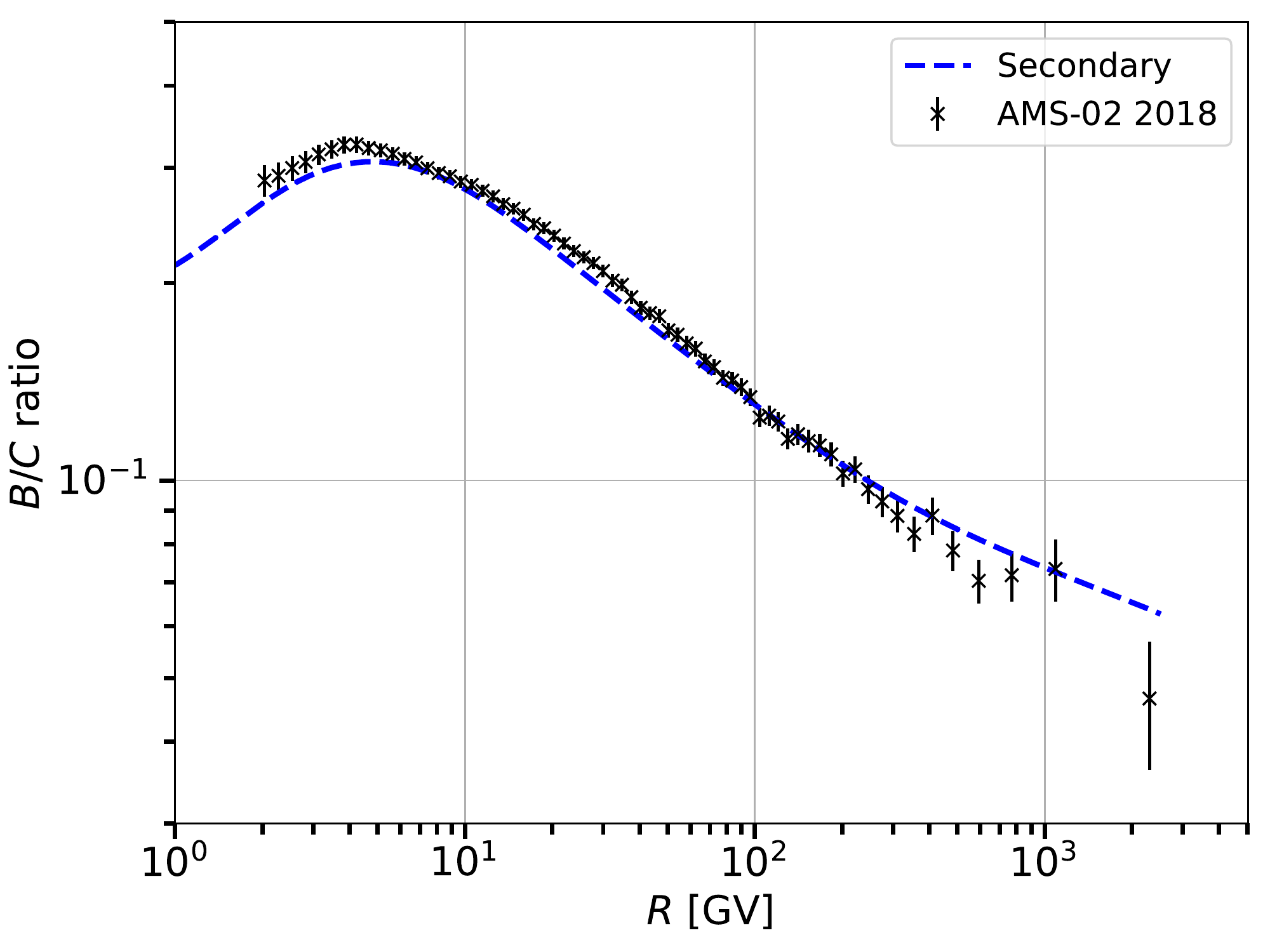}
\caption{Antiproton flux (left column) and B/C (right column) obtained in our fits compared to the AMS-02 data. The upper panels show the best fit fluxes for pure secondary production (i.e.\ no DM contribution). The lower panels show the best fit fluxes if we inject a DM signal compatible with the GCE SED for the $b\bar{b}$ annihilation channel (with the parameters reported in Tab.~\ref{tab:singlefit}). The {\tt MED} DM density model and $L=3\:\text{kpc}$ are assumed. As can be seen, the fit significantly degrades if the DM signal is added.}
\label{fig:CRfit}
\end{figure*}

\begin{figure}
\includegraphics[width=0.49\textwidth]{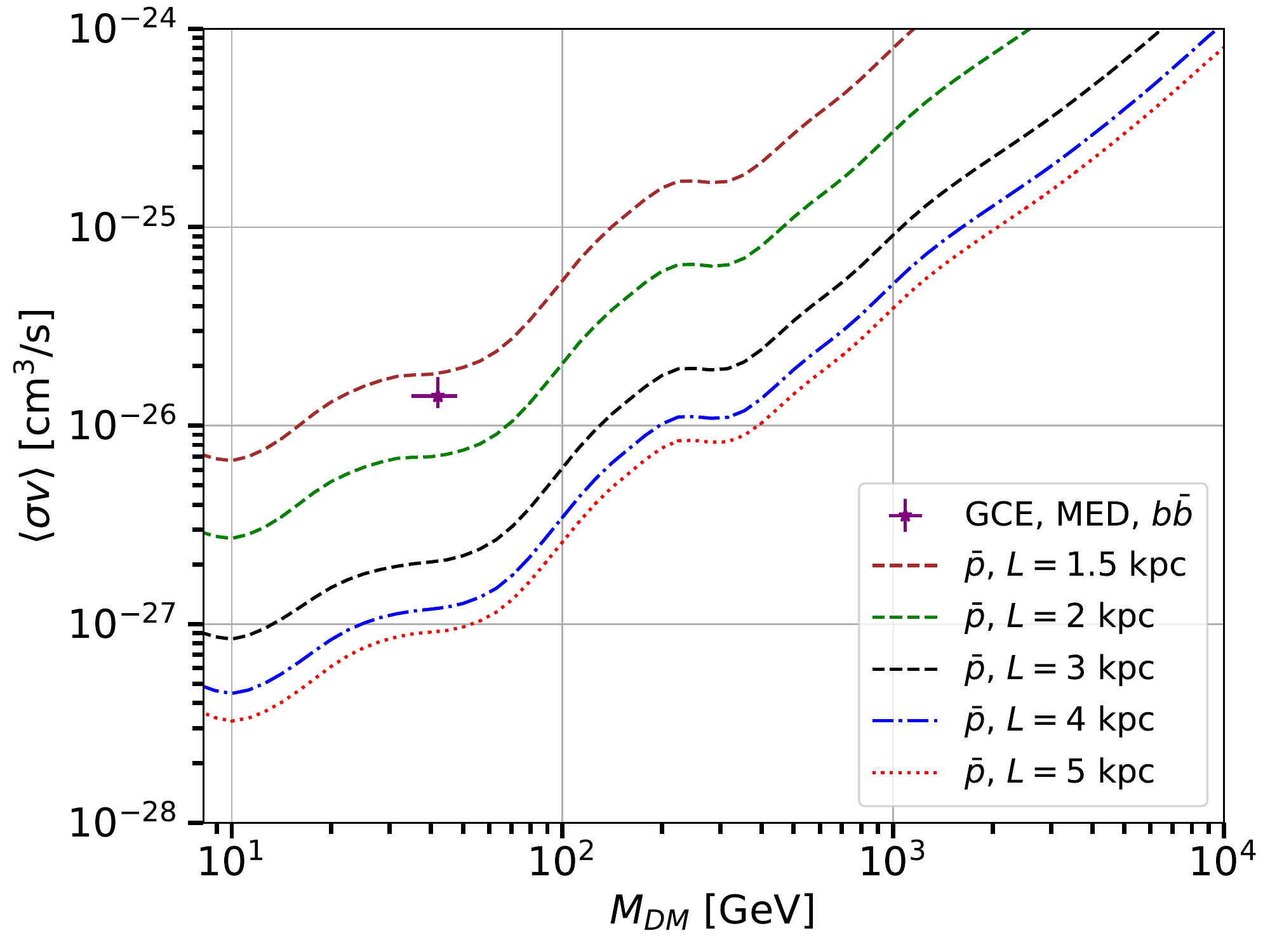}
\includegraphics[width=0.49\textwidth]{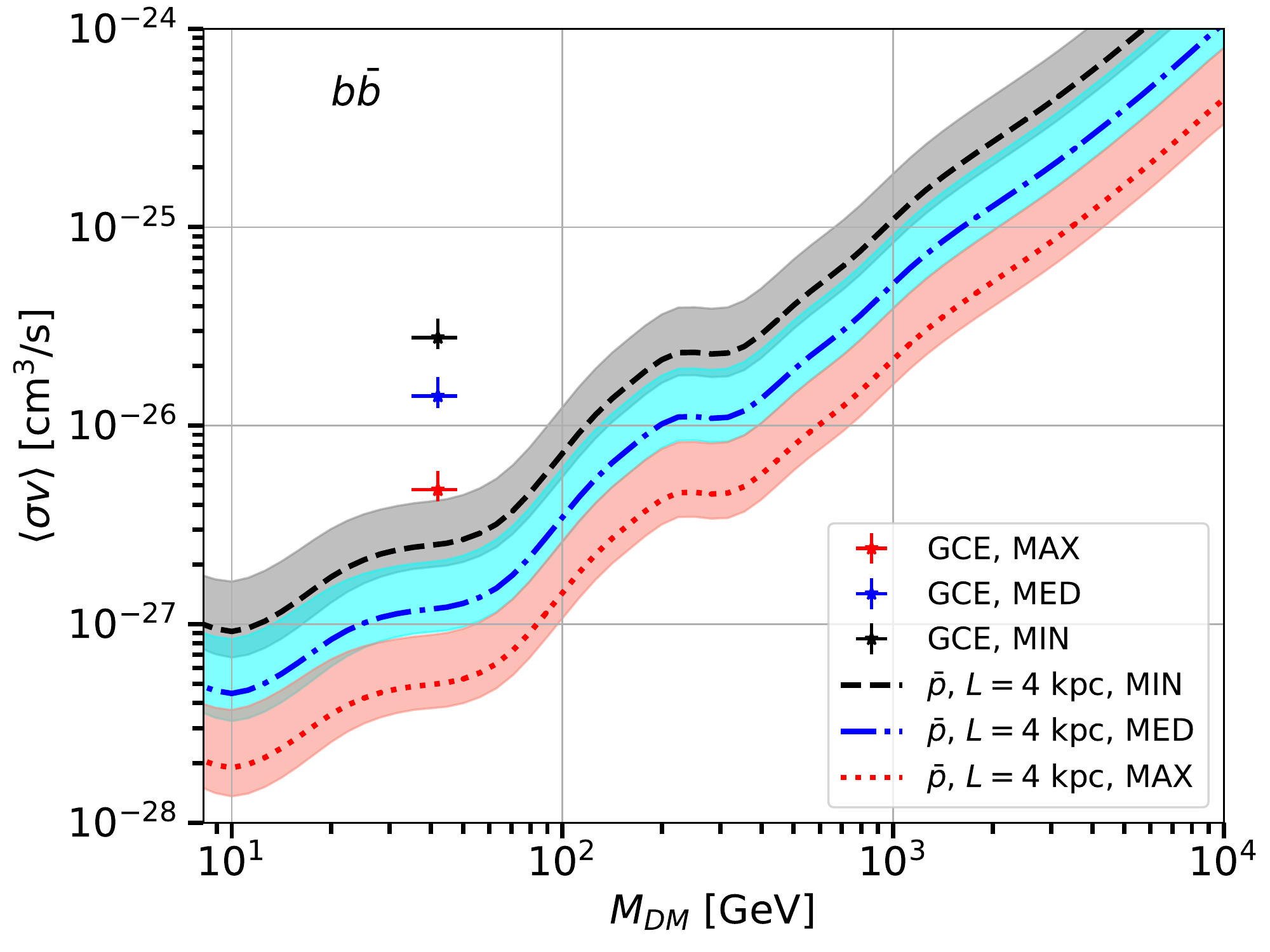}
\caption{Top Panel: $95\%$ CL upper limits on the DM annihilation cross section found by fitting AMS-02 $\bar{p}$ data and assuming different sizes for $L$. In addition we show the best-fit DM parameters we obtain by fitting the GCE. We assume a $b\bar{b}$ annihilation channel and the {\tt MED} DM density model. Bottom panel: Same as top panel changing the DM density model to {\tt MIN}, {\tt MED} and {\tt MAX}. The bands we show for the $\bar{p}$ upper limits include the variation in the results changing $L$ from 3 to 5 kpc.}
\label{fig:antipvariation}
\end{figure}

First, we perform a fit to the AMS-02 $\bar{p}$, B/C data and the antiproton flux ratio between AMS-02 and PAMELA without assuming any DM contribution. The best fit $\bar{p}$ and B/C spectra are shown together with the AMS-02 data in the top panel of Fig.~\ref{fig:CRfit}. The goodness of fit is $\chi^2=173$ on 143 data points with 6 free parameters of the model. Therefore, the result for the reduced $\chi^2$ is 1.26 which indicates that the AMS-02 data are consistent with pure secondary production within $\sim 2 \sigma$. Given some residual uncertainty in our modeling of correlations in the AMS systemtic errors (see above) the secondary hypothesis is definitely in good shape. We report the best-fit propagation parameters in Tab.~\ref{tab:CRparams}. The cross section normalization and solar modulation parameters take values $\mathcal{N}_{\bar{p}}=1.09$ and $\phi_1=0.75\:\text{GV}$ at the best fit point.

The parameters from our fit take values close to those obtained in~\cite{Heisig:2020nse} with previous AMS-02 data sets~\cite{PhysRevLett.117.091103,PhysRevLett.117.231102}. One striking observation is, however, that the residuals between the best-fit model and the newest AMS-02 $\bar{p}$ flux data in the range $\mathcal{R}=10-20\:\text{GV}$ are practically flat. In this rigidity range, previous analyses~\cite{Cuoco:2016eej,Cui:2016ppb}, based on a previous AMS-02 data set for $\bar{p}$ \cite{PhysRevLett.117.091103}, had identified the `antiproton excess' which had tentatively been interpreted as a DM signal (potentially compatible with the GCE). While the excess occured at a much smaller significance ($\sim 1 \sigma$) after including the correlations in the AMS-02 systematic errors~\cite{Heisig:2020nse}, it remained visible in the data. We realized that the complete disappearance of the excess is likely linked to the updated AMS-02 data~\cite{AGUILAR2020} which are systematically lower by $\sim 5\%$ in the rigidity range $\mathcal{R}=10-20\:\text{GV}$ compared to the previous data set~\cite{PhysRevLett.117.091103}.

In the next step, we add a DM contribution with free normalization $\langle \sigma v\rangle$ and mass $M_{\text{DM}}=7-10000\:\text{GeV}$, where we allow the propagation, solar modulation and cross section normalization parameters to float. As final states of the DM annihilation $\bar{b}b$ and $\bar{c}c$ are considered. We note that other two-quark as well as two-gluon final states yield a very similar $\bar{p}$ spectrum as the $\bar{c}c$-channel. Our fits confirm that the previously found $\bar{p}$ excess~\cite{Cuoco:2016eej,Cui:2016ppb} is completely gone in the new AMS-02 data. There is no longer any preference for a DM contribution within the range $M_{\text{DM}}=30-100\:\text{GeV}$. This statement does neither depend on the underlying DM profile nor on the size of the diffusion zone $L$ which mostly affect the normalization of a potential DM signal. The best fit point including a DM contribution is found in the $\bar{b}b$ channel at $M_{\text{DM}}=1.4\:\text{TeV}$. However, this `excess' only reaches significance of $\sim 2\sigma$ ($\sim 1\sigma$) locally (globally). Hence, we do not find any significant preference for a DM signal in the $\bar{p}$ data.

We can then use $\bar{p}$ to provide constraints on DM annihilation. Of particular interest is the DM candidate in the $\bar{b}b$-channel which is compatible with the GCE SED. Employing the parameters reported in Tab.~\ref{tab:singlefit}, we observe that the latter induces a substantial contribution to the the $\bar{p}$ flux. If we keep the propagation parameters fixed, we obtain $\chi^2=238$ for the {\tt MED} DM density model and $L=3\:\text{kpc}$ compared to $\chi^2=173$ without DM. If we allow the propagation, solar modulation and cross section normalization parameters to float, $\chi^2$ is reduced to 217 which, however, still amounts to an exclusion by $>6\sigma$ for the DM contribution. The fit in this case prefers a smaller $\delta=0.36$ and higher $V_a=63\:\text{km/s}$ in order to compensate the DM-induced flux which, however, substantially degrades the fit to B/C. The best fit $\bar{p}$ and B/C spectra including the DM contribution are shown in the lower panels of Fig.~\ref{fig:CRfit}. In the following we wish to investigate, whether the exclusion of the GCE DM canditate is robust with respect to variations of the density profile and $L$.

\begin{table}
\begin{center}
\begin{tabular}{|c|c|c|c|}
\hline
\hline
$K_0\;\,[$kpc$^2$/Myr$]$  & $\delta$  & $\eta$ & $V_a\;\,[$km/s$]$  \\
\hline
0.042   &   0.459  & -1.49 & 52.0  \\
\hline
\end{tabular}
\caption{Best-fit propagation parameters for $L=4\:\text{kpc}$ from the combined fit to $\bar{p}$ and B/C data (assuming pure secondary production of antiprotons). The best fit propagation parameters for different choices of $L$ are obtained by rescaling $K_0$ with $L/4\:\text{kpc}$ and $V_a$ by  $\sqrt{L/4\:\text{kpc}}$.}
\label{tab:CRparams}
\end{center}
\end{table}

We, therefore, derived the 95\% CL upper limits on the DM annihilation cross section within the mass range $M_{\text{DM}}=7-10000\:\text{GeV}$ for values of $L=1.5-5\:\text{kpc}$ and for the {\tt MIN}, {\tt MED}, {\tt MAX} DM profiles. For the purpose of deriving limits we keep the propagation parameters fixed at the values indicated in Tab.~\ref{tab:CRparams}, but fully include the uncertainty in the secondary antiproton production cross section. We tested for a number of parameter points that allowing the propagation parameters to float would only affect the 95\% CL upper limits at the percent level which is negligible for our purposes.

We start by showing the upper limits we find fixing the DM density model to {\tt MED} and testing different values for $L$. We report this result in the top panel of Fig.~\ref{fig:antipvariation} again for the $b\bar{b}$-channel and the {\tt MED} DM density model. We see that the upper limits increase by a factor $\sim 20$ between $L=5$ kpc and 1.5 kpc. This is because for small $L$ a large fraction of the $\bar{p}$ created at the Galactic center escapes through the boundaries of the diffusion zone before reaching the earth. The DM candidate that explains the GCE assuming a $b\bar{b}$ annihilation channel is compatible with the $\bar{p}$ limits only for $L\leq 1.7\:\text{kpc}$. Such a low value of $L$ constitutes a $3\sigma$ deviation from the value preferred by radioactive CRs derived in~\cite{Weinrich:2020ftb}. We also note that another recent evaluation of boron, beryllium and lithium fluxes within a similar propagation setup found $L=6.8\pm 1\:\text{kpc}$ suggesting an even stronger tension of $L\leq 1.7\:\text{kpc}$ with data~\cite{Luque:2021joz}. Further indications against such a small diffusion halo arise from the diffuse gamma ray background~\cite{2012ApJ...750....3A} and from radio observations~\cite{Bringmann:2011py,DiBernardo:2012zu,Orlando:2013ysa}. In the next section, we will show that it is, furthermore, in tension with the low energy $e^+$ spectrum.

In Fig.~\ref{fig:antipvariation} we also show how the upper limits change assuming a different DM density distribution. As expected the {\tt MIN} DM density provides weaker limits with respect to {\tt MED} and {\tt MAX}. However, the limits on $\langle \sigma v \rangle$ scale almost proportionally to the change of $\bar{\mathcal{J}}$, i.e.~the GCE preferred cross section changes in the same way as the $\bar{p}$ limit. Hence, variations in the DM profile do not reconcile the DM interpretation of the GCE with $\bar{p}$ constraints.

We have finally tested, whether our conclusions are affected by the modeling of correlations in the AMS-02 data which we adopted from~\cite{Heisig:2020nse}. For this purposes we recalculated the constraints in the $\bar{b}b$-channel assuming systematic errors are uncorrelated (a common assumption in previous CR analyses). However, we found no significant change in the limit around $M_{\text{DM}}\sim 40\:\text{GeV}$ compared to the case where we include AMS-02 correlations.

\begin{figure}
\includegraphics[width=0.49\textwidth]{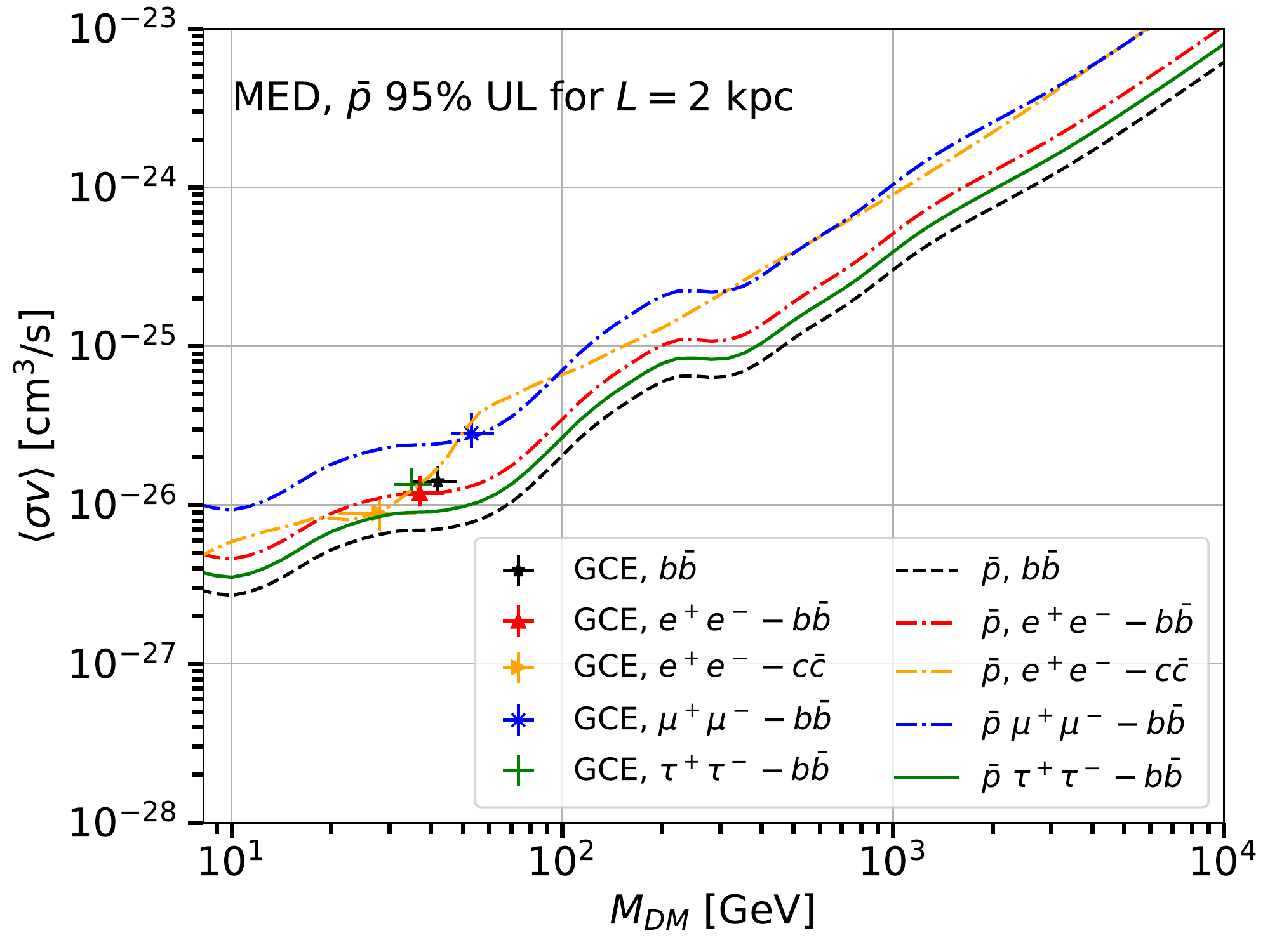}
\includegraphics[width=0.49\textwidth]{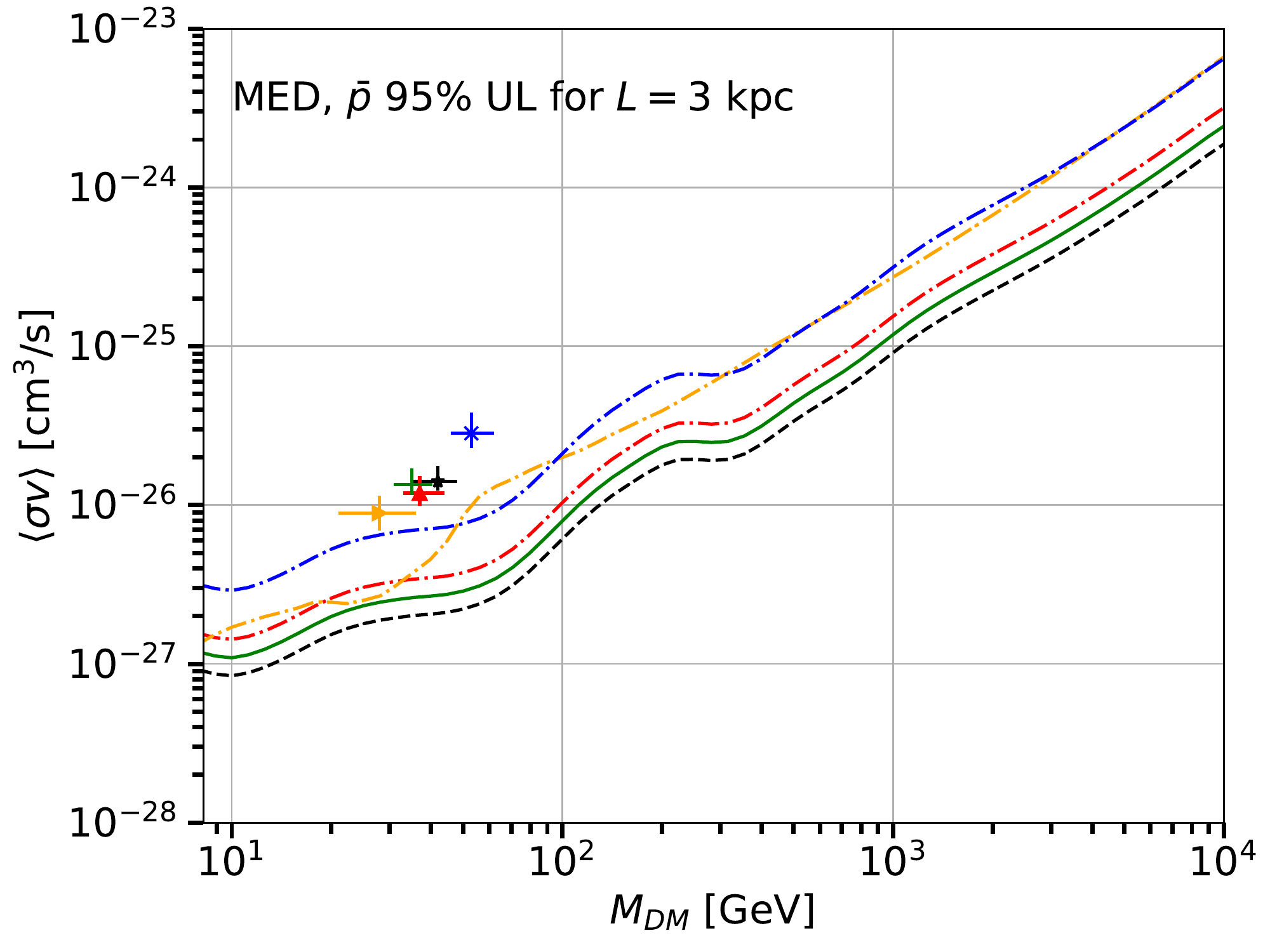}
\caption{Best-fit values for the DM parameters $M_{\rm{DM}}$ and $\langle \sigma v \rangle$ that we find by fitting the GCE SED. We show the cases that best-fit the GCE SED from Sec.~\ref{sec:fitGCE}. We also report the $95\%$ CL upper limits we obtain from $\bar{p}$ flux data for the same DM candidates. We assume the {\tt MED} DM density model and $L=2$ ($L=3$) kpc for the plot in the top (bottom) panel.}
\label{fig:GCEcandidatesantip}
\end{figure}

We now turn to DM models with a significant annihilation fraction into leptons. These should be subject to weaker $\bar{p}$ constraints since the antiproton flux from leptonic final states is practically negligible. In fact, the DM candidates from Tab.~\ref{tab:singlefit} which annihilate into pure $e^+e^-$ and $\mu^+\mu^-$ are not constrained by $\bar{p}$. These channels will be constrained by CR $e^+$ in the next section.

However, $\bar{p}$ are sensitive to the two-channel final states of Tab.~\ref{tab:twochannelsIEM} which are partly leptonic and partly hadronic. In Fig.~\ref{fig:GCEcandidatesantip} we show 95\% CL upper limits for the $e^+ e^- - b\bar{b}$, $e^+ e^- - c\bar{c}$, $\mu^+ \mu^- - b\bar{b}$ and $\tau^+ \tau^- - b\bar{b}$ channels with the branching ratios from Tab.~\ref{tab:twochannelsIEM} for the {\tt MED} DM profile (e.g.\ $e^+ e^- - b\bar{b}$ refers to $50\%$ annihiliation into $e^+e^-$ and $50\%$ into $b\bar{b}$). The limits for the $b\bar{b}$-channel are shown again for comparison. It can be seen that the GCE preferred annihilation cross sections are excluded for all mixed channels with a hadronic component if $L=3\:\text{kpc}$ (lower panel of Fig.~\ref{fig:GCEcandidatesantip}). Reducing the diffusion halo size to $L=2\:\text{kpc}$ reconciles the GCE candidates in the $e^+ e^- - b\bar{b}$, $e^+ e^- - c\bar{c}$ and $\mu^+ \mu^- - b\bar{b}$ channels with the $\bar{p}$ constraints (upper panel of Fig.~\ref{fig:GCEcandidatesantip}). The constraints on the $\tau^+ \tau^- - b\bar{b}$ channel are somewhat stronger due to the larger branching fraction of $80\%$ into $b\bar{b}$ (see Tab.~\ref{tab:twochannelsIEM}). We verified that these findings remain valid for different choice of the DM profile.

To summarize, all GCE DM candidates which annihilate partly or fully hadronically are in some tension with the $\bar{p}$ constraints. A small diffusion halo $L\leq 2\:\text{kpc}$ for the semi-hadronic channels or $L\leq 1.7\:\text{kpc}$ for the $b\bar{b}$-channel appears to be the only possible option to reconcile the GCE DM candidates with $\bar{p}$ constraints. As we noted earlier such a small diffusion halo is compatible with the observed $\bar{p}$-flux, but causes strong trouble with complementary astrophysical probes, in particular with radio data~\cite{Bringmann:2011py,DiBernardo:2012zu,Orlando:2013ysa} and observations of radioactive CRs~\cite{Weinrich:2020ftb,Luque:2021joz}.

\section{Constraints on dark matter using $e^+$ data}
\label{sec:pos}

CR $e^+$ measured by AMS-02 have been used in the past to put severe constraints on the leptonic annihilation channels of DM. 
In Ref.~\cite{Bergstrom:2013jra}, for example, the authors have assumed that the astrophysical background was given by an analytic function that was fitting perfectly the data. 
They calculated upper limits for $\langle \sigma v \rangle$ adding a DM contribution on top of this background model. 
They used this procedure for the leptonic DM channels, $e^{\pm}$, $\mu^{\pm}$ and $\tau^+\tau^-$, for which the $e^+$ flux shape is significantly different from the one of the AMS-02 data. 
However, the resulting constraints can be too optimistic, i.e.~too low, because the astrophysical contribution is modeled by a function that (by construction) perfectly fits the data and thus almost no space is left for a DM contribution.
In Ref.~\cite{DiMauro:2015jxa} the authors have done the more realistic assumption that the $e^+$ flux is given by the following astrophysical contributions: the secondary production of primary CRs interacting with atoms of the interstellar medium and the cumulative flux of PWNe in the ATNF catalog. The upper limits that they found are higher than the ones from Ref.~\cite{Bergstrom:2013jra} but, for the leptonic channels, they are below the thermal cross section up to about $60-100$ GeV.
The ATNF catalog has a large incompleteness for sources farther than a few kpc from the Earth \cite{Manchester:2004bp}. These latter sources would mostly contribute to the $e^+$ flux data below 100 GeV. This energy range is relevant for a possible contribution of $e^+$ from DM particles with masses below a few hundreds of GeV.
In order to account properly for the flux of $e^+$ injected from all Galactic pulsars one should perform simulations based on synthetic pulsar models (see, e.g., Ref.~\cite{FaucherGiguere:2005ny}).
Moreover, the secondary production is affected by systematic due to the modeling of the $e^{\pm}$ production cross sections usually taken from the one in Ref.~\cite{Kamae:2006bf}. This latter reference, as well as others on the same topic, tuned the cross sections for the production of $e^{\pm}$ with Monte Carlo event generators or old particle data taken decades ago and affected by large statistical and systematic errors.

\begin{figure}
\includegraphics[width=0.49\textwidth]{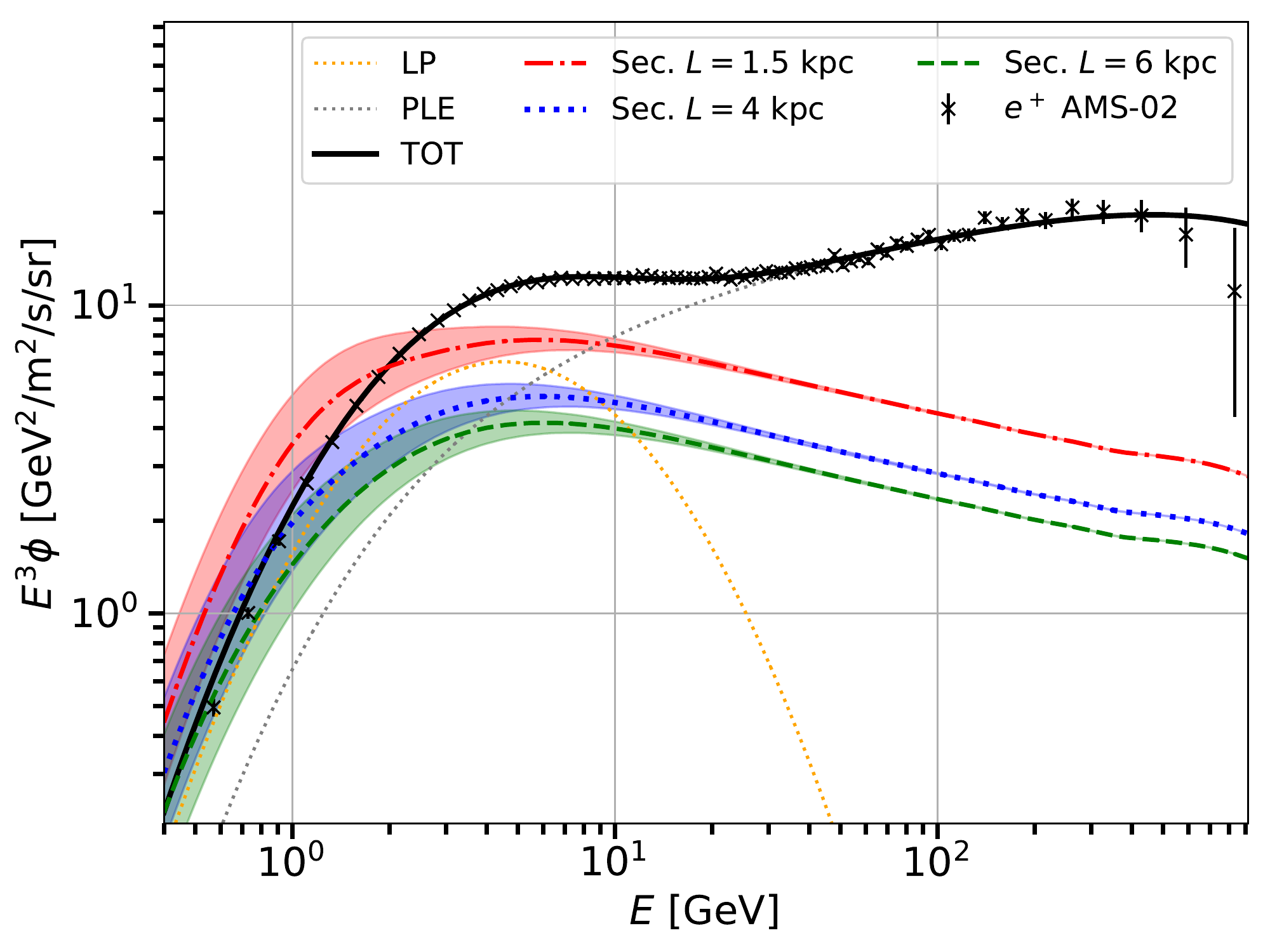}
\caption{AMS-02 $e^+$ flux data (black data points) fitted in the optimistic approach with an analytic function (black solid line) given by sum of a LogParabola (LP, grey dotted line) and a power-law with an exponential cutoff (PLE, orange dotted line). We also show the secondary flux of $e^+$ calculated using the best-fit propagation parameters in Tab.~\ref{tab:CRparams} and $L=1.5,4,6$ kpc (red, blue and green line). The bands for each case represent the variation in the secondary flux by assuming a Fisk potential variation between $0.62-0.82$ GV.}  
\label{fig:sec}
\end{figure}

A more realistic estimation of the pulsar contribution to the $e^+$ as well as the refinement of the $e^+$ production cross sections relevant is beyond the scope of this paper. 
Therefore, we decide to make two simplistic assumptions to derive upper limits on the DM annihilation cross section with AMS-02 $e^+$ data \cite{AGUILAR2020}. 
In the {\it conservative approach} we assume that the astrophysical $e^+$ background is only given by the secondary production, i.e.~there is no PWN contribution.
Then, we add the DM flux of $e^+$ and we use a $\chi^2$ calculation that penalizes models that overshoot the AMS-02 data points. Specifically, if the flux from the secondary production and DM is below the AMS-02 data the $\chi^2$ remains unchanged, instead if it is above the data it is incremented by the typical factor $($model$-$data$)^2/($data error$)^2$.
We show in Fig.~\ref{fig:sec} the comparison between the secondary production calculated for $L=1.5,4,6$ kpc and a Fisk potential between $0.62-0.82$ GV and the $e^+$ data.
We use for this analysis the propagation parameters found in Tab.~\ref{tab:CRparams} and a conservative uncertainty of 0.1 GV on the best-fit value of the Fisk potential obtained by fitting CR data.
The AMS-02 data below 1 GeV rule out vertical sizes of the diffusive halo smaller than 3 kpc. This provides another argument against the small value of $L$ required to reconcile the hadronic GCE DM candidates with $\bar{p}$ constraints (see previous section). We test that the $e^+$ constraints on $\langle \sigma v \rangle$ are similar for all $L>3$ kpc. Therefore, we fix $L=4$ kpc in the following.

The {\it optimistic approach} involves the usage of a smooth analytic function that is able to fit the AMS-02 data. Then, we add the DM contribution and find as 95\% CL upper limit the value of $\langle \sigma v \rangle$ that worsens the $\chi^2$ from the best fit by 2.71.
In calculating the best-fit with DM the free parameters of the analytic functions are left free to float. This approach is thus similar to the one used by Ref.~\cite{Bergstrom:2013jra}.
We use a background model that is given by the superposition of a LogParabola and a power-law with an exponential cutoff. This function fits very well the data above 1 GeV, in fact the reduced $\chi^2$ is $\tilde{\chi}^2=0.62$.
The free parameters of this function are 7 (three for the LogParabola and 4 for the other function).
We show in Fig.~\ref{fig:sec} the comparison between the best-fit model and the AMS-02 data.

\begin{figure}
\includegraphics[width=0.49\textwidth]{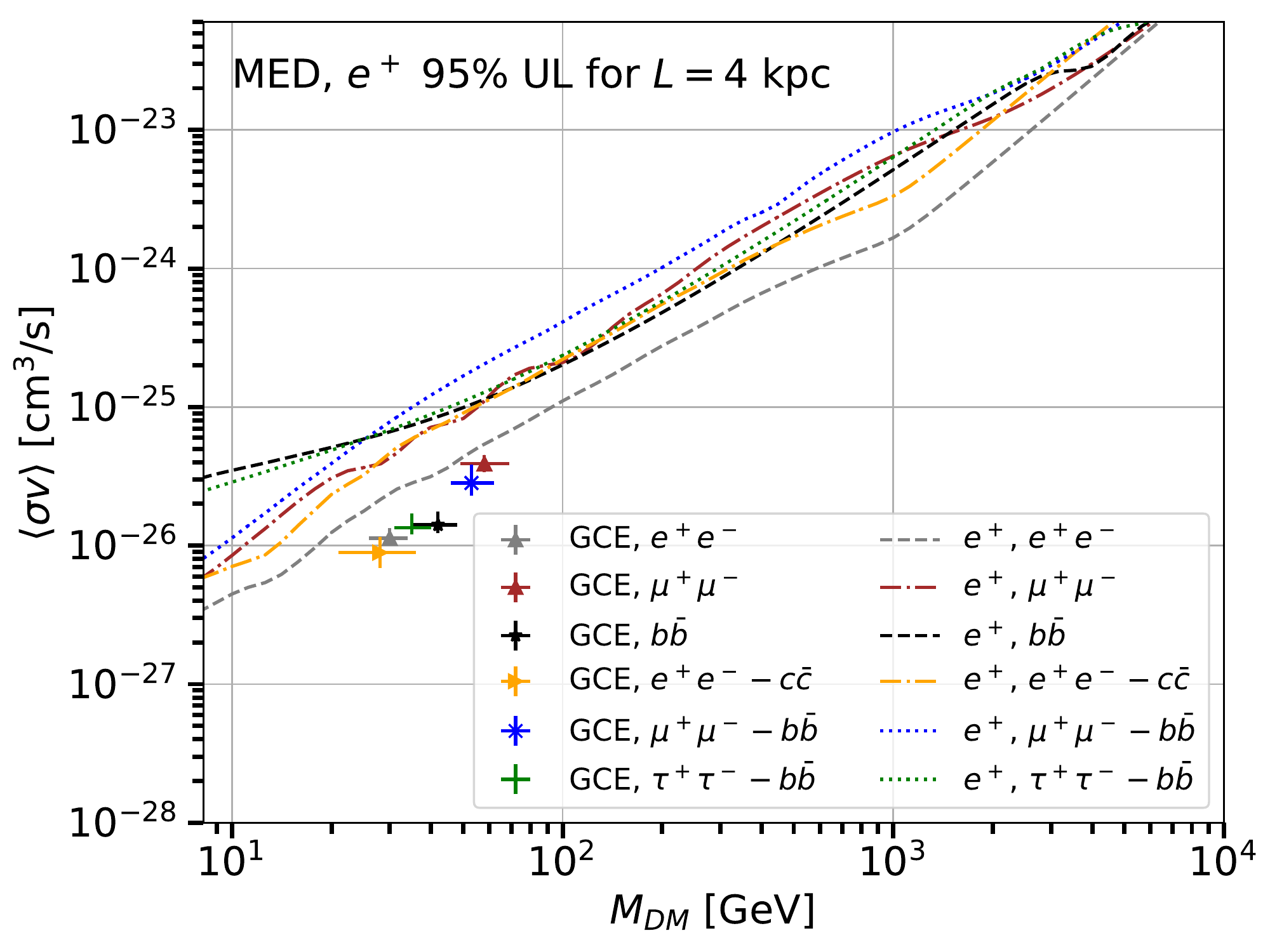}
\includegraphics[width=0.49\textwidth]{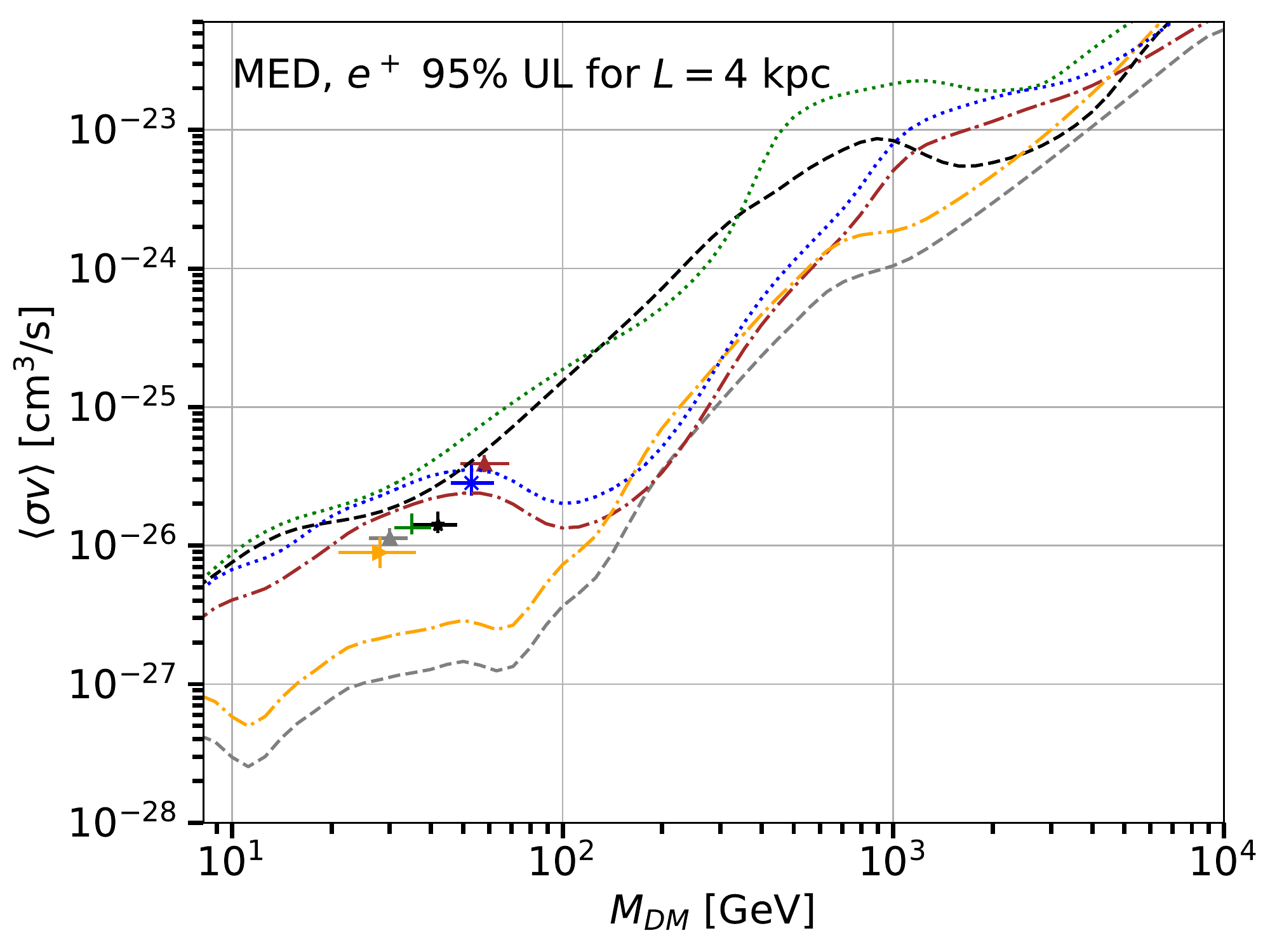}
\includegraphics[width=0.49\textwidth]{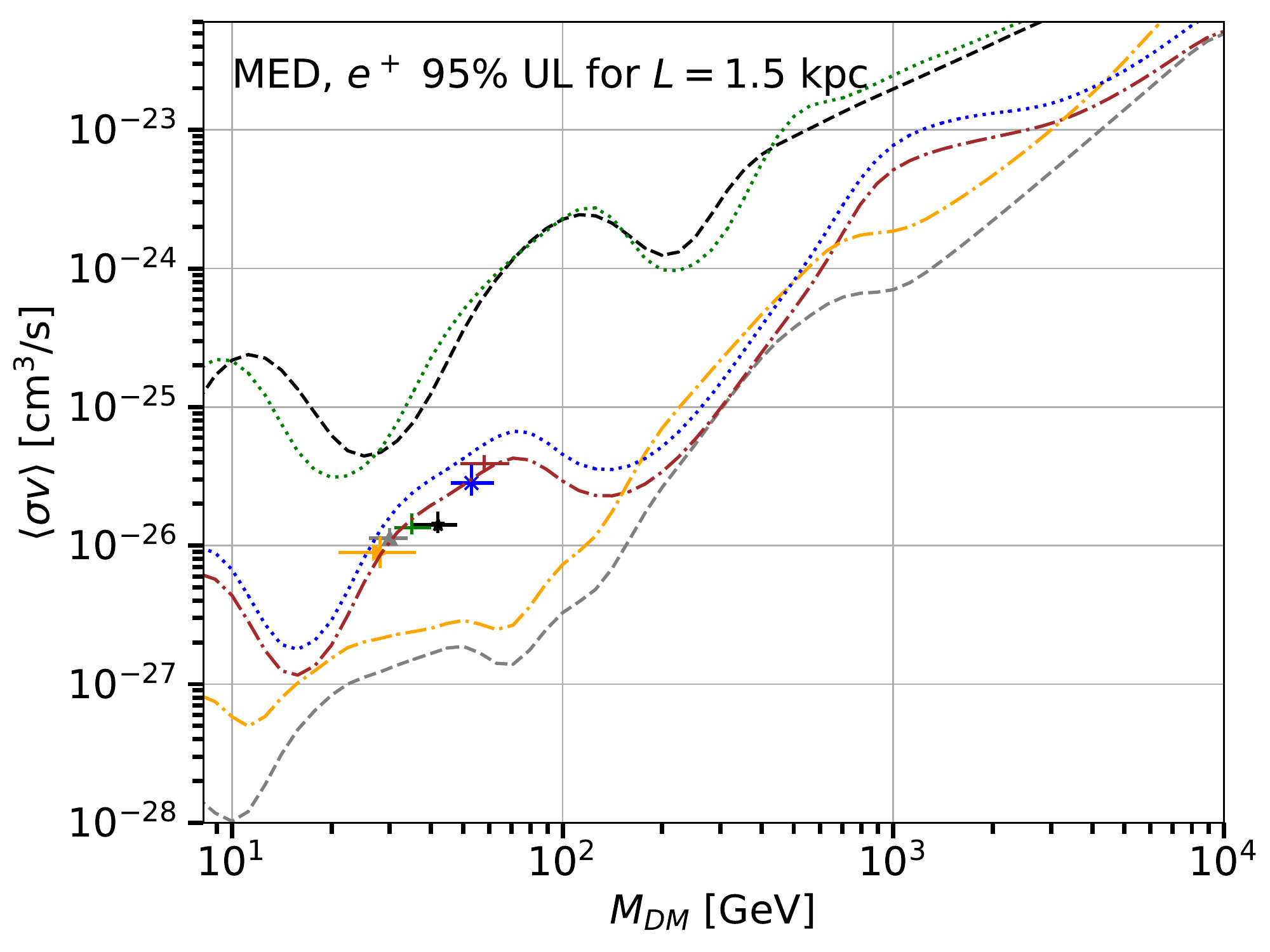}
\caption{Best-fit values for the DM parameters $M_{\rm{DM}}$ and $\langle \sigma v \rangle$ that we find by fitting the GCE SED. We show the cases that best-fit the GCE SED from Sec.~\ref{sec:fitGCE}. We also report the $95\%$ CL upper limits we obtain from $e^+$ flux data for the same DM candidates with the conservative (upper panel) and optimistic methods (central and bottom panels). We assume the {\tt MED} DM density model and $L=4$ kpc for the first two panels and $L=1.5$ kpc for the bottom panel.}
\label{fig:GCEcandidatespos}
\end{figure}

The upper limits that we find with the {\it conservative} and the {\it optimistic approach} are shown in Fig.~\ref{fig:GCEcandidatespos} compared to the best fit of $\langle \sigma v \rangle$ and $M_{\rm{DM}}$ we obtain by fitting the GCE SED.
The constraints we calculate for the channel $e^+e^- - b\bar{b}$ are very similar to the ones obtained for $e^+e^- - c\bar{c}$.
The constraints obtained with the {\it conservative approach} are compatible with the GCE best fit for all tested cases. As expected the DM annihilation channel with the strongest $\langle \sigma v \rangle$ upper limit is the $e^+e^-$ one.
Instead, the results for the {\it optimistic approach} are compatible with the GCE best fit for most single and mixed channels except for the ones with full or partial annihilation into $e^+ e^-$.
In fact, the GCE candidates annihilating into $e^+e^-$, $e^+e^- - b\bar{b}$ or $e^+e^- - c\bar{c}$ have a cross section one order of magnitude higher than allowed by the {\it optimistic} $e^+$ limits. These conclusions do not change if we employ a lower value of the vertical size of the diffusion halo $L=1.5$ kpc as shown in the bottom panel of Fig.~\ref{fig:GCEcandidatespos}.
By using a model for the astrophysical background of $e^+$ given by a refined calculation of the secondary production, tuned on the newest cross section data, and synthetic population of pulsars that account properly for the PWN flux, the upper limits for $\langle \sigma v \rangle$ are expected to be between the ones obtained with the {\it conservative} and the {\it optimistic approach}. Therefore, the tension between any GCE DM channel with an $e^+e^-$ contribution and the AMS-02 $e^+$ data is expected to persist even in such a more complete approach. However, since the {\it optimistic } $e^+$ constraints even for $L=4\:\text{kpc}$ only marginally rule out the dark matter interpretation of the GCE in the $\mu^+\mu^-$-channel, we expect that a more refined analysis with proper modeling of uncertainties will reconcile this channel with $e^+$ constraints.

\section{Conclusions}
\label{sec:conclusions}

In this paper we have shown that the characteristics of the GCE make DM particles annihilating into the Galactic halo of the Milky Way a viable interpretation for explaining the excess.
In fact, the GCE spatial morphology is energy independent and compatible with a NFW profile with $\gamma\sim 1.2-1.3$.
Moreover, the GCE is roughly spherically symmetric and its centroid is located very close to the dynamical center of the Galaxy as expected for DM.
The GCE SED around the peak at a few GeV can be well fitted using a single DM annihilation channel with light quarks, $c\bar{c}$, $b\bar{b}$ or the leptonic channels $e^+ e^-$, $\mu^+\mu^-$ with masses from 20 to 60 GeV and cross sections close to the thermal one.
We demonstrated that the fit to the GCE SED improves significantly in the entire energy range by assuming annihilation into two channels with the best cases that are $\mu^+\mu^- - b\bar{b}$, $\tau^+\tau^- - b\bar{b}$, $e^+e^- - b\bar{b}$, $e^+e^- - c\bar{c}$. We have calculated in the paper the relevant coupling parameters (mass, annihilation cross section and branching ratio) for each of these cases.

Then, we have searched for a cumulative $\gamma$-ray signal in {\it Fermi}-LAT data compatible with DM particles annihilating in the direction of dSphs. We have performed a combined likelihood analysis of LAT data above 0.3 GeV in which we have fully accounted for the uncertainty on the DM density using the information published in Ref.~\cite{2019MNRAS.482.3480P} for 48 dSphs. Since we did not find any significant signal we put upper limits for $\langle \sigma v \rangle$ that are below the thermal cross section up to almost 100 GeV for the $b\bar{b}$ annihilation channel. 
For the first time we tested in a dSphs analysis DM candidates annihilating into two and three channels following the best-fit cases from the fit to the GCE SED.
The upper limits on $\langle \sigma v \rangle$ are compatible with the DM interpretation of the GCE considering the uncertainties present in the DM density distribution.

Following a multimessenger approach we have searched for a possible DM signal also using the recently released 7 years $\bar{p}$ and $e^+$ AMS-02 flux data.
These are among the rarest CRs in the Galaxy and have been widely used in the past as promising cosmic particles for the indirect search for DM.
First, we analyzed $\bar{p}$ data accounting for the uncertainties in the CR propagation, uncertainties in the $\bar{p}$ production cross section and the correlation between AMS-02 data points. Since we did not find any significant preference for a DM contribution we put upper limits for $\langle \sigma v \rangle$. The $\bar{p}$ constraints exclude all GCE DM candidates reported above with hadronic or semi-hadronic final states unless if the vertical size of the diffusive halo is $L<2$ kpc. This value for the vertical diffusive halo size is $2-3\sigma$ below the best fit value obtained in Ref.~\cite{Weinrich:2020ftb} using the latest AMS-02 data on radioactive CRs (see also~\cite{Luque:2021joz}). Moreover, these small values for $L$ are in tension with complementary astrophysical probes, in particular with radio data~\cite{Bringmann:2011py,DiBernardo:2012zu,Orlando:2013ysa}. 
We also showed that variations of the DM density profile cannot reconcile the GCE DM interpretation with $\bar{p}$ constraints.
Instead, pure leptonic channels are compatible with the $\bar{p}$ upper limits regardless of the value of $L$ and the assumed DM density.
Finally, we have calculated upper limits for a DM contribution from the $e^+$ spectrum following a {\it conservative approach} where only secondary $e^+$ were included as background and an {\it optimistic} one where the $e^+$ background is modeled by an analytic function (in order to also include a potential pulsar contribution). In case of the {\it conservative approach} $e^+$ do not provide any further constraints on the DM interpretation of the GCE. In the {\it optimistic approach} all mentioned GCE DM candidates which annihilate purely or partially into $e^+ e^-$ are ruled out.

To conclude DM particles annihilating into $\mu^+ \mu^-$ with a mass of about 60 GeV and a cross section of $4\times 10^{-26}$ cm$^3$/s, which is close to the thermal one, could fit the GCE spectrum. At the same time they are compatible with observations of dwarf spheroidal galaxies and 
would produce a flux of $\bar{p}$ and $e^+$ compatible with the upper limits calculated with the latest AMS-02 data. All other DM annihilation channels we investigated for the GCE are in tension with CR data once we include the latest constraints on the size of the CR diffusion zone. In particular, the two-channel final state $\mu^+\mu^- - b\bar{b}$ ($\tau^+\tau^- - b\bar{b}$) with $M_{\rm{DM}}\sim 50$ (35), $\langle \sigma v \rangle \sim 3\times 10^{-26}$ ($\sim 1.4\times 10^{-26}$) cm$^3$/s and $Br\sim 0.7$ (0.2) would improve the fit to the GCE spectrum, with respect to the $\mu^+ \mu^-$ channel, but is compatible with the $\bar{p}$ upper limits only for an unfavorably small diffusion zone.

\begin{acknowledgments}
MDM research is supported by Fellini - Fellowship for Innovation at INFN, funded by the European Union’s Horizon 2020 research programme under the Marie Skłodowska-Curie Cofund Action, grant agreement no.~754496. MWW\ acknowledges support by the Swedish Research Council (Contract No. 638-2013-8993) and by the Department of Physics of the University of Texas at Austin.
MDM acknowledges support by the NASA Fermi Guest Investigator Program Cycle 12 through the Fermi Program N. 121119 (P.I.~MDM). 
The authors thank Regina Caputo, Judith L. Racusin, Miguel A. Sanchez-Conde, Michael Gustafsson for providing us comments on the paper.

The {\it Fermi} LAT Collaboration acknowledges generous ongoing support from a number of agencies and institutes that have supported both the development and the operation of the LAT as well as scientific data analysis. These include the National Aeronautics and Space Administration and the Department of Energy in the United States, the Commissariat\'a l'Energie Atomique and the Centre National de la Recherche Scientifique / Institut National de Physique Nucl\'eaire et de Physique des Particules in France, the Agenzia Spaziale Italiana and the Istituto Nazionale di Fisica Nucleare in Italy, the Ministry of Education, Culture, Sports, Science and Technology (MEXT), High Energy Accelerator Research Organization (KEK) and Japan Aerospace Exploration Agency (JAXA) in Japan, and the K. A. Wallenberg Foundation, the Swedish Research Council and the Swedish National Space Board in Sweden.
Additional support for science analysis during the operations phase is gratefully acknowledged from the Istituto Nazionale di Astrofisica in Italy and the Centre National d'Etudes Spatiales in France. This work performed in part under DOE Contract DE- AC02-76SF00515.
\end{acknowledgments}

\bibliography{paper}

\end{document}